\documentclass[a4paper,11pt,openbib,oneside]{memoir}

\usepackage{graphicx}
\usepackage{subcaption}
\usepackage{hyperref}
\usepackage{amsmath,amsfonts,amssymb,amsthm}
\usepackage{datetime}
\usepackage{gensymb}
\usepackage{algorithm,algpseudocode}
\usepackage{adjustbox}
\usepackage{enumitem}
\usepackage{glossaries}
\usepackage{amsmath}
\usepackage{amsfonts}
\usepackage{amssymb}
\usepackage{tikz}
\usetikzlibrary{arrows,shapes,snakes,automata,backgrounds,petri,topaths}

\setsecnumdepth{subsection}
\maxsecnumdepth{subsubsection}

\usepackage{fouriernc}
\usepackage[T1]{fontenc}

\OnehalfSpacing

\makepagestyle{myvf} 
\makeoddfoot{myvf}{}{\thepage}{}
\makeevenfoot{myvf}{\thepage}{}{}
\makeheadrule{myvf}{\textwidth}{\normalrulethickness}
\makeevenhead{myvf}{\small\textsc{\leftmark}}{}{}
\makeoddhead{myvf}{}{}{\small\textsc{\rightmark}}
\settrimmedsize{297mm}{210mm}{*}
\settypeblocksize{696.24pt}{451.44pt}{*} 
\setulmargins{1in}{*}{*} 
\setlrmargins{*}{*}{1} 
\setmarginnotes{17pt}{51pt}{\onelineskip} 
\setheadfoot{\onelineskip}{2\onelineskip} 
\setheaderspaces{*}{\onelineskip}{*} 
\checkandfixthelayout

\begin{document}

\frontmatter

\begin{titlingpage}
\begin{SingleSpace}
\calccentering{\unitlength} 
\begin{adjustwidth*}{\unitlength}{-\unitlength}
\begin{center}

\textbf{\LARGE Performance Analysis of 6G Multiuser Massive MIMO-OFDM THz Wireless Systems with Hybrid Beamforming under Intercarrier Interference}

\vspace{20mm}
\includegraphics[scale=0.1]{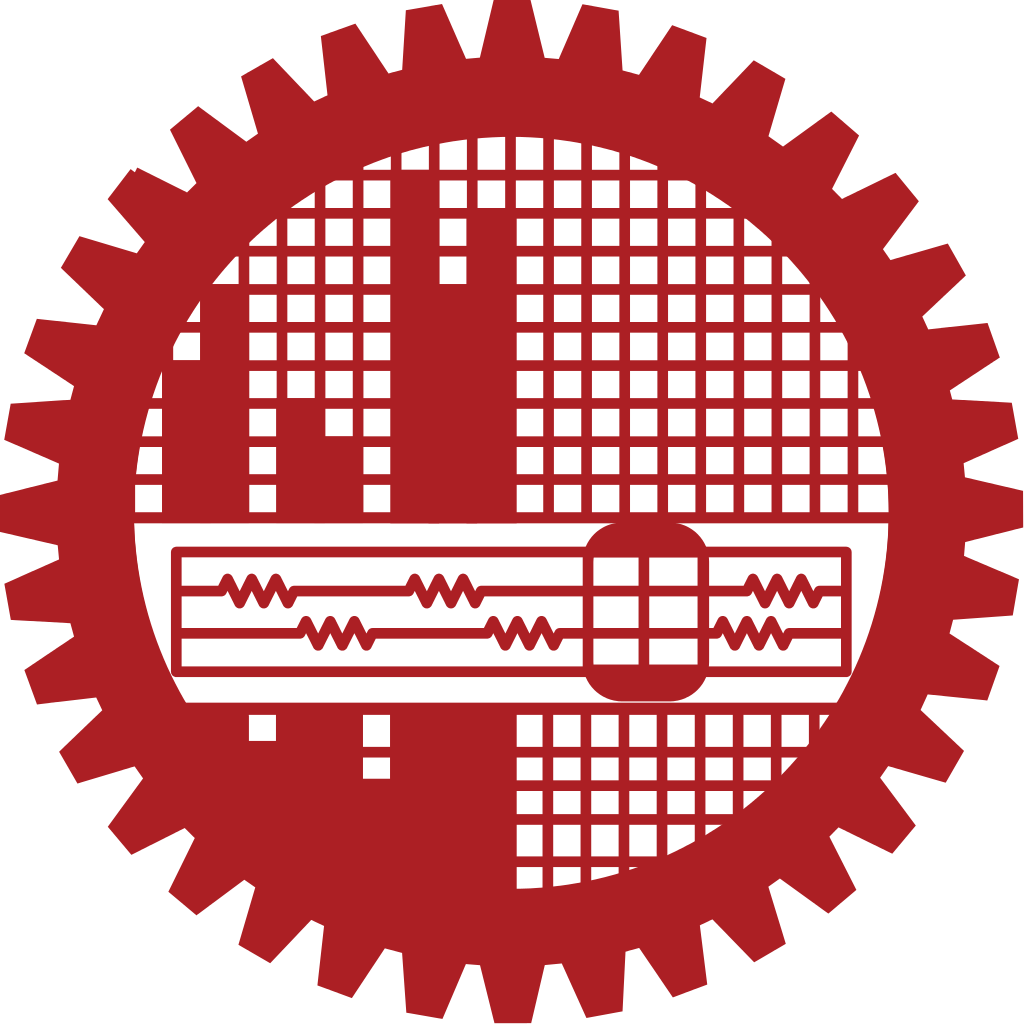}\\
\vspace{12mm}
{\large \textbf{Course No.: EEE 400}}\\
\vspace{12mm}
{\large \textbf{Submitted By}}\\
\vspace{5mm}
{\large{Md Saheed Ullah}}\\
{\large{saheed@udel.edu}}\\
\vspace{1mm}
{\large{Zulqarnain Bin Ashraf}}\\
{\large{zulqarnainbinashraf@gmail.com}}\\
\vspace{1mm}
{\large{Sudipta Chandra Sarker }}\\
{\large{sudiptabkhs4@gmail.com}}\\
\vspace{12mm}
{\large \textbf{Submitted To} }\\
\vspace{3mm}
{\large {Department of Electrical and Electronic Engineering}}\\
{\large {in partial fulfillment of the requirements for the degree of
Bachelor of Science in Electrical and Electronic Engineering.}}\\
\vspace{12mm}
{\large \textbf{Supervised by} }\\
\vspace{3mm}
{\large{Dr. Md. Forkan Uddin}}\\
{\large {Professor}}\\
{\large {Department of Electrical and Electronic Engineering}}\\

\large {Bangladesh University of Engineering and Technology (BUET)}\\
{\large Dhaka-1205, Bangladesh}\\
\vspace{6mm}

{\large May, 2022}\\
\end{center}
\end{adjustwidth*}
\end{SingleSpace}
\end{titlingpage}
%
%
%
%
%
%
\chapter*{Certification}
\begin{SingleSpace}
\large {This is to certify that the work presented in the thesis is an outcome of the investigation carried out by the authors under the supervision of Professor Dr. Md. Forkan Uddin, Department of Electrical and Electronic Engineering, Bangladesh University of Engineering and Technology (BUET), Dhaka. It is declared that this thesis has been submitted only for the award of graduation.} 
\vspace{1cm}
\begin{center}
\large{Authors}
\end{center}
\vspace{15mm}
\begin{minipage}{4cm}
    \centering
    \line(1,0){120}\\
    \large {Md Saheed Ullah}\\
    
\end{minipage}
\hspace{1.5cm}
\begin{minipage}{5cm}
    \centering
    \line(1,0){150}\\
    \large {Zulqarnain Bin Ashraf}\\
    
\end{minipage}
\hspace{1.5cm}
\begin{minipage}{5cm}
    \centering
    \line(1,0){150}\\
    \large {Sudipta Chandra Sarker}\\
    
\end{minipage}
\vspace{40mm}
\begin{center}
\large{Signature of the Supervisor}\\
\vspace{10mm}
\line(1,0){150}\\
\large {Dr. Md. Forkan Uddin}\\
\vspace{2mm}
\large {Professor}\\
\large {Department of Electrical and Electronic Engineering }\\
\large {Bangladesh University of Engineering and Technology (BUET) }\\
\end{center}
%

\end{SingleSpace}
\chapter*{Dedication}

\begin{quote}
    \vspace{2\baselineskip}
    \usefont{\encodingdefault}{pzc}{m}{n}
    \begin{center}
    \Large To the treasured people of our country, 
    
    with the hope that this endeavor would one day be to their well-being.\\
    To our beloved parents. Your endless support and encouragement will never go unnoticed.\\
    And to true friends. May we all find a few!
    \end{center}
\end{quote}

\clearpage

%
%

\chapter*{Acknowledgments}
We would like to express our deepest gratitude to our supervisor Dr. Md. Forkan Uddin for his guidance on this project, showing us the path of conducting successful research and above all, for always being there as our mentor. He shared his wisdom with us in analyzing subject matters and, at the same time, valued our thinking approach to synthesizing those topics. His suggestions drove us towards better ways of thinking, his reviews enriched us in solving problems, and his support gave us strength at the time of our disappointment. We shall forever cherish the memories of working with him. 

We want to thank the board members of our thesis committee, Dr. Md. Farhad Hossain and Dr. Lutfa Akter for their invaluable time and constructive suggestion on our work. We are grateful to all of our teachers, as well as our friends, classmates, and seniors, for their valuable input and consistent encouragement. Last but not least, we would like to express our gratitude to our families for their unwavering love and unending support.
\clearpage
\chapter*{Abstract}
6G networks are expected to provide more diverse capabilities than their predecessors and are likely to support applications beyond current mobile applications, such as virtual and augmented reality (VR/AR), AI, and the Internet of Things (IoT). In addition, it is anticipated that mobile network operators will employ flexible, decentralized approaches for 6G, including local spectrum licensing, spectrum sharing, and infrastructure sharing. 6G will utilize all cutting-edge technologies, including MIMO, OFDM, THz, and Hybrid Beamforming. THz communication has been hailed as a significant enabler for 6G networks, offering orders of magnitude more spectrum than existing cellular bands. In contrast to typical multiple-input multiple-output (MIMO) systems, THz MIMO precoding cannot be conducted totally at baseband using digital precoders due to the restricted number of signal mixers and analog-to-digital converters that can be supported due to their cost and power consumption. A hybrid precoding transceiver design, combining a digital precoder with an analog precoder, has recently attracted substantial interest as a cost-effective option, particularly when massive MIMO is proposed for 6G. In this thesis, we analyzed the performance of multiuser massive MIMO-OFDM THz wireless systems with hybrid beamforming. Carrier frequency offset (CFO) is one of the most well-known disturbances for OFDM. For practicality, we accounted for CFO, which results in Intercarrier Interference. For the terahertz (THz) wireless system, a multi-antenna base station (BS) uses a frequency-selective fading channel to connect with a single-antenna user. We assumed that the base station implements fully-connected hybrid beamforming and multi-carrier modulation to provide ultra-high data rates. Incorporating the combined impact of molecular absorption, high sparsity, and multi-path fading, we analyzed a three-dimensional wideband THz channel and the carrier frequency offset in multi-carrier systems. With this model, we first presented a two-stage wideband hybrid beamforming technique comprising Riemannian manifolds optimization for analog beamforming and then a zero-forcing (ZF) approach for digital beamforming. We adjusted the objective function to reduce complexity, and instead of maximizing the bit rate, we determined parameters by minimizing interference. Numerical results demonstrate the significance of considering ICI for practical implementation for the THz system. We demonstrated how our change in problem formulation minimizes latency without compromising results. We also evaluated spectral efficiency by varying the number of RF chains and antennas. The spectral efficiency grows as the number of RF chains and antennas increases, but the spectral efficiency of antennas declines when the number of users increases.
\\

\clearpage

\tableofcontents*
\listoftables
\listoffigures

\printglossary

\mainmatter

\chapter{Introduction} \label{intro}

\section{The Road Towards 6G: An Overview on Evolution of Mobile Wireless Communication Networks}
The exponential growth of connected devices and the growing demand for high-speed services have been the driving factors for the evolution of wireless technologies. Since 1980, approximately every ten years, a new generation of wireless cellular communication has emerged, including the first generation analog FM cellular systems in 1981, the second generation in 1992, the third generation (3G) in 2001, the fourth generation (4G) (often referred to as long-term evolution [LTE]) in 2011, and the fifth generation (5G) in 2020 \cite{salih2020evolution}. According to the Ericsson Mobility Report, global mobile data traffic, excluding traffic generated by fixed wireless access (FWA), is expected to reach 288EB per month in 2027. By 2027, overall mobile network traffic will reach approximately 370EB per month, including FWA traffic. By the end of 2027, the average monthly global smartphone consumption is expected to reach 41GB \cite{ericsson_mobility_report_2020}.
\begin{figure}[h]
  \centerline
  {\includegraphics[width=\textwidth]{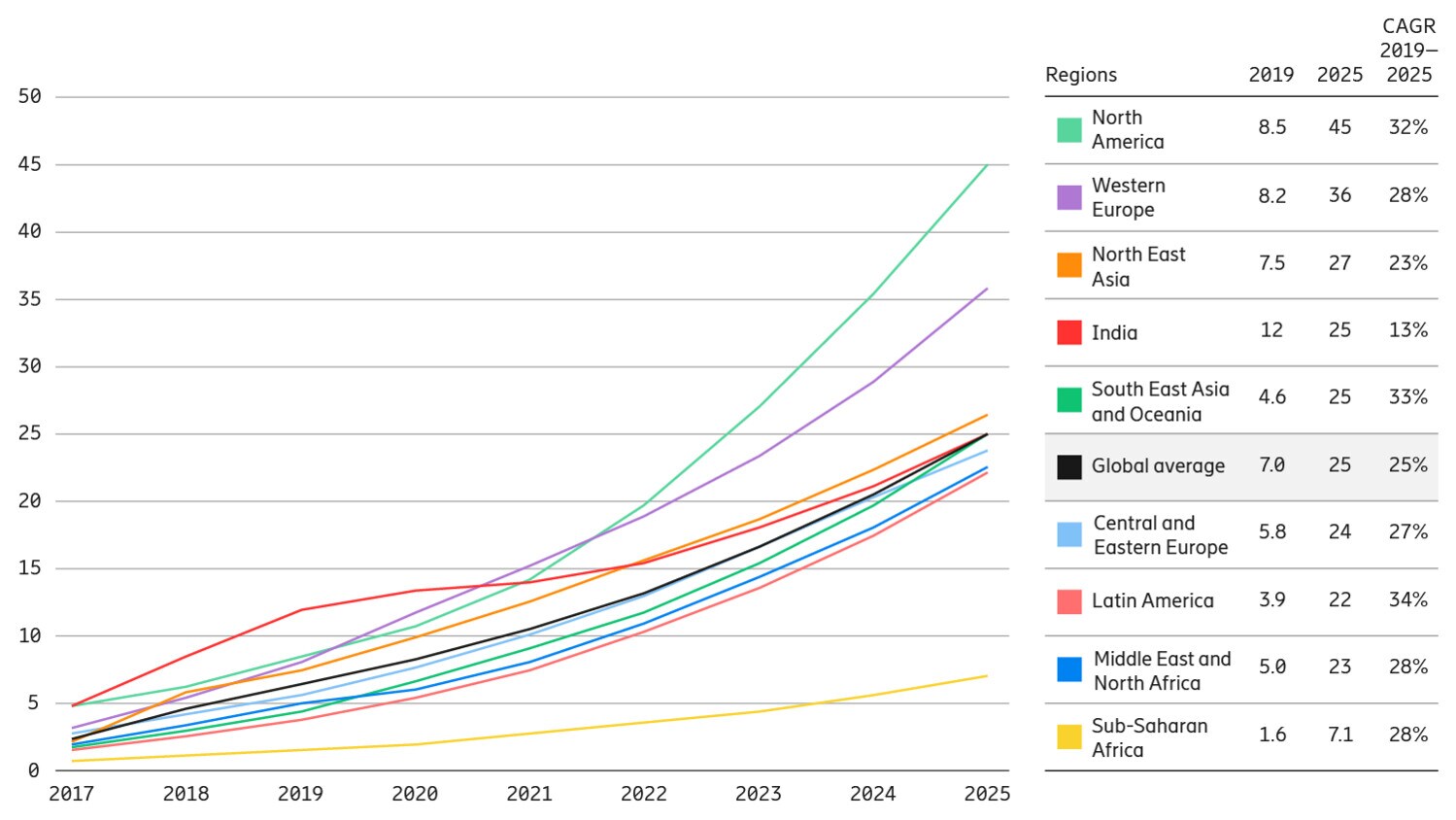}}
  \caption[Mobile data traffic per smartphone (GB per month)]{Mobile data traffic per smartphone (GB per month) \cite{ericsson2020ericsson}}
\end{figure}

The remarkable upsurge of Internet of Everything (IoE)-based smart applications has paved the way for the evolution of existing wireless networks. The emergence of the Internet-of-Everything (IoE), Extended Reality, Connected Autonomous Systems, Telemedicine, and Industrial Internet of Things (IIoT) enable seamless connectivity of smart homes, smart cities, multisensory virtual experiences and e-health applications by connecting billions of people and devices via a unified communication interface. Besides generating massive data, the upsurge of IoE will naturally give rise to a myriad of new traffic and data service types, leading to diverse communication requirements. This grand vision requires a radical departure from the conventional 'one-size-fits-all' network model of fourth-generation systems.

\begin{figure}[h]
  \centerline
  {\includegraphics[width=\textwidth]{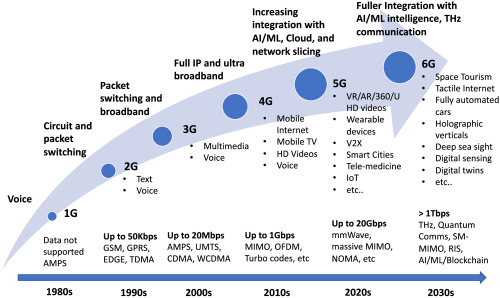}}
  \caption[Evolution of Wireless Mobile Generations]{Evolution of Wireless Mobile Generations \cite{ranaweera20224g}.} 
\end{figure}

The fifth generation (5G) of wireless technology is a technological leap forward, providing a better quality of service (QoS) than preceding communication networks \cite{shafi20175g, zhang2016one,jaber20165g,andrews2014will}. Along with reducing latency, improving connection and reliability, and enabling gigabit-per-second speeds, 5G is expected to enable a range of service kinds, including new frequency bands (e.g., millimeter-wave (mmWave) and optical spectra), advanced spectrum usage and management, and the simultaneous integration of licensed and unlicensed bands - which are frequently characterized by conflicting requirements and diverse sets of key performance indicators \cite{tsiropoulos2017cooperation}. These features position 5G as a critical enabler for Internet-of-Things (IoT) application environments in which machine-to-people (M2P)  communications (e.g., industrial automation, smart cities, and intelligent mobility) and machine-to-machine (M2M) communications (e.g., autonomous communications between sensors and actuators) are expected to coexist with people-to-people communications (e.g., voice over internet protocol (IP), video conferencing, video streaming, and web browsing).
\begin{figure}[h]
  \centerline
  {\includegraphics[width=0.9\textwidth]{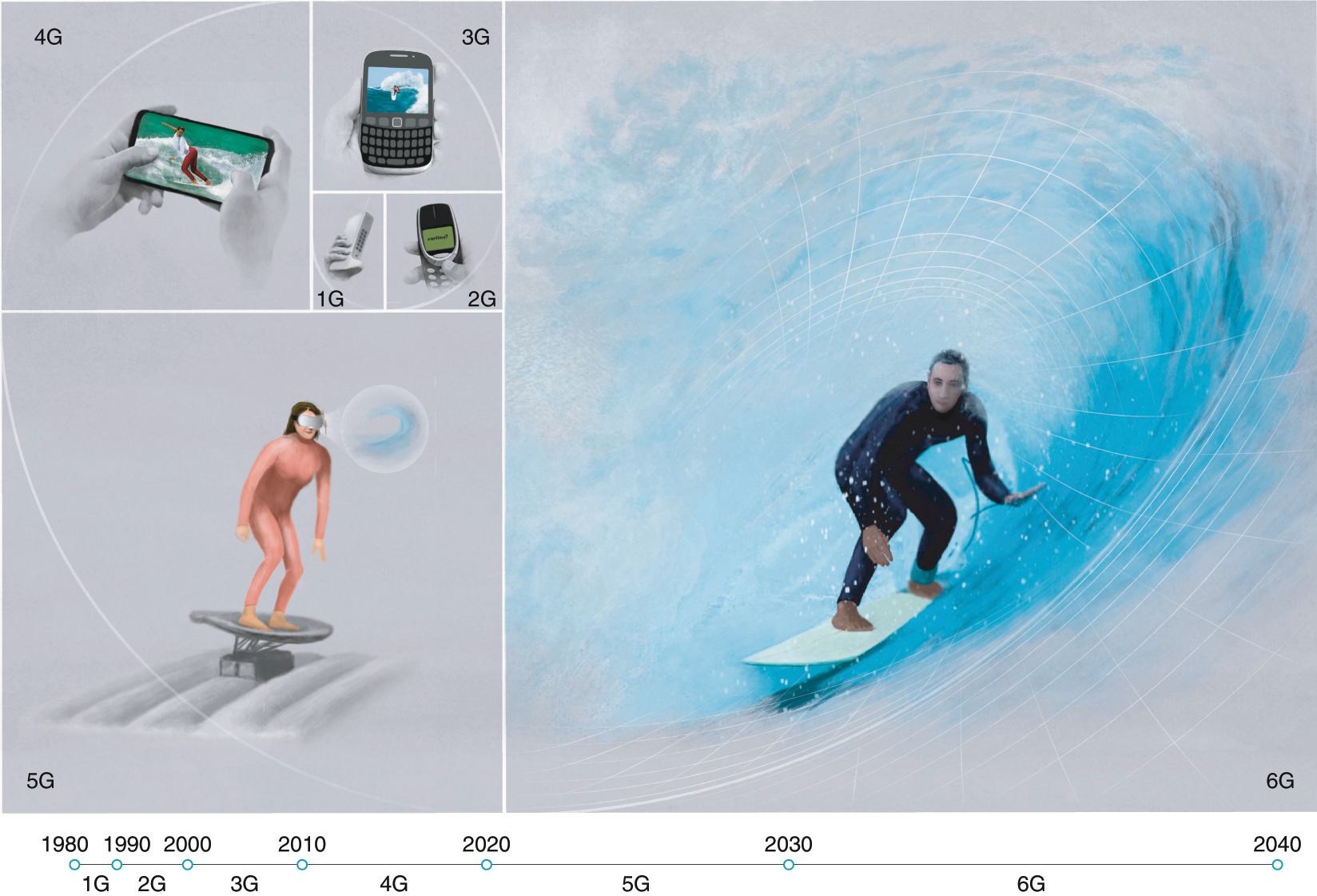}}
  
\caption[A user's perception of the different communications networks, from 1G to the hypothetical 6G. In 1G and 2G, voice and text are available. In 3G and 4G, pictures and videos become commonplace. In 5G, live ultra-high-definition three-dimensional data can be employed. In 6G, it is expected that we could have a ubiquitous virtual existence]{A user's perception of the different communications networks, from 1G to the hypothetical 6G. In 1G and 2G, voice and text are available. In 3G and 4G, pictures and video become commonplace. In 5G, live ultra-high-definition three-dimensional data can be
employed. In 6G, it is expected that we could have a ubiquitous virtual existence \cite{dang2020should}.} 
\end{figure}

5G networks will deliver an extensive variety of services comprising enhanced mobile broadband (eMBB), ultra-reliable and low-latency communications (uRLLC), and massive machine type communications (mMTC) \cite{patzold20195g}. However, it is debatable if they will be able to fulfill the demands of future emerging intelligent and automation systems after ten years \cite{giordani2020toward}. The volume of wireless data traffic and the magnitude of connected things are predicted to double by a factor of a hundred in a cubic meter. Additionally, data-hungry applications such as sending holographic videos require a bandwidth that the mm-wave spectrum lacks. This situation poses significant issues in terms of area or spatial spectral efficiency, as well as the required frequency ranges for connectivity. As a result, a broader radio frequency spectrum bandwidth has been necessary, which can be found exclusively in the sub-terahertz and terahertz bands. Moreover, the recent upsurge of diversified mobile applications, especially those supported by Artificial Intelligence (AI) technology, is spurring heated discussions on the future evolution of wireless communications \cite{alsharif2020machine}. Emerging IoE services will necessitate an end-to-end co-design of communication, control, and computing features, which has largely been disregarded. These challenges have prompted industry and academia to start conceptualizing the next generation of wireless communication systems (6G), intending to meet future communication demands in the 2030s while also ensuring the systems' sustainability and competitiveness \cite{letaief2019roadmap}. Thus, 6G is expected to provide a broader coverage that enables subscribers to communicate with one another anywhere at a high data rate speed, owing to the unconventional technologies adopted by 6G communication systems, such as extremely large bandwidth (THz waves) and sophisticated artificial intelligence (AI) that encompasses operational and environmental aspects, as well as network services. Table 1.1 compares the main specifications and technologies in 5G and 6G \cite{elmeadawy20196g}.

\begin{table}[hbt]
\caption{Comparison between 5G and 6G}\label{my-label}
\centering
\begin{tabular}{|c|| c| c|} 
 \hline
 \textbf{Characteristic} & \textbf{5G} & \textbf{6G} \\ [0.5ex] 
 \hline\hline
 \textbf{Operating frequency} & $3 - 300$ GHz & up to $1$ THz \\[0.25ex] 
 \hline
 \textbf{Uplink data rate} & $10$ Gbps & $1$ Tbps \\[0.25ex]
 \hline
 \textbf{Downlink data rate} & $20$ Gbps & $1$ Tbps \\[0.25ex]
 \hline
 \textbf{Spectral efficiency} & $10 \, \text{bps/Hz/m}^2$  & $1000 \, \text{bps/Hz/m}^2$ \\[0.25ex]
 \hline
\textbf{Reliability} & $10^{-5}$ & $10^{-9}$ \\[0.25ex]
 \hline
 \textbf{Maximum mobility} & $500$ km/h & $1000$ km/h \\ [0.25ex]
 \hline
 \textbf{U-plane latency} & $0.5$ msec & $0.1$ msec \\[0.25ex]
 \hline
 \textbf{C-plane latency} & $10$ msec & $1$ msec \\[0.25ex]
 \hline
 \textbf{Processing delay} & $100$ ns & $10$ ns \\[0.25ex]
 \hline
\textbf{Traffic capacity} & $10$ Mbps/m$^2$ & $1-10$ Gbps/m$^2$ \\[0.25ex]
 \hline
 \textbf{Localization precision} & $10$ cm on 2D & $1$ cm on 3D \\[0.25ex]
 \hline
 \textbf{Uniform user experience} & $50$ Mbps 2D & $10$ Gbps 3D \\[0.25ex]
 \hline
\textbf{Time buffer} & not real-time & real-time \\[0.25ex]
 \hline
 \textbf{Center of gravity} & user & service \\ [0.25ex]
 \hline
 \textbf{Satellite integration} & No & Fully \\[0.25ex]
 \hline
 \textbf{AI integration} & Partially & Fully \\[0.25ex]
 \hline
 \textbf{XR integration} & Partially & Fully \\[0.25ex]
 \hline
 \textbf{Haptic communication integration} & Partially & Fully \\[0.25ex]
 \hline
\textbf{Automation integration} & Partially & Fully \\ [1ex]
 \hline
\end{tabular}
\end{table}

\section{6G Networks: Vision, Requirements, Key Technologies, and Applications}

5G marked a significant advance in developing low-latency tactile access networks by introducing new wireless pathways through innovative techniques. These included the use of new frequency bands (e.g., millimeter-wave (mmWave) and optical spectra), advanced spectrum utilization and management, integration of licensed and unlicensed bands, and a complete redesign of the core network. However, as data-centric and automated processes rapidly evolve, demanding terabit-per-second data rates, latency in the hundreds of microseconds, and 107 connections per $km^2$, there's a realization that even the capabilities of emerging 5G systems may be surpassed\cite{9733558}. 

5G communication notably overlooked the convergence of communication, intelligence, sensing, control, and computing functionalities. This convergence is imperative for future Internet of Everything (IoE) applications. Specific devices, such as Virtual Reality (VR), need capabilities beyond 5G (B5G) due to their requirement for a minimum of 10 Gbps data rates \cite{mumtaz2017terahertz}. With 5G reaching its limits by 2030, the design goals for its successor are already explored in the literature, gradually transforming 5G into an enhanced Beyond-5G or the sixth generation (6G) system.

6G is inherently designed to meet the performance requirements of next-generation connectivity, evolving from connected everything to connected intelligence, enabling "Human-Thing-Intelligence" inter-connectivity. Additionally, it will support high-precision communications for tactile and haptic applications, providing sensory experiences including smell, touch, vision, and hearing \cite{8613209}. 

\subsection{Technical Requirements and Key Performance Metrics}
The key technical requirements characterizing the 6G communication system include \cite{9178307}:

\begin{itemize}
 \item \textbf{High Energy Efficiency:} Offering extreme data rates to address massive-scale connectivity and provide ultra-high throughput, even in extreme or emergency scenarios with varying device densities, spectrum and infrastructure availability, and traffic patterns.
 \item \textbf{Low Backhaul and Access Network Congestion:} Achieving the quality of immersion and per-user capacity required by Augmented Reality (AR) and Virtual Reality (VR) applications, impacting retail, tourism, education, etc.
 \item \textbf{Tactile Internet:} Delivering real-time tactile feedback with sub-millisecond (ms) latency to fulfill the requirements of haptic applications, such as e-health.
 \item \textbf{AI Integrated Communication:} Incorporating artificial intelligence (AI) to support seamless data-centric context-aware communications for controlling environments like smart structures, autonomous transportation systems, and smart industry \cite{peltonen20206g}.
 \item \textbf{Enhanced Data Security:} Meeting the extremely high levels of communication reliability and low end-to-end latency to support ultra-high mobility scenarios, such as flying vehicles.
\end{itemize}

\begin{figure}
  \centerline
  {\includegraphics[scale=0.6,width=\textwidth]{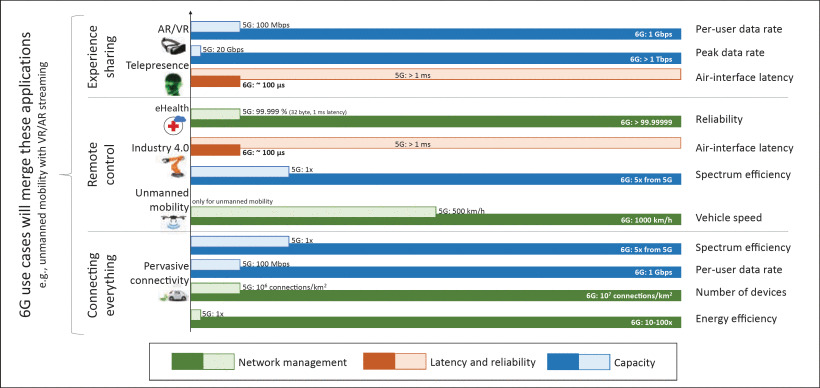}}
 \caption[Representation of multiple KPIs of 6G use cases, together with the improvements with respect to 5G networks]{Representation of multiple KPIs of 6G use cases, together with the improvements with respect to 5G networks \cite{9040264}.} 
\end{figure}

The 6G communication systems are expected to be featured by the following types of Key Performance Factors (KPI) associated services \cite{8579209}:
\begin{itemize}
\item Ubiquitous mobile ultra-broadband (uMUB)
\item Ultra-high-speed with low-latency communications (uHSLLC)
\item Massive machine-type communication (mMTC)
\item Ultra-high data density (uHDD)
\end{itemize}

The 6G system is expected to offer 1000 times the simultaneous wireless connectivity of the 5G system. In comparison to 5G's enhanced mobile broadband (eMBB), 6G is intended to incorporate ubiquitous services or uMUB. Ultra-reliable low-latency communications, a key trait of 5G, will be an essential driver in 6G communication providing uHSLLC by adding features such as E2E delay of less than 1 ms \cite{tariq2020speculative}, more than 99.99999\% reliability, and 1 Tbps peak data rate. Massively connected devices (up to 10 million/$km^{2}$) will be provided in the 6G communication system. 6G is projected to give Gbps coverage everywhere, including novel environments like the sky (10,000 kilometers) and the sea (20 nautical miles) \cite{nakamura20205g}. As opposed to the often-used area spectral efficiency, volume spectral efficiency will be much better in 6G. The 6G system will feature ultra-long battery life and advanced energy harvesting technology. Mobile devices will no longer require separate charging in 6G systems.

\subsection{Prospects and Applications of 6G}
5G presents latency, energy, costs, hardware complexity, throughput, and end-to-end reliability trade-offs. For example, the requirements of mobile broadband and ultra-reliable low-latency communications are addressed by different configurations of 5G networks. 6G, on the contrary, will be developed to meet stringent network demands jointly (e.g., ultra-high reliability, capacity, efficiency, and low latency) in a holistic fashion because of the expected economic, social, technological, and environmental context of the 2030 era. Fully AI will be integrated into the 6G communication infrastructure. Integrating AI will incorporate all the network instrumentation, management, physical-layer signal processing, resource management and service-based communications. It will fuel the Industry 4.0 revolution, the digital transformation of industrial manufacturing. Fig. 2 illustrates the communication architecture scenario toward conceptualizing the 6G communication systems. The 6G applications can be characterized under uMUB, uHLSLLC, mMTC, and uHDD services. Some significant prospects and applications of 6G wireless communication are briefly outlined here.
\subsubsection*{Internet of Everything (IoE)}
IoE is an extended version of IoT that includes things, data, people, and processes \cite{vaya2020internet}. The central concept of IoE is to integrate various sensing devices that can be related to 'everything' for identifying, monitoring the status, and making decisions in an intelligent manner to create new prospects. Sensors in the Internet of Things can monitor various parameters, including velocity, position, light, bio-signals, pressure, and temperature. These sensors are deployed in several applications, including healthcare systems, smart cities, traffic management, and the industrial realm, to aid decision support systems \cite{nayak20216g}.

5G has revolutionary aims for IoE by transforming the traditional mobile communication layout. However, 5G is considered the beginning of IoE and addresses many challenges from standardization to commercialization. The 6G system will provide full IoE support. It is a kind of IoT, an umbrella term that integrates the four properties in one frame: data, people, processes, and physical devices. The integration of 6G and IoE will aid in advancing services related to the Internet of Things, the Internet of Medical Things, robotics, smart grids, smart cities, and body sensor networks, among others.
Additionally, it is anticipated that numerous new applications will arise due to merging IoE with 6G communication. However, IoE is predicted to rely on 6G as it requires the ability to connect N-intelligent devices, where N is a scalable number that can reach billions. Moreover, IoE requires high data rates in order to support and facilitate the N number of devices with minimal latency \cite{nayak20216g}. Therefore, IoE and 6G can facilitate the business processes to create magnitudes of data and re-invent digitisation with improved and agile data analytics.

\begin{figure}[]
  \centerline
  {\includegraphics[width=\textwidth]{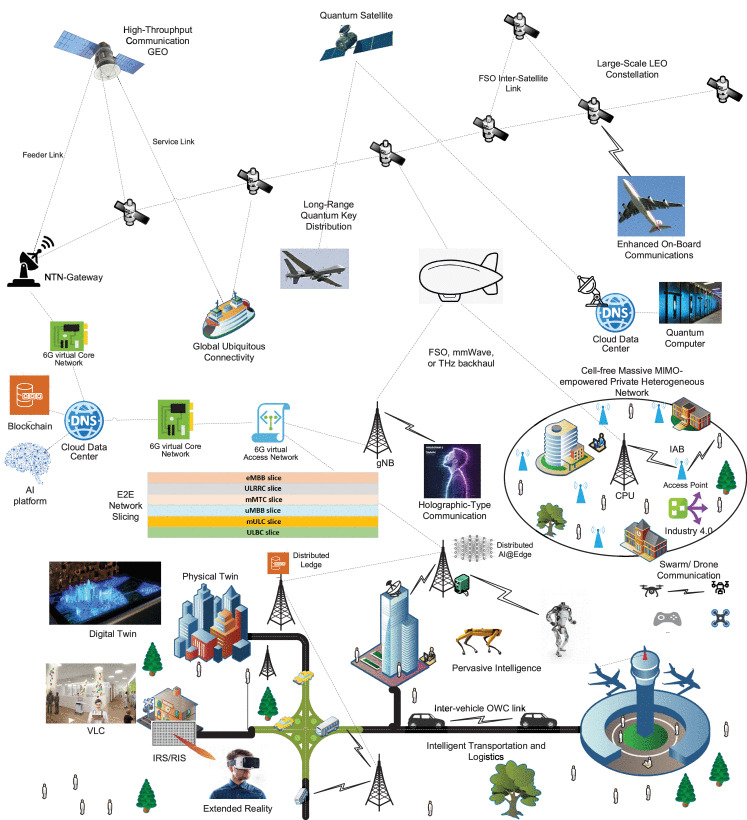}}
 \caption[Possible 6G Communication Architecture Scenario.]{Possible 6G Communication Architecture Scenario \cite{chowdhury20206g}.} 
\end{figure}

\subsubsection*{Super-Smart Society}
The superior features of 6G will accelerate the building of smart societies leading to life quality improvements, environmental monitoring, and automation using AI-based M2M communication and energy harvesting. This application can be characterized under all uMUB, uHLSLLC, mMTC, and uHDD services. The 6G wireless connectivity will make our society super-smart through smart mobile devices, autonomous vehicles, and so on. Besides, many cities will deploy flying taxis based on 6G wireless technology. Smart homes become a reality because any device in a remote location can be controlled by using a command given from a smart device \cite{9144301}.
\subsubsection*{Immersive Extended Reality (IXR)}
Extended reality (hereinafter referred to as XR) is a new immersive technology that combines the physical and virtual worlds through wearables and computers to facilitate human-machine interactions \cite{tariq2020speculative, siriwardhana2021survey}. The fab four (AR, VR, MR, and XR) technologies use different sensors to collect location, orientation and acceleration data. Similar to video traffic that saturates the 4G networks, the proliferation of ER devices will be blocked by the limited capacity of 5G with a peak rate of 20 Gbps, especially at the cell edge. In addition to bandwidth, interactive ER applications such as immersive gaming, remote surgery, and remote industrial control, low latency and high reliability are mandatory, all of which are expected to be offered by 6G.
\subsubsection*{Smart Grid 2.0}
Smart Grid 2.0 integrates cognitive decision-making systems with smart meters to monitor precise and reliable consumption. Additionally, smart grid 2.0 aims to detect outages, monitor power quality, enable demand response, and improve network connectivity in order to meet ever-increasing energy demand.

Communication has been a concern since the inception of smart grid development, as a significant number of devices must be connected to monitor and control electrical equipment from remote locations. The system requires high-quality transmission, security, and resource management to realize such a control strategy. 5G communication satisfies the low-latency and high-bandwidth requirements for commercializing the smart grid project with a small number of devices. However, if we are concerned about the diversification and the ramifications of climate change, 6G communication systems are required to fulfill the need \cite{khan20206g}.
\subsubsection*{Industry 5.0}
Industry 5.0 refers to people working alongside robots and smart machines to add a personal human touch to Industry 4.0 pillars of automation and efficiency \cite{nahavandi2019industry}. Similar to Industry 4.0, cloud/edge computing, big data, AI, 6G, and IoE are expected to be the key enabling technologies of Industry 5.0 \cite{demir2019industry}. In particular, a massive number of things in Industry 5.0 are connected using wired or wireless technologies to provide various applications and services that are enabled by a complete integration of cloud/edge computing, big data, and AI \cite{9397776}.
\subsubsection*{Holographic Telepresence (Teleportation)}
 With significant advances in holographic display technology in recent years, such as Microsoft's HoloLens, it is envisioned that its application will become a reality in the next decade. Compared to traditional 3D videos using binocular parallax, HT can project realistic, full-motion, real-time three-dimensional (3D) images of distant people and objects with a high level of realism rivaling the physical presence \cite{xu20113d}. Remote rendering high-definition holograms will bring a truly immersive experience through a mobile network. For example, holographic telepresence will allow remote participants to be projected as holograms into a meeting room or allow the attendee of online training or education to interact with ultra-realistic objects. However, even with image compression, HTC results in massive bandwidth requirements on the order of terabits per second.
Along with frame rate, resolution, and color depth in two-dimensional (2D) video, the hologram's quality is determined by volumetric data such as tilt, angle, and position. There are some road-blocking factors in the path of adopting HT technology. HTC requires ultra-low latency for proper immersion and high-precision synchronization across massive bundles of interrelated streams for reconstructing holograms, which can be achieved via 6G with a latency of 0.1 milliseconds and data rates of multiple gigabits per second \cite{giordani2020toward}.
\subsubsection*{Connected Robotics and Autonomous Systems}
Several automotive technology researchers are now investigating next-generation transportation systems, including autonomous driving, cooperative vehicle networks, the Internet of Vehicles (IoV) \cite{raja2020ai}, vehicular ad-hoc networks (VANETs), air-to-ground networks, and space-air-ground interconnected networks. At the moment, the evolution of data-centric automation systems has surpassed the capabilities of 5G. In some application domains, it is demanding even more than 10 Gbps transmission rates, such as XR devices. 6 G systems will accelerate the implementation of connected robotics and autonomous systems. The drone-delivery unmanned aerial vehicle (UAV) system is an example of this type of system. The 6G system facilitates the commercialization of self-driving cars (autonomous or driverless cars). A self-driving car perceives its surroundings through the use of multiple sensors, including light detection and ranging (LiDAR), radar, GPS, sonar, odometry, and inertial measurement units. In addition, the 6G system's mMUB and uHLSLLC services will verify the reliability of vehicle-to-everything (V2X) and vehicle-to-server connectivity. Unmanned aerial vehicle (UAV) is a type of unmanned aircraft. 6G networks will connect the ground-based controller and the UAV to the ground. UAVs are beneficial in a variety of fields, including military, commerce, science, agriculture, entertainment, law and order, product delivery, surveillance, aerial photography, disaster management, and drone racing \cite{maddikunta2021unmanned}. Additionally, the UAV will be used to enable wireless broadcast and high-rate transmissions in the absence or failure of a cellular base station (BS) \cite{li2018uav}.
\subsubsection*{Wireless Brain-Computer Interactions}
The brain-computer interface (BCI) is an approach to controlling the appliances used daily in smart societies, especially home appliances and medical apparatus. It is a direct communication path between the brain and external devices \cite{jafri2019wireless}. BCI acquires the brain signals transmitted to a digital device and analyzes and interprets the signals into further commands or actions. Wireless BCI requires guaranteed high data rates, ultra-low-latency, and ultra-high reliability. BCI services necessitate higher performance metrics compared to what 5G delivers \cite{saad2019vision}. The features of uHLSLLC and uMTC in 6G wireless communication will facilitate the practical implementation of BCI systems for a smart life \cite{9144301}.

\subsubsection*{Intelligent Healthcare and Biomedical Communication}
Similar to the industrial revolution from Industry 1.0 to Industry 5.0, there have been diverse evolution in healthcare development, such as AR/VR, holographic telepresence, mobile edge computing, and AI, which is now Healthcare 5.0 with the advent of digital wellness. AI-driven intelligent healthcare will be developed based on many innovative approaches, including Quality of Life (QoL), Intelligent Wearable Devices (IWD), IIoMT and H2H services \cite{mucchi20206g}. The 6G network will facilitate a reliable remote monitoring technology for healthcare systems. Even remote surgery will be made possible by using 6G communication. High-data-rate, low latency, ultra-reliable (zero-error) 6G network will assist the transfer of massive volumes of medical data swiftly and reliably, enhancing access to care and the quality of care.
On the other hand, THz, one of the critical driving technologies of 6G, has growing potential uses in healthcare services, such as terahertz pulse imaging in dermatology, oral healthcare, the pharmaceutical industry, and medical imaging. Also, biomedical communication is a crucial prospect of the 6G wireless communication system. The in-body sensors with the delivery of battery-less communication technologies are particularly desirable for reliable and long-term monitoring \cite{lin2020wireless, gravina2017multi}. Body sensors can afford precise and continuous monitoring of human physiological signals for applications in clinical diagnostics, athletics, and human-machine interfaces. The uMUB and uHLSLLC services of 6G can characterize these applications.

\subsection{Fundamental Enabling Technologies of 6G}
Like previous generations, 6G has the potential to fundamentally alter how consumers and businesses communicate, paving the way for a new generation of use cases that can take advantage of the increased speed, capacity, latency, and flexibility, as well as parallel computing, visualization, and artificial intelligence advancements. Only a combination of heterogeneous technologies integrated into a comprehensive system is expected to meet the needs adequately. A few vital technologies anticipated for 6G are discussed below:
\subsubsection*{Artificial Intelligence}
The most critical and newly introduced technology for 6G communication systems is AI \cite{mahmood2020six,murshed2023cnn}. There was no involvement of AI in 4G communication systems. The 5G communication systems support partial or very limited AI. However, it is evident that AI and its variants will be at the core of 6G and thus act as the most important enabling technology for 6G.

 Due to the advances in AI techniques, especially deep learning and the availability of massive training data, in recent years, there has been an overwhelming interest in using AI for the design and optimization of wireless networks. AI is expected to play a key role in all areas. Those can be broadly classified into three levels: device or end-user equipment, localized network domain, and overall network level. This will transform the 6G network from a self-organizing regime to a self-sustaining regime \cite{saad2019vision}. AI will also play a vital role in M2M, machine-to-human, and human-to-machine communications. It also prompts communication in the BCI. AI-based communication systems will be supported by metamaterials, intelligent structures, intelligent networks, intelligent devices, intelligent cognitive radio, self-sustaining wireless networks, and machine learning. Hence, AI technology will help reach the goals of uMUB, uHSLLC, mMTC, and uHDD services in 6G communication. Recent progress makes it possible to apply machine learning to RF signal processing, spectrum mining, and mapping.

\subsubsection*{Terahertz Communications}
Spectral efficiency can be improved by increasing the bandwidth. Looking forward to the 6G era, wireless technologies operating at higher frequencies, such as THz or optical frequency bands, are likely to play a key part in the next generation RAN, delivering exceptionally high bandwidth.

The crucial aspects of THz interfaces include (i) widely available bandwidth to support very high data rates and (ii) high path loss originating from the high frequency (highly directional antennas will most probably be indispensable) \cite{mumtaz2017terahertz}. The narrow beamwidths generated by the highly directional antennas reduce the interference. THz communication technologies will accelerate the deployment of uMUB, uHSLLC, and uHDD services in 6G communication. It will enhance the 6G potentials by providing wireless cognition, sensing, imaging, communication, and positioning. Due to the shorter wavelength, the THz frequency band in 6G is beneficial for incorporating a vast number of antennas to deliver hundreds of beams compared to the mmWave band. Later chapters will discuss Terahertz communications in further detail.

\subsubsection*{Visible Light Communications}
Visible Light Communications (VLC) is a subset of optical wireless communications operating in the 400 THz to 800 THz frequency range. In contrast to lower THz radio frequency (RF) technologies that rely on antennas, VLC transceivers are implemented using illumination sources - particularly light-emitting diodes (LEDs) and image-sensor or photodiode arrays. With these transceivers, a large bandwidth may be easily achieved while consuming very little power (100 mW for 10 Mbps to 100 Mbps) and without generating electromagnetic or radio interference \cite{sevincer2013lightnets}. Along with the high power efficiency, long life (up to ten years), and low cost of common LEDs, VLC is an attractive solution for use cases that require extended battery life and low access costs, such as massive IoT and wireless sensor networks (WSN). VLC outperforms RF technologies in various non-terrestrial circumstances, including aerospace and undersea, which could be vital aspects of the future 6G ecosystem. According to certain studies, each VLC link can attain a data rate of up to 0.5Gb/s, making it a viable choice for meeting the 6G data rate requirement \cite{chen2018omnidirectional}.

\subsubsection*{Massive MIMO and Intelligent Reflecting Surfaces}
One fundamental way to improve spectral efficiency is the application of the multiple-input multiple-output (MIMO) technique \cite{8695254, attarifar2019modified}. When the MIMO technique is developed, spectral efficiency is also improved. Therefore, the massive MIMO will be integral to both 5G and 6G systems due to the need for better spectral and energy efficiency, higher data rates, and higher frequencies. The massive MIMO technology will be crucial in the 6G system to support uHSLLC, mMTC, and uHDD services. Compared to 5G, we expect to shift from traditional massive MIMO toward IRS in 6G wireless systems to offer large surfaces for wireless communications and heterogeneous devices. IRS is a recent hardware technology with immense potential for energy-efficient green communication. It is also known as meta-surface, consisting of many reflecting diode units that can reflect any incident electromagnetic signals with an adjustable phase shift. Reconfigurable intelligent surfaces are envisaged as the massive MIMO 2.0 in 6G. These materials can integrate index modulation to increase the spectral efficiency in 6G networks \cite{basar2019wireless}. Gradient descent and fractional programming significantly optimize the intelligent surface phase shifts and transmit power. With that adjustable reflected phase-shifted signal and the transmitted signal, we can improve the energy efficiency of the system as well. This technology will be a great solution to maximize the data rate and minimize the transmit power in upcoming 6G networks.
\subsubsection*{Blockchain and Distributed Ledger Technologies (DLT)}
Blockchains and DLT will be one of the most disruptive IoE technologies. It can be considered the next generation of distributed sensing services whose demand for connectivity will require a synergistic blend of URLLC and massive machine type communications (mMTC) to guarantee low-latency, reliable connectivity, and scalability \cite{dai2019blockchain}. The inherent qualities of blockchain, such as decentralized tamper resistance and secrecy, create the opportunity to make it appropriate for multiple applications in 6G communication. Blockchain combines a distributed network structure, consensus mechanism, and advanced cryptography to represent prospective capabilities not available in the existing structures. It establishes a secure and verifiable approach for spectrum management by establishing transparency, verifying transactions, and preventing unauthorized access. The distributed nature eliminates the single point of failure concern and boosts security. The biggest problem of blockchain networking in 5G is the throughput (10-1000 transactions per second). Another challenge is the necessity for local and international standardization and regulation of the enormous use of blockchain in 5G. These limitations can be mitigated in 6G by leveraging consensus algorithms, implementing unique blockchain design and sharing strategies, and increasing block size of the network.

\subsubsection*{Quantum Communications}
Quantum communication is another promising technology that has the potential to significantly contribute to two critical 6G criteria, namely extremely high data rates and security. Quantum entanglement's intrinsic security feature, which cannot be cloned or accessed without tampering, makes it an ideal technology for 6G and beyond networks. Furthermore, quantum communications improve throughput due to the superposition nature of qubits. Quantum communication is advantageous for long-distance communication as well. However, the present repeater paradigm does not work for quantum communication since entanglement cannot be cloned \cite{huang2019survey}. Satellites, high-altitude platforms, and unmanned aerial vehicles (UAVs) can be utilized as trusted nodes for key regeneration and redistribution. In terms of quantum device design, single-photon emitter devices that operate at temperatures just above absolute zero have already been produced. Much work is still needed to make them operate in average temperatures. It may be a long shot to see much of an impact of quantum communication in 6G \cite{tariq2020speculative}.

\subsubsection*{Cell-Free Communications}
The tight integration of multiple frequencies and communication technologies will be crucial in 6G systems. As a result, the user will be able to switch smoothly between networks without requiring any manual configurations to the device \cite{giordani2020toward}. Cell-free and non-orthogonal communications in 6G will replace cellular and orthogonal communications concepts. The entire potential of UAV wireless networks or drone cells will be realized, and their application will be widely expanded to mobilize network resources and integrate them with cell-free massive MIMO to achieve truly cell-free networks with arbitrarily low latency. Currently, user movement between cells causes excessive handovers in dense networks, resulting in handover failures, delay, data loss, and the ping-pong effect. All of these issues will be addressed by 6G cell-free communications, which will deliver a higher quality of service. Cell-free communication will be achieved through a combination of multi-connectivity and multi-tier hybrid approaches and diverse and heterogeneous radios within the devices.

\subsubsection*{Integration of Wireless Information and Energy Transfer}
Wireless energy transfer will be involved in 6G, providing suitable power to the batteries in devices such as; smartphones and sensors. The base stations in 6G will be used for transferring power as Wireless Information and Energy Transfer (WIET) \cite{mao2020ai}. WIET is an innovative technology that will allow the development of batteryless smart devices, charging wireless networks and saving the battery lifetime of other devices. Energy harvesting (EH) techniques may be costly and impossible in hazardous environments, building structures, or the human body. Therefore, EH is foreseen as a key component of future IoT networks since it allows: 1) wireless charging, which significantly simplifies the servicing and maintenance of IoT devices while increasing their durability thanks to contact-free features and 2) enhanced energy efficiency and network-wide reduction of emissions footprint \cite{9319211}.

\subsubsection*{Holographic Beamforming}
Beamforming is a signal processing technique that enables the steering of an array of antennas to send radio signals in a specific direction. It is a subset of smart antennas or advanced antenna systems. Numerous advantages of beamforming technology include a high signal-to-noise ratio, interference prevention and rejection, and high network efficiency. Holographic beamforming (HBF) is a novel technique for beamforming that differs significantly from MIMO systems in that it uses software-defined antennas. HBF is a beneficial approach in 6G for signal transmission and reception in multi-antenna communication devices. In addition, HBF can play a significant role in physical-layer security, wireless power transfer, enhanced network coverage, and positioning.
\subsubsection*{Mobile Edge Computing}
Edge computing is critical for improving the performance of network services, maximizing the efficiency of network resources (both physical and virtual), lowering a mobile operator's CAPEX/OPEX, and reducing network complexity (both in the control plane and user plane) \cite{nasimi2018edge}. However, a huge number of end-users, each with a distinct set of business and technological requirements, forces the network operator to consider alternative approaches to overcome edge computing limitations by employing cutting-edge AI tools and modern machine learning methodologies. To accomplish this, the edge intelligence (EI) concept is introduced to integrate AI and machine learning techniques at the edge of mobile networks to deliver automation and intelligence. Furthermore, due to the growing number of intelligent portable devices, user equipment, the Internet of Intelligent Things (IoIT), and the expansion of intelligent services, there is an increasing interest for EI at the edge of 6G mobile networks to automate their respective tasks \cite{habibi2021towards}.

\section{TeraHertz Communication}
With the exponential growth of data traffic as projected by Edholm's law \cite{cherry2004edholm}, the wireless communication systems show a tendency of explosive growth for ultra-high data rates. While advanced physical layer solutions and improved hardware components (e.g., sources, detectors, and antennas) might help meet the ever-increasing data rate requirement, it is still impractical for conventional communication systems to reach 100 Gbps and even several Terabits per second (Tbps) \cite{song2011present}. The most apparent distinction between 5G and earlier generations of cellular systems is the realization that the conventional sub-6 GHz spectrum will not be sufficient to fulfill the demand of emerging applications. All these challenges stimulated the investigation of a suitable regime in the frequency spectrum capable of satisfying users' escalating requirements and enabling massive capacity and connectivity utilization.

\begin{figure}[!ht]
  \centerline
  {\includegraphics[width=\textwidth]{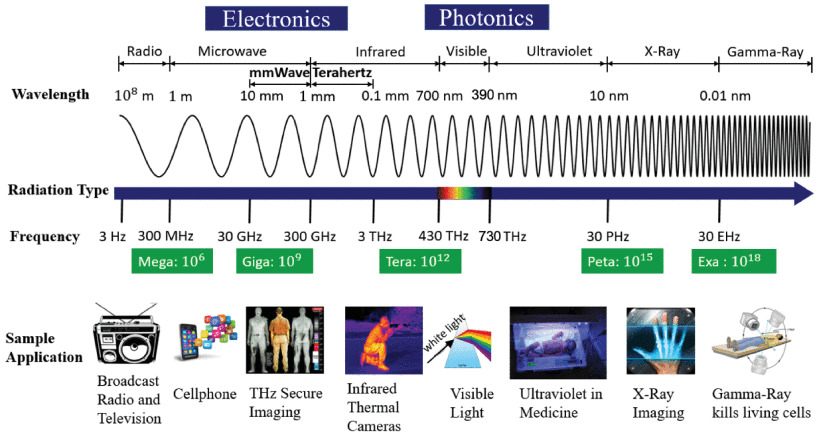}}
 \caption[The electromagnetic spectrum, and applications as a function of frequency.]{The electromagnetic spectrum, and applications as a function of frequency \cite{8732419}.} 
\end{figure}
Naturally, the millimeter wave (mmWave) spectrum emerged as a potential solution due to its wide bandwidth ranging from hundreds of MHz to several GHz \cite{huang2017multi}. Although these bands were formerly considered inappropriate for mobile operations due to their adverse propagation characteristics, device and antenna technology advancements have made them viable for commercial wireless applications \cite{rappaport2013millimeter}. However, as we look forward, it is clear that we are gradually moving toward applications such as virtual and augmented reality, ultra-high-definition video conferencing, 3D gaming, and the use of wireless for brain-machine interfaces, all of which will impose even stricter constraints on throughput, reliability, and latency requirements. For example, future wireless local area networks (WLAN) and wireless personal area networks (WPAN) systems will require data rates of at least a few tens of gigabits per second. Moreover, uncompressed ultra-high-definition films,3D videos and virtual reality (VR) will support data rates of up to 10 Gbps, 24 Gbps and 100 Gbps, respectively \cite{mumtaz2017terahertz}. As a result, there is an urgent need to investigate higher-frequency resources considerably.

Except for the mmWave communication, free-space optical (FSO) communication system has also attached great attention owing to its unique advantages, including the large bandwidth, license-free spectrum, less power consumption and so on \cite{chan2006free}. Due to the excellent features of optical carrier, the architecture of FSO communication can be applied to satellite-to-ground and inter-satellite optical communication links. However, the performance of FSO communication system processes can be heavily influenced by unpredictable environmental conditions like clouds, fog, rain, haze, etc., as well as limitations of difficult beam tracking and high background noise \cite{kaushal2016optical}. Also, employing optical waves for indoor wireless optical communications is not a viable solution due to the low sensitivity of incoherent receivers, significant diffuse reflection losses, and a limited power budget due to eye-safety constraints \cite{gfeller1979wireless}. 

Next to the mm-wave spectrum, the Terahertz (THz) frequency band (0.1-10 THz) has begun to garner significant global interest, demonstrating its potential as a critical wireless technology for meeting future demands for 6G wireless systems. Seamless data transfer, unlimited bandwidth, microsecond latency, and ultra-fast download are all characteristics of THz technology, which is expected to revolutionize the telecommunications landscape and alter how people communicate and access information. As a result, numerous exciting applications are anticipated, including Tbps WLAN systems (Tera-WiFi), Tbps Internet-of-Things (Tera-IoT) in wireless data centers, and wireless networks with Tbps integrated access backhaul (Tera-IAB), and ultra-broadband THz space communications (Tera-SpaceCom). Apart from these macro/micro-scale applications, the THz band can be used for wireless connections in nanomachine networks, enabling wireless networks-on-chip (WiNoC) communications and the Internet of Nano-Things (IoNT) \cite{chen2019survey}.

The THz frequency spectrum offers wide bandwidth, potentially ranging from tens to several THz, implying a prospective capacity of Terabits per second \cite{akyildiz2014teranets}. As a result, the bandwidth provided is one order of magnitude more than that of millimeter-wave (mmW) systems. THz signals also provide more excellent link directionality and are less susceptible to eavesdropping than millimeter-wave signals \cite{federici2010review}.

THz waves are being investigated as potential possibilities for uplink communication. THz band analysis suggests that these frequencies also have a number of advantages over optical frequencies. They enable non-line-of-sight (NLoS) propagation and function well in inclement climate circumstances such as fog, dust, and turbulence. Additionally, the THz frequency band is unaffected by ambient noise emitted by optical sources and is not subject to any health or safety regulations \cite{su2012experimental}. The comparison of the THz frequency band to other existing technologies in Table 1.2 demonstrates the ultimate promise of this band for the next generation of wireless networks \cite{elayan2018terahertz}.

\begin{table}[htbp]
\caption{Comparison between wireless communication technologies}\label{my-label1}
\begin{adjustbox}{width=\columnwidth,center}
\begin{tabular}{|c|c|c|c|c|c|} 
 \hline
 \textbf{Technology} & \textbf{mmW} & \textbf{THz Band} & \textbf{Infrared} & \textbf{VLC} & \textbf{Ultra-Violet} \\ [0.5ex] 
 \hline\hline
 \textbf{Frequency Range} & 30 GHz - 300 GHz & 100 GHz - 10 THz & 10 THz - 430 THz & 430 THz - 790 THz & 790 THz - 30 PHz \\[0.25ex] 
 \hline
 \textbf{Range} & Short range & Short/Medium range & Short/Long range & Short range & Short range \\[0.25ex]
 \hline
 \textbf{Power Consumption} & Medium & Medium & Relatively low & Relatively low & Expected to be low \\[0.25ex]
 \hline
 \textbf{Network Topology} & Point to Multi-point & Point to Multi-point & Point to Point & Point to Point & Point to Multi-point \\[0.25ex]
 \hline
\textbf{Noise Source} & Thermal noise & Thermal noise & Sun/Ambient Light & Sun/Ambient Light & Sun/Ambient Light \\[0.25ex]
 \hline
 \textbf{Weather Conditions} & Robust & Robust & Sensitive & - & Sensitive \\ [0.25ex]
 \hline
 \textbf{Security} & Medium & High & High & High & To be determined \\[1ex]
 \hline
\end{tabular}
\end{adjustbox}
\end{table}

\section{MIMO Systems}
The performance limitation of any wireless network will always be at the physical layer because, fundamentally, the amount of information that can be transferred between two locations is limited by the availability of spectrum, the laws of electromagnetic propagation, and the principles of information theory. All proposed solutions for improving a wireless network's efficiency appear to fit into one of three categories: 1) exploitation of currently available or underutilized spectrum; 2) deployment of an increasing number of access points, each covering a proportionately smaller area; and 3) utilization of multiple-antenna access points or terminals. While future wireless systems and standards are anticipated to employ a growing number of access points and additional spectral bands, the imperative of enhancing spectral efficiency in a given band will always remain. Using multiple antennas, sometimes referred to as multiple-input, multiple-output (MIMO) technology, is the only feasible method for significantly increasing spectral efficiency.

MIMO technology can be categorized logically into three categories, each evolving during a distinct epoch: Point-to-Point MIMO, Multi-User MIMO, and Massive MIMO \cite{marzetta2016fundamentals}.

\subsubsection*{Point-to-Point MIMO}

\begin{figure}[!ht]

   \centering
 
  \includegraphics[width=0.8\textwidth]{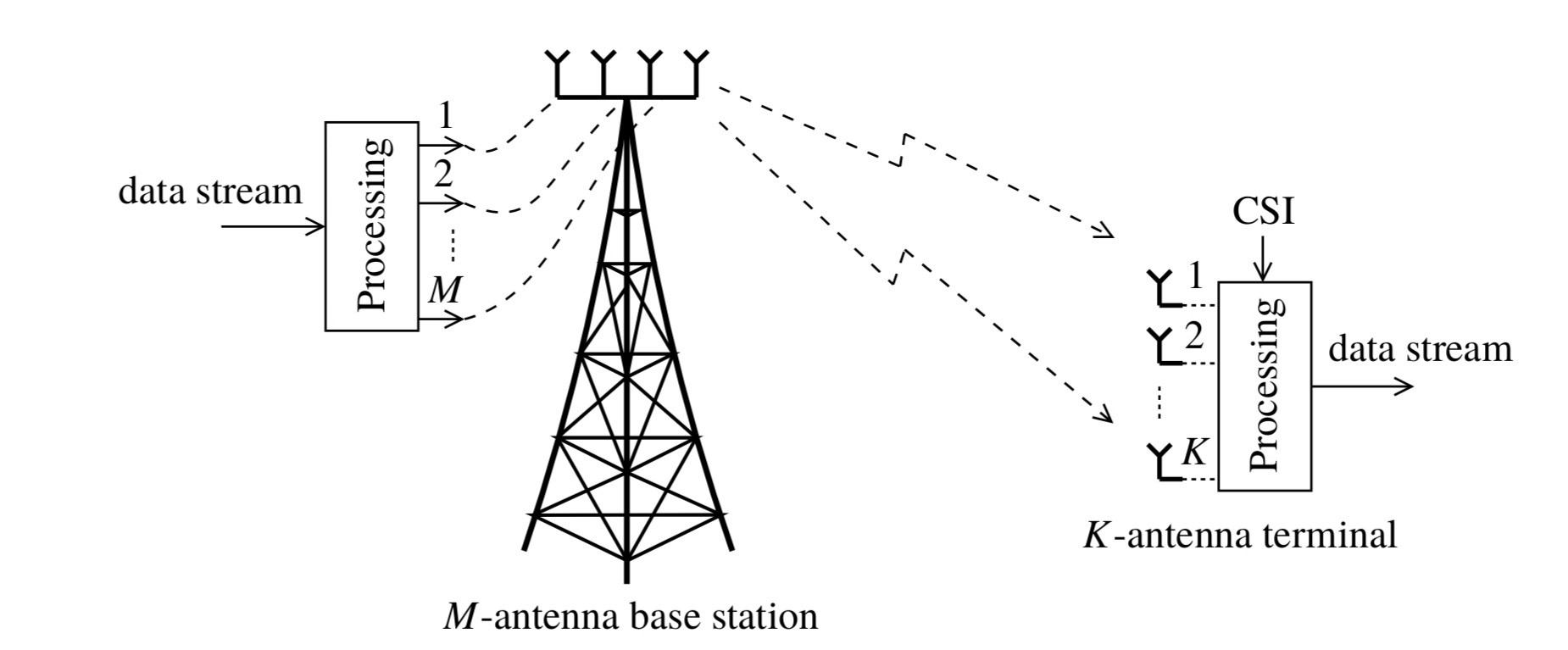}
 \caption[Uplink and Downlink Transmission in a Point-to-Point MIMO System]{Uplink and Downlink Transmission in a Point-to-Point MIMO System}
 
\end{figure}

Point-to-Point Multiple Input, Multiple Output (MIMO) emerged in the late 1990s, representing the simplest form of MIMO: a base station equipped with an antenna array serves a terminal equipped with an antenna array. Different terminals are multiplexed orthogonally, for example, by combining time- and frequency-division multiplexing. Each time a channel is used, a vector is transmitted and received.

In Figure 1.7, a Point-to-Point MIMO link is illustrated, where a base station with a concentrated array of M antennas transmits data to a user with a concentrated array of K antennas. Assuming Gaussian transmit signals are independent and identically distributed (i.i.d.), and perfect Channel State Information (CSI) is available at the receiver, the instantaneous achievable rate can be defined as:

\begin{align}
  C = \log_2\det\left(\mathbf{I}_{K} + \frac{\rho}{N_{t}} \mathbf{H} \mathbf{H}^{\textrm{H}}\right) {\frac{{\textrm{bits}}}{{\textrm{s}}} \over {\textrm{Hz}}}.
\end{align}
where $\mathbf{H}$ is the $M\times K$ frequency response of the matrix-valued channel connecting the base station and the user antennas, $\mathbf{I}_K$ denotes the $K\times K$ identity matrix, the scalar $\rho$ represents the scalar power, and the superscript "H" denotes "conjugate transpose."

For various reasons, Point-to-Point MIMO is not readily scalable beyond ${8\times 8}$. Firstly, the propagation environment may not support eight data streams, as line-of-sight conditions present a particular challenge. Secondly, scaling up the number of antennas requires proportional amounts of time for training. Thirdly, the user equipment becomes complicated, requiring independent electronics chains for each antenna. Finally, achieving performance close to the Shannon limit necessitates rather involved signal processing for both the base station and the user.

\subsubsection*{Multiuser MIMO}
\begin{figure}[!ht]

   \centering
 
  \includegraphics[width=0.9\textwidth]{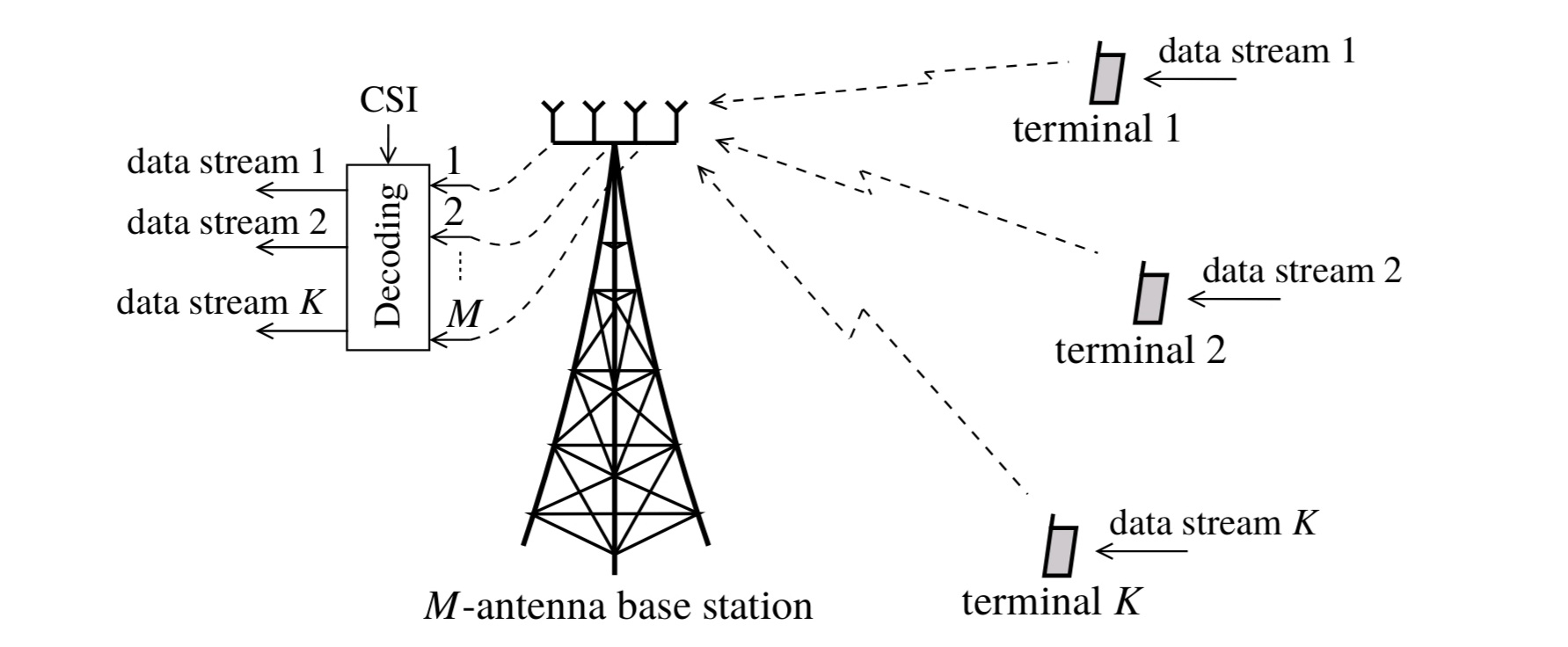}
  \includegraphics[width=\textwidth]{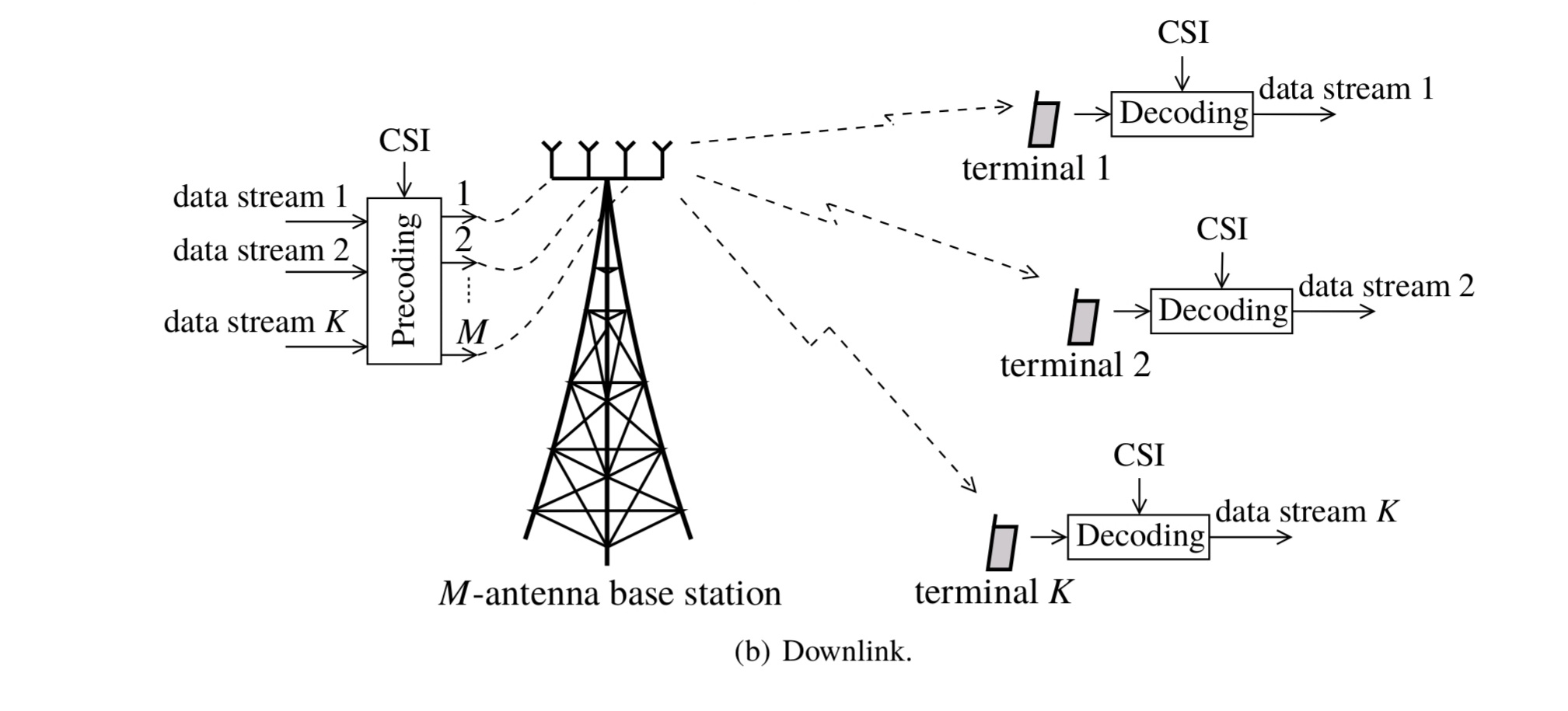}
 \caption[Uplink and Downlink Transmission in a Multiuser MIMO System]{Uplink and Downlink Transmission in a Multiuser MIMO System}

\end{figure}
MU-MIMO systems can obtain the promising multiplexing gain of massive point-to-point MIMO systems while eliminating problems due to unfavorable propagation environments. Effectively, the Multiuser MIMO scenario is obtained from the Point-to-Point MIMO setup by breaking up the K-antenna terminal into multiple autonomous terminals. 

We assume that terminals in Multiuser MIMO have a single antenna.
Hence, in the setup in Figure 1.8, the base station serves K terminals. Let G be an ${M\times K}$ matrix corresponding to the frequency response between the base station array and the K terminals. The uplink and downlink sum spectral efficiencies are given by:
\begin{equation}
    C_{\mathrm{sum\, up}} = \log_{2}\det\left(\mathbf{I}_{K} + \frac{\rho_{\mathrm{u}}}{K}\mathbf{G}_{\mathrm{u}}^{\mathrm{H}}\mathbf{G}_{\mathrm{u}}\right),
\end{equation}

\begin{equation}
    C_{\mathrm{sum\, down}} = \sup\nolimits_{\mathbf{a}}\left\{\log_{2}\det\left(\mathbf{I}_{M} + \rho_{\mathrm{d}}\mathbf{G}_{\mathrm{d}}\mathbf{D}_{\mathbf{a}}\mathbf{G}_{\mathrm{d}}^{\mathrm{H}}\right)\right\}, \quad \mathbf{a}\geq\mathbf{0},\ \mathbf{1}^{\mathrm{T}}\mathbf{a}=1,
\end{equation}
where ${D_a}$ is a diagonal matrix whose diagonal elements comprise the ${M\times1}$ vector, \textbf{a}, and \textbf{1} denotes the ${M\times1}$ vector of ones.

On a positive note, Multi-User MIMO has two advantages over Point-to-Point MIMO. To begin with, it is more resistant to the propagation environment. It can operate effectively even under line-of-sight conditions if the typical angular separation between users is greater than the angular resolution of the base station array. Second, only single-antenna terminals are required.

Two factors prevent the Shannon-theoretic version of Multi-User MIMO from being scalable: first, the exponentially increasing complexity of dirty paper coding/decoding, and second, the time required to acquire CSI, which increases exponentially with both the number of service antennas and the number of users.
\subsubsection*{Massive MIMO}
\begin{figure}[!ht]

   \centering
 
  \includegraphics[width=0.9\textwidth]{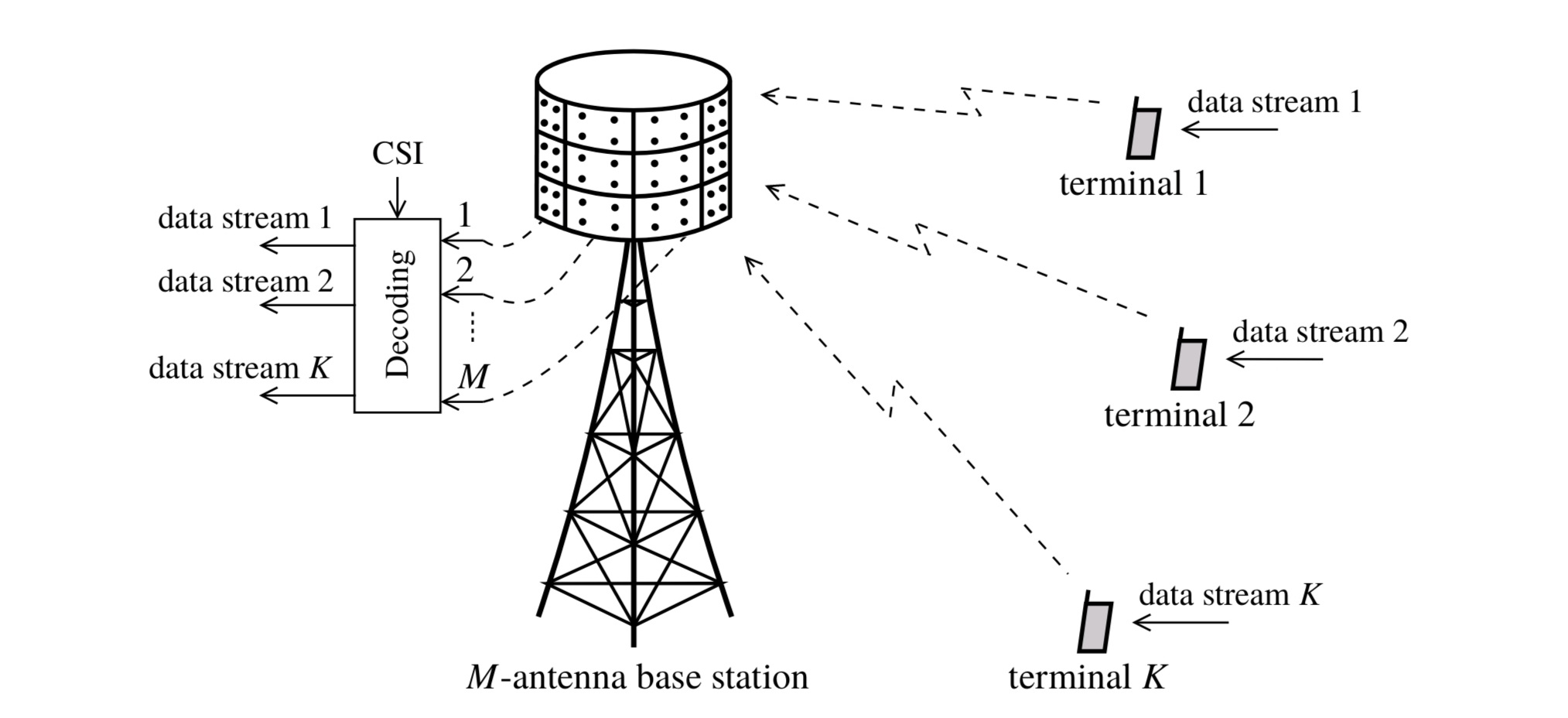}
  \includegraphics[width=0.9\textwidth]{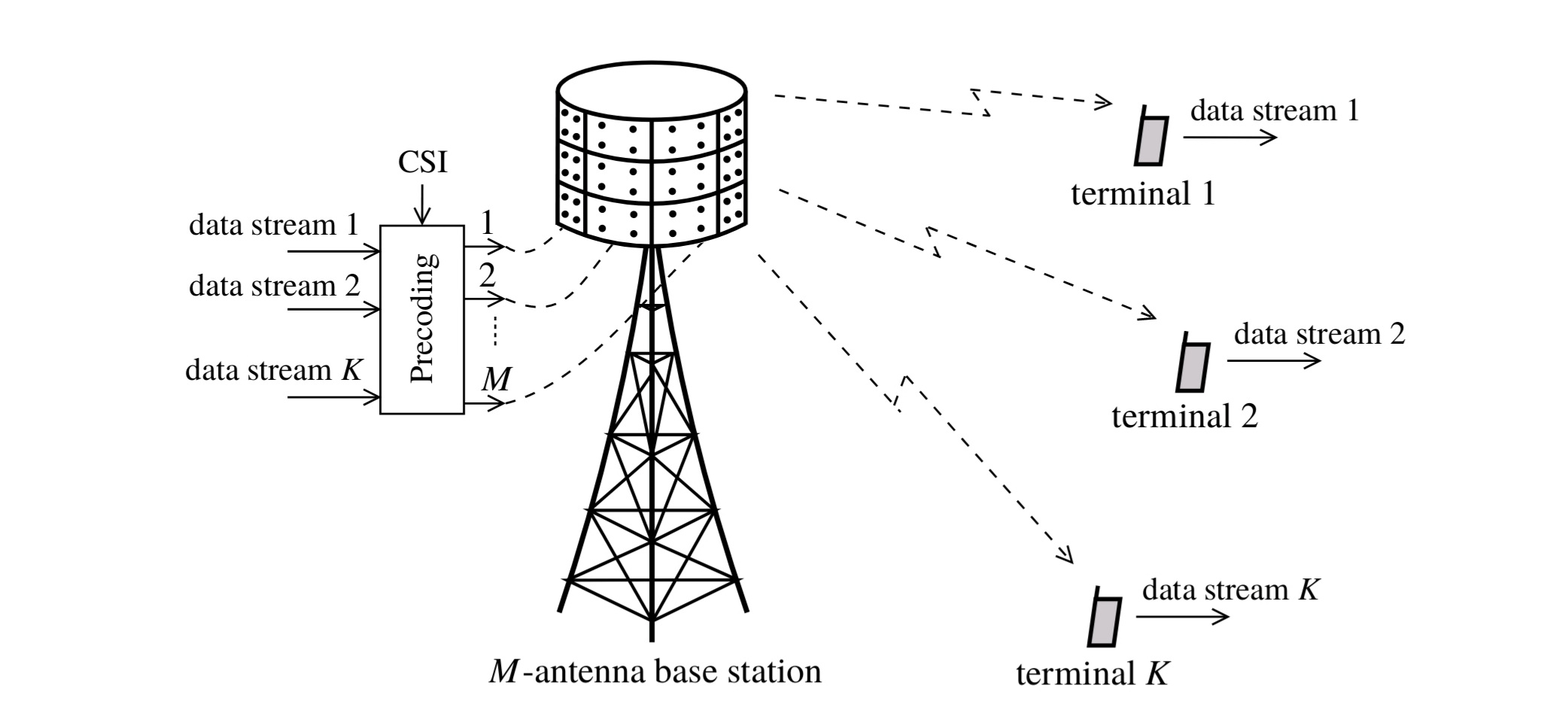}
 \caption[Uplink and Downlink Transmission in a Massive MIMO System]{Uplink and Downlink Transmission in a Massive MIMO System}
 
\end{figure}
Massive MIMO (multiple-input multiple-output) improves the sum SE of cellular networks by spatial multiplexing of a large number of user equipments (UEs) per cell. It is therefore considered a key time-division duplex (TDD) technology for the next generation of cellular networks. The main difference between Massive MIMO and classical multi-user MIMO is the large number of antennas at each base station (BS) whose signals are processed by individual radio-frequency chains. Massive MIMO breaks the scalability barrier by not attempting to achieve the full Shannon limit and, paradoxically, by increasing the size of the system. It departs from Shannon-theoretic practice in three ways. First, only the base station learns the downlink channel. In a TDD system, the time required to acquire CSI is independent of the number of base station antennas. Second, the number of base station antennas is typically increased to several times the number of users. Third, a simple linear precoding multiplexing is employed on the downlink, coupled with linear decoding demultiplexing on the uplink. As the number of base station antennas increases, linear precoding and decoding performance can approach the Shannon limit. 

Some of the benefits of massive MIMO technology are \cite{chataut2020massive}:

\noindent \textbf{Spectral Efficiency:} Massive MIMO maximizes spectral efficiency by allowing its antenna array to focus narrow beams on a user. It is possible to attain a spectral efficiency that is more than ten times that of the present MIMO system used for 4G/LTE.

\noindent \textbf{Energy Efficiency:} As antenna array is focused in a small specific section, it requires less radiated power and reduces the energy requirement in massive MIMO systems.

\noindent \textbf{High Data Rate:} The array gain and spatial multiplexing provided by massive MIMO increases the data rate and capacity of wireless systems.

\noindent \textbf{User Tracking:} Since massive MIMO uses narrow signal beams toward the user, user tracking becomes more reliable and accurate.

\noindent \textbf{Low Power Consumption:} Massive MIMO is built using ultra-low-power linear amplifiers, obviating the need for bulky electronic equipment in the system. As a result, power consumption can be significantly lowered.

\noindent \textbf{Less Fading:} A Large number of the antenna at the receiver makes massive MIMO resilient against fading.

\noindent \textbf{Low Latency:} Massive MIMO significantly reduces air interface latency.

\noindent \textbf{Robustness:} Massive MIMO systems are impervious to unintended interference and internal jamming. Also, these systems are resilient to a single or a few antenna failures as a result of the large antennas.

\noindent \textbf{Reliability:} The use of a large number of antennas in massive MIMO results in a higher diversity gain, which improves link reliability.

\noindent \textbf{Enhanced Security:} Massive MIMO improves physical security by utilizing orthogonal mobile station channels and narrow beams.

\noindent \textbf{Low Complex Linear Processing:} More number of base station antennas makes the simple signal detectors and precoders optimal for the system.

\section{Beamforming}
Future networks are anticipated to ease the burden on the current infrastructure by providing much higher data rates via enlarged channel bandwidths. Multiple beam antennas have been identified as a critical technology for the fifth generation (5G), the sixth generation (6G), and more generally, beyond 5G (B5G) wireless communication links in both terrestrial networks (TNs) and non-terrestrial networks (NTNs) to support the ever-increasing demand for connectivity and data-rates. Considering the shortage of available frequencies traditionally used for mobile communications, mm-wave bands became a suitable alternative. The large bandwidth available at these frequencies facilitates the provision of data rates that meet future requirements. However, the mobile environment at these mm-wave bands is significantly more complex than at current frequencies. Higher propagation losses that greatly vary depending on the environment require an updated network infrastructure and new hardware concepts.

Beamforming antenna arrays will play an important role in 5G and beyond wireless network implementations since even handsets can accommodate a larger number of antenna elements at mm-wave frequencies. Beamforming is a technique in which each received signal is multiplied by complex weight vectors that adjust the magnitude and phase of the signal from each antenna element \cite{mohammad2006mi}. Aside from having a higher directional gain, these antenna types offer advanced beamforming capabilities. This increases cellular network capacity by improving the signal-to-interference ratio (SIR) through direct targeting of user groups. In addition, the narrow transmit beams concurrently lower the degree of interference in the radio environment and make it possible to retain sufficient signal power at the receiver terminal over wider distances in rural areas.

Beamforming techniques can be grouped into three categories: Analog, Digital and Hybrid Beamforming.
\subsubsection*{Analog Beamforming}
Figure 1.10 depicts a basic analog beamforming transmitter architecture. This architecture includes a single RF chain and multiple phase shifters that feed an antenna array.
\begin{figure}[!ht]

   \centering
 
  \includegraphics[width=0.75\textwidth]{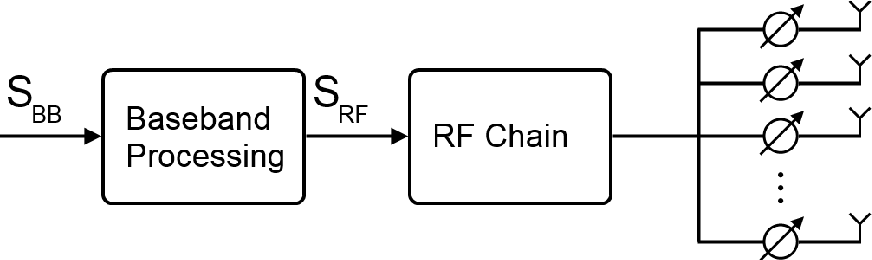}
  
 \caption[Analog Beamforming]{Analog Beamforming \cite{white2016millimeter}}
 
\end{figure}

The first practical analog beamforming antennas date back to 1961. The steering was attained with a selective RF switch and fixed phase shifters. The basics of this technique are still used today, but with advanced hardware and precoding algorithms \cite{butler1961beam}. These improvements permit separate phase control for each element.
Unlike earlier passive architectures, active beamforming antennas allow the beam to be directed virtually to any angle. This type of beamforming, true to its name, is accomplished in the analog domain at RF or intermediate frequencies \cite{powell2014technical}. This architecture is being employed in a variety of high-end millimeter-wave systems, including radar and short-range communication systems such as IEEE 802.11ad. 

Analog beamforming architectures are less expensive and less complicated than alternative approaches. The performance of the analog architecture can be enhanced further by adjusting the signal magnitude incident on the radiators. However, implementing a multi-stream transmission with analog beamforming is an extremely difficult task \cite{alkhateeb2014mimo}.
\subsubsection*{Digital Beamforming}
While analog beamforming is generally restricted to one RF chain even when using a large number of antenna arrays, digital beamforming, in theory, supports as many RF chains as there are antenna elements. Transmission and reception flexibility are improved if suitable precoding is implemented in the digital baseband. The additional degree of freedom can be utilized for advanced techniques such as multi-beam MIMO. In comparison to alternative beamforming systems, these advantages result in the highest theoretical performance \cite{roh2014millimeter}. Figure 1.11 depicts the general architecture of a digital beamforming transmitter with multiple RF chains.
\begin{figure}[!ht]

   \centering
 
  \includegraphics[width=0.7\textwidth]{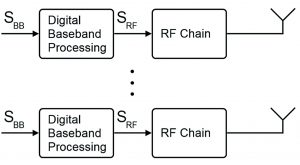}
  
 \caption[Digital Beamforming]{Digital Beamforming \cite{white2016millimeter}}
 
\end{figure}
Beam squint is a well-known issue with phase offset-based analog beamforming architectures. This is a major drawback in light of future network plans to utilize large bandwidths in the mm-wave band. Digital control of the RF chain permits phase optimization according to frequency across a broad spectrum.
Digital beamforming is ideally suited for usage in base stations, where performance is more important than mobility. Nevertheless, digital beamforming may not always be optimal for practical 5G application implementations. The hardware's extremely high complexity and requirements may considerably raise the cost, energy consumption, and integration in mobile devices. The fact that digital beamforming can allow multi-stream transmission and simultaneously serve multiple users is a great catalyst of the technology \cite{white2016millimeter}.
\subsubsection*{Hybrid Beamforming}
A Hybrid system, first described in \cite{el2012capacity}, enables a Base Station (BS) to have a huge number of antennas without being overly difficult to implement. This complexity stems mostly from the difficulty of integrating many RF chains in mmWave, although studies such as \cite{swindlehurst2014millimeter,ullah2023beyond} have found promising results.

A significant cost reduction can be achieved by reducing the number of complete RF chains. This also results in a reduction in overall energy consumption. Digital baseband processing offers fewer degrees of freedom since the number of converters is substantially lower than the number of antennas. Consequently, the number of concurrently supported streams is limited compared to full digital beamforming. Due to specific channel characteristics in millimeter-wave bands \cite{alkhateeb2014mimo}, the resulting performance gap is comparatively low.

Figure 1.12 depicts the architecture of a hybrid beamforming transmitter. Theoretically, it is conceivable for each amplifier to be connected to each radiating element. The precoding is divided between the analog and digital domains.
\begin{figure}

   \centering
 
  \includegraphics[scale=0.5]{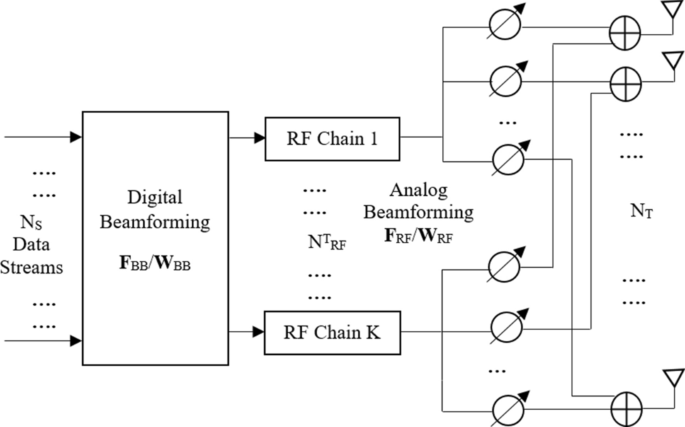}
  
 \caption[Hybrid Beamforming]{Hybrid Beamforming \cite{dilli2021performance}}
 
\end{figure}
\section{Orthogonal Frequency-Division Multiplexing (OFDM)}
OFDM is a specific type of multicarrier transmission in which a single data stream is sent over a number of lower-rate subcarriers (SCs). It is important to note that OFDM can be regarded as either a modulation or multiplexing technique. OFDM is utilized primarily to boost resilience against frequency-selective fading and narrowband interference. In a single-carrier system, a single fade or interference can cause the failure of the entire link. In contrast, only a tiny fraction of the SCs will be affected in a multicarrier system. The few erroneous SCs can then be corrected using error-correction coding. Figure 1.13 compares the conventional nonoverlapping multicarrier technique to the overlapping multicarrier modulation technique. By utilizing the technique of overlapping multicarrier modulation, we save about 50 percent of bandwidth \cite{prasad2004ofdm}.

\begin{figure}[!ht]
\centering
 
  \includegraphics[width=0.8\textwidth]{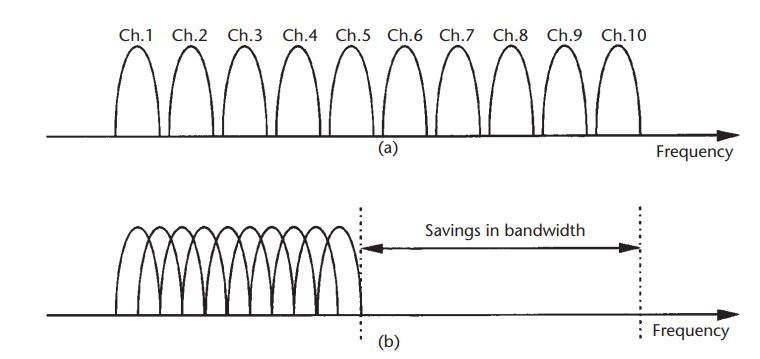}
  
 \caption[Concept of the OFDM signal: (a) conventional multicarrier technique, and (b) orthogonal multicarrier modulation technique.]{Concept of the OFDM signal: (a) conventional multicarrier technique, and (b) orthogonal multicarrier modulation technique. \cite{prasad2004ofdm}}
 
\end{figure}

Due to the critical demand for channel-state-information (CSI) in spatially-coded OFDM systems, channel estimation/prediction has become an integral part of such systems, allowing for the reduction of fading, interference, and noise/disturbances. It also enhances the receiver's information-symbol detection rate. By improving the signal-to-noise ratio (SNR) at the receiving end, high data-rate services can be provided to wireless communication users utilizing spatially-coded OFDM systems without additional bandwidth. Therefore, the spatially-coded OFDM strategies stand as a promising choice for the 5G high data-rate communication techniques.
\section{Intercarrier Interference}
Massive multiple-input multiple-output orthogonal frequency division multiplexing (MIMO-OFDM) is an approach for enhancing the data rate and spectral efficiency of 5G wireless communication systems. With the assumption of immobile users, it is simple to analyze the performance of any system. However, if the user moves at different velocities, it is difficult to maintain contact between the user and the base station (BS). Under high mobility conditions, rapid fading, double dispersion, and nonstationary environment, the conventional system fails to adapt and intercarrier interference (ICI) becomes a concern. 
\begin{figure}[!ht]
\centering
 
  \includegraphics[scale=0.7]{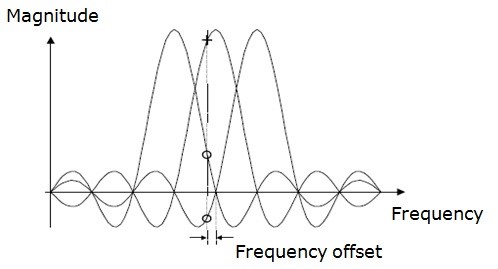}
  
 \caption[ICI due to frequency offset]{ICI due to frequency offset}
 
\end{figure}

Intercarrier Interference (ICI) is an impairment well known to degrade the performance of OFDM transmissions. It arises from carrier frequency offsets (CFOs), from the Doppler spread due to channel time-variation, and, to a lesser extent, from sampling frequency offsets (SFOs) \cite{chawla2020intercarrier}. ICI has yet to be considered a significant issue in OFDM systems such as New Radio (NR). Still, it could be a problem in future communication systems under high mobility, THz, or ultra-reliable low-latency communication (URLLC) scenarios.
\section{Motivation}
Terahertz Wireless Communication will play a significant role in deploying Beyond 5G multi-gigabit short-range communication systems. However, research in this area is still nascent, and only a few studies have explored the prospective communication systems in this frequency band. In the prior generations, ICI was insignificant due to the limited subcarriers and bandwidth. With the expansion of THz technology, however, CFO, which creates ICI, must be taken into account. There was a dearth of work on ICI, and the primary focus was on the complete elimination of ICI. However, it is exceedingly difficult or impractical to eradicate ICI in THz systems. In some other works, ICI was considered, but neither the requisite technologies for THz communication nor a multi-user system was investigated. However, designing a system based on a single user will take much work as there will be a massive number of users in practice. In this study, we have addressed each of the challenges mentioned above: (1) We illustrated the effect of ICI on the system, (2) we considered the multi-user scenario and (3) THz frequency band was considered as the system.

\section{Contributions}
We integrated all modern 6G technologies into our system. As we considered THz, ICI will be a big factor. So we demonstrated the significance of ICI in MIMO-OFDM wireless communication systems. To make the optimization approach more time efficient, we updated an existing algorithm for beamforming optimization to optimize the data rate.
\newline
\newline
The research objectives of this thesis are as follows:
\begin{itemize}
\item Integration of all modern 6G technologies in a single system
\item Demonstration of the significance of ICI in MIMO-OFDM wireless communication systems
\item Development of an algorithm for beamforming optimization to maximize the data rate
\item Making the optimization method time efficient 
\item The problem has been solved for several different scenarios and provided a number of insights on how spectral efficiency varies with different parameters.

\end{itemize}

\section{Outline of the Thesis}

The thesis is organized as follows: Section \ref{intro}  provides the background material and represents the basics of the following arguments. Section \ref{related} discusses related works in THz Communications and Massive MIMO-OFDM THz Wireless Systems with Hybrid Beamforming. In Section \ref{model}, system and channel models are described. Then, Section \ref{probform} describes the problem formulation and solution approach. To verify the success rate in solving the problem, Section \ref{num} presents the numerical analysis. Finally, Section \ref{conclu} summarizes the findings and makes recommendations for future research.

%
%
\let\textcircled=\pgftextcircled
\chapter{Related Work}
\label{related}
There is substantial experience with wireless communication systems operating at frequencies lower than 6 GHz. The transition from sub-6 GHz to millimeter-wave presented a variety of technological hurdles, ranging from initial access to beamforming deployment, owing to the time required to construct completely digital solutions \cite{lopez2019opportunities}. The growth of 5G has resulted in several notable improvements. However, it becomes significantly more difficult as the frequency increases. The primary objective of moving to higher frequency bands is to considerably increase the amount of accessible spectrum \cite{tripathi2021millimeter}. The MIMO THz communication has become a research hotspot in recent years.
\section{THz Communications}
Terahertz (THz) communications is a cutting-edge wireless technology for the future beyond fifth-generation (5G+) and sixth-generation (6G) wireless networks \cite{federici2010review}, owing to its enormous potential for ultra-high-data-rate transmission \cite{akyildiz2014terahertz}, the Internet of nano-things \cite{akyildiz2014teranets} and massive connectivity \cite{jornet2012phlame,ullah2022spectral} in the THz band $(0.1-10$ THz). Specifically, the rapidly growing traffic demand, which is predicted to exceed terabits per second (Tbps) within the following $ (5–10) $ years \cite{akyildiz2014terahertz}, can scarcely be met by present wireless technologies operating in the microwave spectrum below six gigahertz (GHz). Due to the lack of accessible spectrum resources in the microwave band, possible data rates are severely constrained, resulting in a surge in research in higher-frequency bands \cite{heath2016overview,lin2016terahertz}, such as the millimeter wave (mmWave) band spanning 30 to 300 GHz. However, the data throughput given by the mmWave band is in the order of ten gigabits per second (Gbps) \cite{rappaport2011state}, which is still much less than the predicted traffic demand. In light of this, THz communication has emerged as a critical enabler for Tbps networks in future wireless data applications, owing to its ultra-large useable bandwidth, which spans from tens of GHz to several THz \cite{liu2010broadband}. In the sub-THz to THz range, hardware constraints such as data converter speed and computing complexity will make optimum usage of broad bandwidths problematic \cite{rajatheva2020white}.
Confronting hardware constraints is a significant difficulty when developing THz communications with massive MIMO. Research into innovative waveforms, hardware impairment mitigation, and novel materials for constructing devices operating in that band is also required. This establishes a new realm with a number of outstanding research questions spanning from hardware to physical layer protocols operating in the sub-THz to THz range. Notably, the fast advancement of THz device technologies, such as novel graphene-based THz transceivers \cite{ju2011graphene} and ultra-broadband THz antennas \cite{liu2010broadband}, enables THz communications to become a reality. Plasmonic nano-antennas for THz communications have been proposed based on the extraordinary properties of nanomaterials such as graphene  \cite{jornet2013graphene}. These nano-antennas are a few nanometers broad and a few micrometers long and may be integrated into nanodevices. Additionally, novel methods based on hybrid electronic-photonic approaches have been presented for realistic THz transceiver systems \cite{sengupta2018terahertz}. With the rapid development of THz circuitry and novel material antennas, THz communications are turning into reality \cite{akyildiz2014terahertz}.

According to the Friis transmission formula,
though, the path loss becomes more severe as the frequency
increases, and thus THz signals undergo higher attenuation
than their mmWave and microwave counterparts. So, one of the major issues affecting THz wave propagation is significant path loss, limiting THz transmission distances to a few meters. As a result, THz communication systems appear to be extremely promising for use in an indoor setting \cite{ju2011graphene}. However, the incredibly short wavelength has an advantage in that a huge number of antennas may be packed firmly into a tiny space at a transceiver. \cite{han2018ultra} develops an end-to-end channel model for ultra-massive multiple-input multiple-output (MIMO) communications in the THz band, taking into consideration the features of graphene-based plasmonic micro antenna arrays. The created model is then used to study the massive MIMO system's ability to operate in both spatial multiplexing and beamforming modes.
Reflection and diffraction dominate signal transmission in microwave-frequency wireless systems, but scattering is frequently ignored. Scattering-induced path components may compensate for radio channel degradation effects, particularly at 30 GHz carrier frequencies. However, at mmWave and THz frequencies, the signal wavelength is equivalent to, or even smaller than, suspended particles in the air (dust, snow, and rain) or surface flaws and roughness. Additionally, conventional statistical models based on droplets \cite{3gpp2018study} are incapable of capturing the mobility of objects in highly directional antenna designs. As a result, it is critical to build new channel models capable of accurately capturing the consistency of propagation properties.
The authors of \cite{docomo20165g} outlined pertinent channel modeling factors and highlighted that accurately defining the spatial consistency of radio channels was the most difficult modification to integrate into classic drop-based simulators. In \cite{3gpp2018study}, the 3GPP standards body described a spatial consistency technique for drop-based simulations that may be used with both cluster and ray-random variables. However, practical specifics such as collecting correlation distances in highly directive narrow-band antenna arrays and beam steering while considering high-resolution scales still need to be included.

One of the difficulties in implementing THz communications with massive MIMO is overcoming hardware limits. By leveraging the distinctive properties of nanomaterials, such as graphene, plasmonic nano-antennas for THz communications have been proposed \cite{jornet2013graphene}. These nano-antennas are just a few micrometers in width and a few micrometers in length and may be incorporated into nanodevices \cite{jornet2013graphene}. Additionally, innovative methods combining electrical and photonic techniques have been established for the creation of practical THz transceivers \cite{sengupta2018terahertz}. A notable benefit of photonic components is the ease with which frequency-division multiple-carrier systems may be produced utilizing the multi-wavelength laser source designed for optical multiplexing networks. However, radio frequency (RF) chain components use significantly more power and are more complicated \cite{zhang2018mixed} than THz nano-antennas, placing severe limits on massive MIMO systems. To address this issue, \cite{lin2015indoor} studied a low-complexity indoor THz wireless system with hybrid beamforming, in which the number of RF chains is significantly fewer than the number of antennas. Additionally, spatial modulation approaches based on densely packed reconfigurable antenna arrays are being researched to boost capacity and spectral efficiency \cite{sarieddeen2019terahertz}. Thus, the high beamsteering gain, multiplexing gain, and spatial diversity gain provided by massive multiple-input multiple-output (m-MIMO) methods may be used to overcome severe route loss \cite{lin2016terahertz}.

\section{Massive MIMO-OFDM THz Wireless Systems with Hybrid Beamforming}
The requirement of orthogonality is used to define the Massive MIMO (m-MIMO) Network. Due to this function, the impacts of fading can be eliminated. It should be emphasized that m-MIMO deployment results in significant interference due to the huge number of antennas mounted on a tiny array. Notably, beamforming is used to address this issue \cite{yang2018digital,hassan2019edge,dutta2019case}. There has been much research on the design of efficient and resilient MIMO systems. For example, in \cite{el2020large}, research on beamforming is undertaken under various conditions, including inter-user interference and thermal noise, which occur when many antennas are employed at the same base station. It has been demonstrated that these issues can be overcome by using hybrid beamforming at the base station.

Existing hybrid beamforming systems frequently assume that infinite-resolution phase shifters will be used to construct analog beamformers. However, the components necessary to construct precise phase shifters can be rather costly \cite{montori2010design,el2020large}. In practice, low-resolution phase shifters with a lower cost are frequently utilized. The simplest method for designing beamformers with limited resolution phase shifters is first to build the RF beamformer with infinite resolution and then quantize the value of each phase shifter to a finite set \cite{liang2014low}. However, this approach is not useful in systems with low-resolution phase shifters \cite{sohrabi2015hybrid}. Additionally, a low-complexity beamforming approach was presented in \cite{zhu2016novel} for multiuser MIMO systems.

Hybrid beamforming design for OFDM systems is tremendously demanding as its analog precoders and combiners are shared across all subcarriers. By utilizing Jensen's inequality, a heuristic approach is presented for OFDM-based multiuser hybrid precoding in \cite{sohrabi2017hybrid}. However, approaches in \cite{zhu2016novel} and \cite{sohrabi2017hybrid} are based on the average channel information, which decreases computing complexity but causes performance degradation. \cite{zilli2021constrained} presents a unique two-stage joint hybrid precoder and combiner architecture for optimizing the average attainable sum -rate of frequency-selective millimeter-wave massive MIMO-OFDM systems.

For channels with enormous transmitting or receiving arrays or surfaces, a special model is required. Such channel models must be consistent across frequencies and space at high frequencies while also capturing radio channel characteristics, material properties, and radio access interface features. However, because some characteristics, such as path loss and angle of arrival (AoA), cannot be assumed to remain constant between the antennas, the channel model becomes non-stationary in space \cite{de2020non}. Consistency across frequency bands is critical for successful localization when the user may simultaneously use many bands, such as when the network implements the control and user plane separation (CUPS) concept. Conversely, the channel coherence time will be significantly reduced, demanding faster and more frequent updates to capture radio changes completely.

Unfortunately,  considering the cost and energy consumption, conventional digital beamforming requires a  dedicated  RF chain for each antenna element, which is no longer suitable for the THz massive  MIMO  systems. A hybrid structure as a  new MIMO architecture can effectively reduce the required number of RF  chains. In this structure,  the analog and digital precoder provide the beamforming and multiplexing gains, respectively. To date,  most research focuses on two hybrid beamforming (HBF) structures, such that fully- \cite{el2014spatially} and partially- \cite{gao2016energy} connected architecture. \cite{lin2016energy} studied fully connected and partially connected structures and demonstrated that the fully connected structure has a more excellent spectral and energy efficiency than the subconnected structure when insertion loss is included. Many topologies have been thoroughly researched to overcome the limitation of the number of RF chains. Analog RF beamforming schemes implemented using analog circuitry are introduced in \cite{venkateswaran2010analog,chen2011multi,tsang2011coding,hur2013millimeter}. They commonly employ analog phase shifters, which limit the beamformer's components to have a constant modulus. As a result, analog beamforming performs poorly in comparison to completely digital beamforming methods. Another method for restricting the number of RF chains is using simple analog switches to achieve antenna subset selection \cite{sanayei2004antenna,molisch2005capacity,sudarshan2006channel}. They cannot, however, achieve complete diversity gain in correlated channels since the antenna selection strategy uses just a subset of channels \cite{molisch2003reduced,molisch2004fft}. Although \cite{lin2015indoor,lin2016energy} are self-contained works, none address frequency selective fading in wideband THz systems.
Given that THz communications are expected to operate over broadband channels \cite{han2015multi}, designing frequency-selective hybrid beamforming techniques for wideband THz systems is critical. Notably, research into frequency-selective hybrid beamforming design in THz systems is still in its infancy. Recent research has focused on hybrid beamforming for mmWave systems operating across broad channels.

With massive antennas in wideband mmWave systems, the transmit signals will be sensitive to the spatial-wideband effect, i.e., the physical propagation delay of electromagnetic waves traveling across the array becomes large \cite{wang2018spatial}. Taking the spatial-wideband effect into account, \cite{wang2019beam} proposed a channel estimation scheme for frequency-division duplex mmWave systems with hybrid beamforming architecture, in which the frequency-insensitive parameters of each uplink channel path are extracted using a super-resolution compressed sensing approach. In \cite{alkhateeb2016frequency}, an effective hybrid beamforming technique was suggested for orthogonal frequency-division multiplexing (OFDM)-based mmWave system with restricted feedback channels that optimizes possible mutual information under total power and unitary power restrictions. \cite{alkhateeb2016frequency} also established a helpful criterion for hybrid codebook creation, in which both the baseband and RF precoders are obtained from quantized codebooks. Following \cite{alkhateeb2016frequency, park2017dynamic} investigated the dynamic subarrays design for wideband hybrid beamforming, proposing a criterion for creating the best subarrays that optimize spectral efficiency. 

The important findings in \cite{alkhateeb2016frequency, park2017dynamic, sohrabi2017hybrid} provide valuable insights and guidelines to the design of frequency-selective hybrid beamforming while having several limitations, as follows:

1) The impact of molecule absorption on signal transmission in the mmWave and THz bands was not considered. Furthermore, \cite{sohrabi2017hybrid} showed that hybrid beamforming with a small number of RF chains could asymptotically approach the performance of fully digital beamforming for a sufficiently large number of transceiver antennas.

2) Researchers did not make use of the strong channel correlation between THz subcarriers, resulting in an increase in computing complexity.

(3) Researchers assumed a completely digital receiver and perfect carrier frequency offset (CFO) synchronization, which may not be feasible. However, CFO is a widespread issue in THz communications, as the Doppler shift of THz channels is orders of magnitude greater than that of conventional microwave channels \cite{you2017bdma}.

4) No consideration was given to resilient design against imprecise channel state information (CSI). Perfect CSI at the transmitter is exceedingly difficult to obtain in massive MIMO systems, especially for frequency-selective channels.
Due to the constraints mentioned above, the approaches presented in \cite{alkhateeb2016frequency,park2017dynamic, sohrabi2017hybrid} cannot be directly applied to wideband THz wireless systems.

In \cite{yu2016alternating}, a new structure employing the fully digital block diagonalization (BD) algorithm for mmWave MIMO-OFDM  systems is proposed. However,  this structure requires more phase shifters with infinite precision,  which is unrealistic for practical systems. Moreover, BD or zero-forcing  (ZF)  schemes are commonly used to eliminate multiuser Interference (MUI). Since MUI can be canceled entirely, the BD or ZF  schemes result in a  good performance.

For the downlink of K-user multiple input single output (MISO) systems, \cite{liang2014approach, liang2014low} demonstrate that hybrid beamforming with K RF chains at the base station may obtain an acceptable sum rate in comparison to the sum-rate achieved by completely digital zero-forcing (ZF) beamforming, which is near ideal for large MIMO systems \cite{rusek2012scaling}. The architecture of \cite{liang2014approach, zhu2016novel} entails matching the RF precoder to the channel's phase and configuring the digital precoder to act as the effective channel's ZF beamformer.

Intercarrier interference (ICI) is induced by analog channelization failure. ICI arises as a result of the varied subcarrier spacings \cite{marijanovic2017intercarrier, schaich2015subcarrier}. While maintaining an equal level of service and accounting for ICI, variable channel conditions, and erroneous channel information, it is possible to establish the appropriate bandwidth allocation for each user \cite{sathananthan2001probability}. Sathananthan et al. \cite{sathananthan2001probability} discussed carrier frequency offset (CFO) in OFDM transmissions since it results in ICI. CFO occurs as a result of carrier frequency mismatch or Doppler shift. OFDM is susceptible to CFO and phase noise; for a high number of subcarriers, these defects will result in subcarrier orthogonality being lost and intercarrier interference being introduced. Numerous techniques for mitigating the CFO's influence on OFDM have been devised \cite{zhao1996sensitivity, zhao1998intercarrier, moose1994technique, armstrong1999analysis}. Pollet et al. \cite{pollet1995ber} use basic mathematical manipulation to calculate the approximate decrease in signal-to-noise ratio (SNR) caused by the CFO.

Certain hybrid precoder (HP) architectures have been designed for single-user and multiuser massive MIMO systems for narrow-band flat fading channels \cite{alkhateeb2014channel,alkhateeb2015limited,liang2014low}. The broadband frequency-selective fading (FSF) channel is more practicable and demanding because of the vast bandwidth and varying delays of the numerous scattering paths. Hybrid precoding based on orthogonal frequency division multiplexing (OFDM) is recommended for converting the FSF channel into several parallel narrow-band flat fading channels \cite{gao2016channel}. \cite{park2017dynamic} and \cite{sohrabi2017hybrid} offered a single-user system architecture method.

Specific hybrid precoding methods for multiuser systems across FSF channels have been proposed. An advanced concept of alternating optimization to design hybrid precoding was proposed in \cite{du2018hybrid}. Several low-complexity approaches have also been examined in \cite{du2018hybrid} and \cite{zhu2016novel} to achieve a trade-off between performance and computing complexity. \cite{liu2019low} proposed a low-complexity hybrid precoding architecture that decouples digital and analog precoding and is suited for mmWave applications. However, they did not consider ICI. \cite{du2018hybrid} discussed the construction of a hybrid analog-digital beamforming system for a multiuser m-MIMO system based on OFDM. They suggested an efficient alternating algorithm designed for hybrid beamforming based on manifold optimization. To be more precise, analog beamforming is optimized by manifold optimization, whereas digital beamforming is constructed using the MMSE criterion. Additionally, two simpler designs are presented to reduce computational complexity without sacrificing performance significantly. However, their technique was limited to mmWave frequencies and did not consider ICI. \cite{yuan2018hybrid, yuan2020hybrid} proposed a hybrid beamforming system for MIMO-OFDM Terahertz wireless systems over frequency-selective channels. They considered ICI, but their technology was restricted to a single user. Since they were working with a single user, they designed the system to mitigate the effect of ICI on the user's other sub-carriers.\cite{yuan2018hybrid} suggested a beam steering codebook searching technique for designing analog BF based on the propagation model in the THz band. This approach takes into account the channel state information of all subcarriers. They then proposed designing digital BF using the regularized
channel inversion (RCI) approach in conjunction with ICI removal. The orthogonality requirement defines the Massive MIMO (m-MIMO) Network. Due to this function, the impacts of fading can be eliminated. It should be emphasized that m-MIMO deployment results in significant interference due to the huge number of antennas mounted on a tiny array. Notably, beamforming is used to address this issue \cite{yang2018digital,hassan2019edge,dutta2019case}. There has been much research on the design of efficient and resilient MIMO systems. For example, in \cite{el2020large}, research on beamforming is undertaken under various conditions, including inter-user interference and thermal noise, which occur when many antennas are employed at the same base station. It has been demonstrated that these issues can be overcome by using hybrid beamforming at the base station.

\section{Summary of Literature Review}
So, hybrid beamforming with OFDM massive MIMO multiuser system for THz communication considering ICI was missing. In \cite{du2018hybrid}, they considered hybrid beamforming design for multiuser massive
MIMO-OFDM System, but ICI was missing in their system and in \cite{yuan2018hybrid, yuan2020hybrid} they considered hybrid beamforming design for massive MIMO-OFDM system considering ICI but for a single user. In our studied system, a multi-antenna base station (BS) employing the fully-connected hybrid architecture broadcasts utilizing multi-carrier modulation to a single-antenna user equipment (UE). For simplicity, we consider the limitation that each subcarrier must have equal power. The adaptive power restriction over subcarriers is a fascinating topic, but it is not the main emphasis of this study. We discuss the carrier frequency offset (CFO) in OFDM transmission \cite{sathananthan2001probability} because it results in inter-carrier interference (ICI).

\space The goal of this thesis was to analyze the performance of multiuser massive MIMO-OFDM THz wireless systems with hybrid beamforming under ICI.

\clearpage 

%
%
\let\textcircled=\pgftextcircled
\chapter{System And Channel Models}
\label{model}
This section demonstrates the detailed system model of this research. The ICI model, which is crucial to this work, is also discussed here. Later in this section, the channel model is shown. The path gain, azimuth, and elevation angle related to the channel model are also addressed later.
\section{System Model}
\begin{figure}[!ht]
  \centerline
  {\includegraphics[scale=0.5]{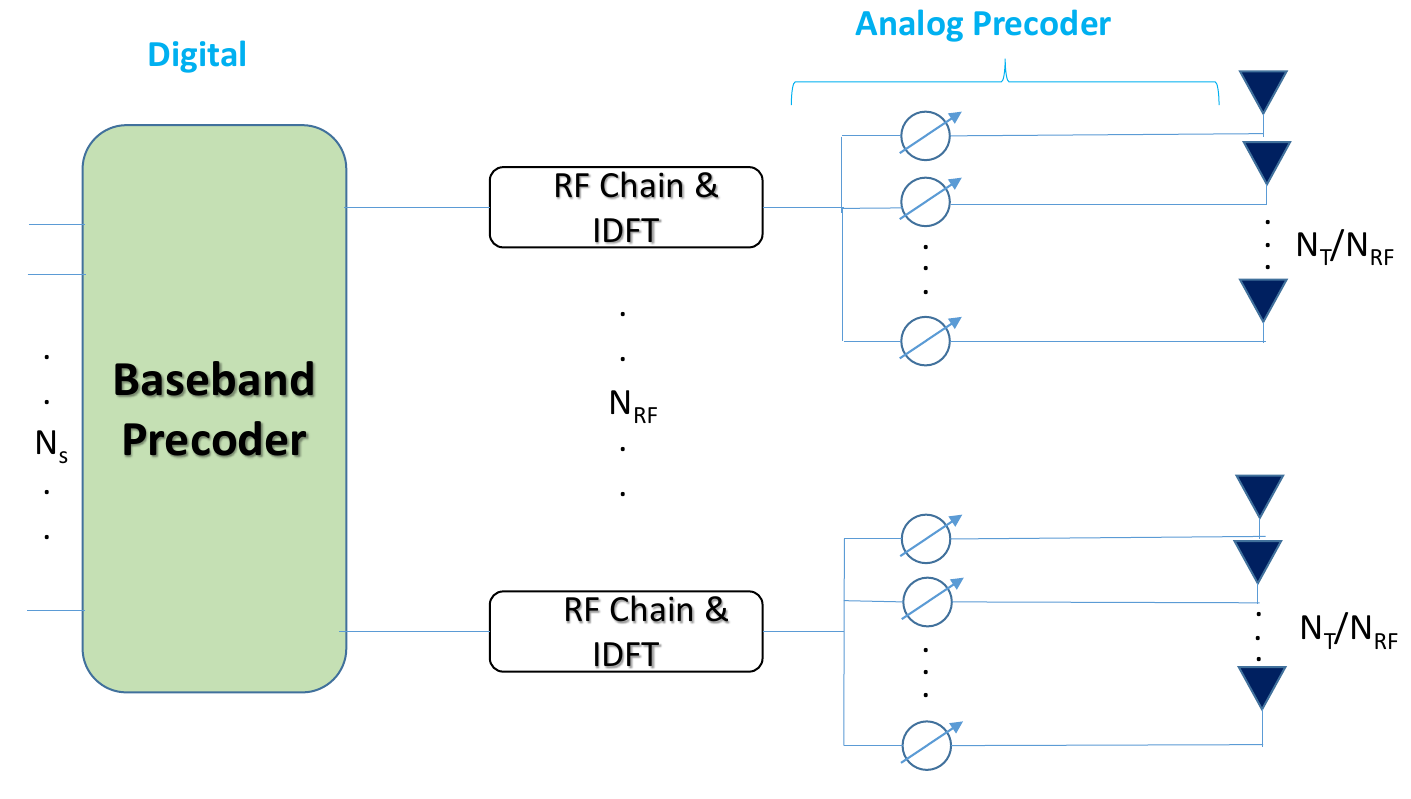}}
 \caption[Possible 6G Communication Architecture Scenario.]{Sub-connected architecture of the base station} 
 \label{partially_connected}
\end{figure}

\begin{figure}[!ht]
  \centerline
  {\includegraphics[scale=0.5]{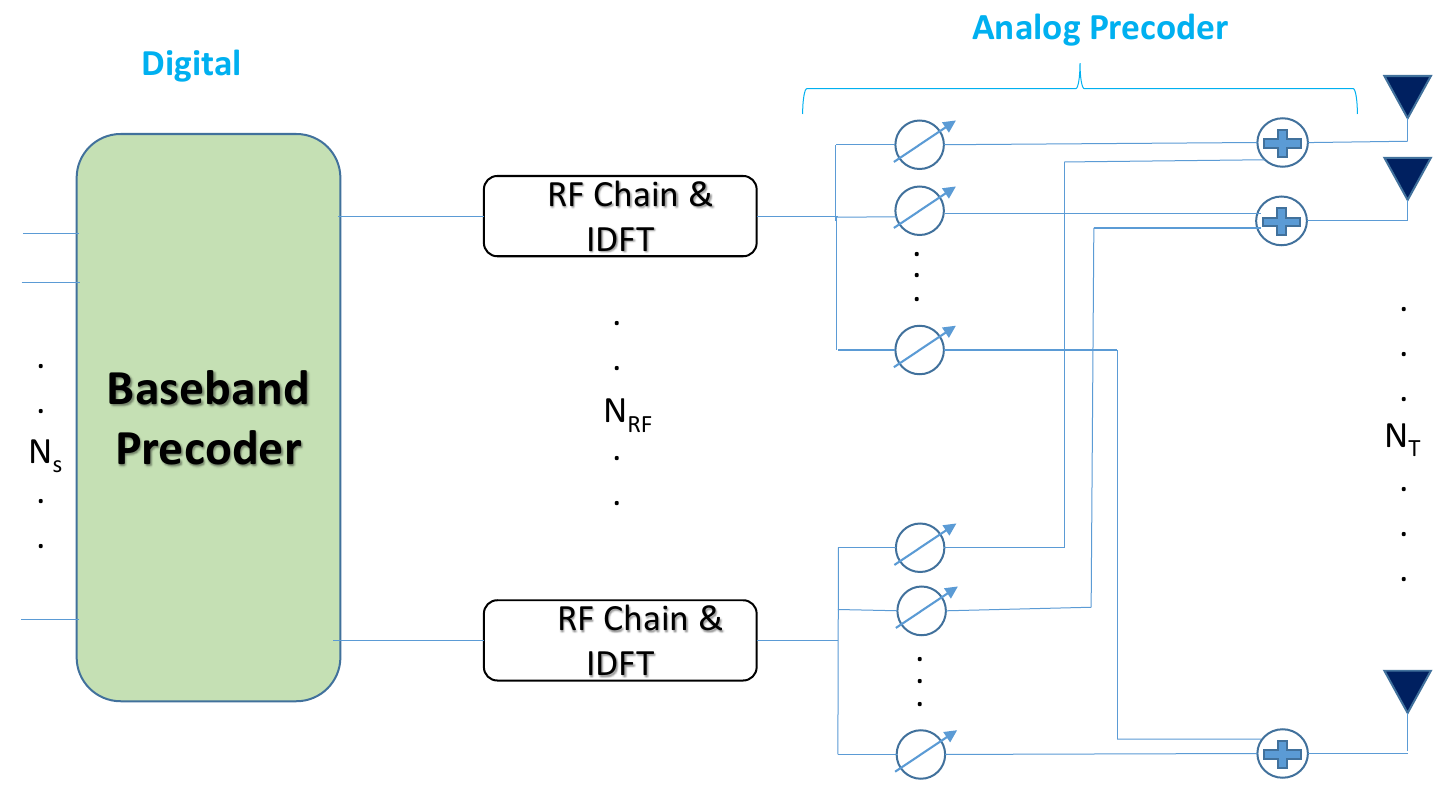}}
 \caption[Possible 6G Communication Architecture Scenario.]{Fully connected architecture of the base station} 
 \label{fully_connected}
\end{figure}

We consider a multi-user communication system for OFDM-based massive MIMO using the THz frequency band. The base station is equipped with $N_{RF}$ RF chains and $N_t$ transmit antennas, allowing it to simultaneously serve U number of single-antenna users. We assume the BS transmits using OFDM. The OFDM base station uses K number of subcarriers where all subcarriers simultaneously serve all single-antenna users. In the THz range, the channel performance diminishes as the distance increases. So designed system prototype can be employed in indoor and other short-distance applications. 
\\
There are two common approaches for linking antennas to RF chains: the completely connected architecture and the sub-connected architecture. In partially connected architecture like figure ~\ref{partially_connected}, disconnected subarrays are driven by the individual RF chain. The antenna is divided into many smaller groups, each of which will be coupled to a distinct RF chain through a sub-connected structure \cite{liu2021initial}. With this configuration, each RF chain is limited to driving a single disjoint subarray of $\frac{N_T}{N_{RF}}$ antennas. A total of $N_T$ antennas can be driven by such $N_{RF}$ RF links.
The performance metrics like spectral efficiency and energy efficiency of a terahertz system with two distinct structures are largely determined by the beamforming approach  \cite{lin2016terahertz}. Because of the completely integrated architecture, antennas are shared throughout multiple radio frequency chains. The result is that in a fully integrated design, one radio frequency chain is capable of operating the whole antenna array. In contrast, in a fully connected architecture, an RF chain is connected to all antennas by a series of phase shifters unique to each antenna, as shown in figure ~\ref{fully_connected}. In this architecture, all $N_T$ transmit antennas are connected to each of all $N_{RF}$ Chains.
\\
\cite{lin2016energy} examined fully connected and sub-connected structures and proved that, when insertion loss is included, the fully connected structure has superior spectral and energy efficiency to the sub-connected structure. As a result, the system model used in this thesis is a fully linked architecture. Different subcarrier signals are digitally precoded and then transformed to the time domain using Inverse Fast Fourier transforms (IFFTs) at the Base stations  \cite{du2019weighted}. \\
This fully connected architecture transmits $N_s$ data streams over $N_t$ transmit antennas to serve $U$ single antenna users simultaneously. $N_t$ transmit antennas connected to $N_{RF}$ RF chains at the base stations. The number of RF chains at the BS is subject to the constraints $N_s \le N_{RF} \le N_t$ \cite{7397861}. And Constraints for transmit antenna is $N_t \gg N_{RF}$ \cite{8491217}.

 At the BS, $S = [s[1], s[2], \cdots, s[K]]^T$ is a $N_s \times K$ vector, where $s[k], k \in \{1, 2,\cdots, K\}$, denotes the
data transmitted to the users at the $k^{th}$ subcarrier and $[\cdot]^T$
denotes the transpose. The symbol vector $s[k]$ is a $N_s \times 1$ such that, $s[k] = [s_1,s_2,s_3,.....,s_{N_s}] $. s[k] is precoded by the $N_{RF} \times U$
digital baseband beamformer F[k] in the frequency domain and
then the precoded signal is transformed into the time domain
utilizing the K-point IFFT. Since the RF BF is implemented after the IFFT, the analog RF beamformer is fixed for all subcarriers. Hence,  the transmitted symbol at $k^{th}$ subcarrier can be expressed as $x[k] = W\; F[k]\; s[k]$. 
 W is a $N_t \times N_{RF}$ analog precoding matrix which can be written as, 
 \begin{equation*}
     W = \;\;
     \begin{bmatrix}
w_{11} & w_{12} & ...& w_{1{N_{RF}}}
\\
w_{21} & w_{22} & ...& w_{2{N_{RF}}}
\\
.&.&.&.
\\
.&.&.&.
\\
.&.&.&.
\\
w_{N_t1} & w_{N_t2} & ...& w_{N_t{N_{RF}}}
\end{bmatrix}
 \end{equation*}
 where each entry in $W$ is limited by a constant modulus constraint such that
$|w_{nm}|=1 / \sqrt{N_t}$, $m \in\{1,2, \ldots, N_{RF}\}$ , $ n \in\{1,2, \ldots, N_t\}$ and $| \cdot |$ denotes the norm. Furthermore,
we consider the power constraint for each subcarrier which is imposed by normalizing $f_k$ such that $\left\|\mathbf{W} \mathbf{f}_{k}\right\|_{F}^{2}=L$, where
$||\cdot ||_F$ denotes the Frobenius norm and L indicates number of data streams available in the system.
\\
Moreover, F is a $N_{RF} \times U \times K$ sized baseband digital beamforming matrix and F[k] is a $N_{RF}\times U$ matrix, which can be expressed as,
 \begin{equation*}
 F[k] = \;\;
     \begin{bmatrix}
f_{11} & f_{12} & ...& f_{1{U}}
\\
f_{21} & f_{22} & ...& f_{2{U}}
\\
.&.&.&.
\\
.&.&.&.
\\
.&.&.&.
\\
f_{N_{RF}1} & f_{N_{RF}2} & ...& f_{N_{RF}U}
\end{bmatrix}
 \end{equation*}
 H is a $U \times N_t \times K$ channel matrix where K is the total number of subcarriers and k is the indexing parameter of subcarrier. $H[k]$ is a $U \times N_t$ matrix which can be further expressed as,
 \begin{equation*}
 H[k] = \;\;
     \begin{bmatrix}
h_{11} & h_{12} & ...& h_{1{N_t}}
\\
h_{21} & h_{22} & ...& h_{2{N_t}}
\\
.&.&.&.
\\
.&.&.&.
\\
.&.&.&.
\\
h_{U1} & h_{U2} & ...& h_{U{N_t}}
\end{bmatrix}
 \end{equation*}
 
At the UE, the received signal at the $k^{th}$ subcarrier is given
by $ y_k =H_k x_k+n_k $, where $H_k$ is the $U \times N_t$ THz channel matrix between the
BS and the UE, and $n_k$ denotes the additive Gaussian white
noise (AWGN) at the UE with the power of $\sigma_n^2$.

This article considers the CFO in the OFDM system for practicality. CFO is induced by carrier frequency mismatch or Doppler shift in practice. As a result, after FFT at the receiver, the combined signal at the $k^{\text{th}}$ subcarrier is written as \cite{wu2007signal} :
\begin{equation}\label{eq3.1}
     y[k] = S_0\; H\left[k\right]\;W\; F[k]\;s[k] + \Delta + n[k]
\end{equation}

Here, $ \Delta=\sum_{u=1}^U \sum_{\substack{i=0 \\ \text{when } u=q,i\neq k}}^K S_{i-k} h_q[i] W F_u[i]S_u[i] $ denotes the interference caused by other sub-carriers of the same and other users, and $S$ indicates Inter-Carrier Interference (ICI) coefficient. And $n[k]$ denotes the additive Gaussian white noise (AWGN) at the UE with the power of $\sigma_n^2$. 

\section{ICI model}
The sequence $S_i , i \in \{1-K,\cdots,0,\cdots,K-1\}$ denotes the ICI coefficient which depends on the CFO and is given by \cite{sathananthan2001probability}
\begin{equation}
    S_{i} = \frac{\sin \pi(i+\varepsilon)}{K \sin \frac{\pi}{K}(i+\varepsilon)} e^{j \pi\left(1-\frac{1}{K}\right)(i+\varepsilon)}
\end{equation}

Here, $\varepsilon $ is the ratio between the CFO and the subcarrier spacing. For zero CFO, $S_i$ reduces to the unit impulse sequence. We clarify that ICI mainly captures the power leakage from neighboring subcarriers. Thus, the ICI decreases when the bandwidth of subcarriers increases. With the employment of directional antennas, the path loss and delay spread of channels are reduced, but the ICI also increases \cite{sathananthan2001probability}.
\section{Channel Model}
In this thesis, the Saleh-Valenzuela model is applied to generate the channel model. The S-V model is a statistical, cluster-based channel model extensively employed in mmWave indoor applications. It connects the clustering phenomena with the stochastic angles of departure/arrival for each beam. 
The S-V model, which has been adjusted for the THz band, is employed in this research.
The channel vector for $q^{th}$ user can be written as
\begin{align}
    H_q(f, d) &= \sqrt{\frac{N_t}{N_c \cdot N_{\text{ray}}}}
    \sum_{i=1}^{N_{c}} \sum_{j=1}^{N_{\text{ray}}} \alpha_{ij}^q(f, d) a^q(\phi_{ij}, \theta_{ij}) G_t G_r
\end{align}
Here, $\alpha^q_{ij}$ $\in$ $\mathbb{C}$ which is the channel gain of $i^{th}$ ray and $j^{th}$ cluster. $N_c$ and $N_{ray}$ are used to denote the number of clusters and the number of rays in that cluster. $a^q(\phi_{ij},\theta_{ij})$ are used to denote the transmit antenna array response at the base station where $\phi_{ij}$ and $\theta_{ij}$ are the azimuth and elevation angles of the angle of departure (AOD). The transmit antenna array response of $i^{th}$ ray of $j^{th}$ cluster and $u^{th}$ user can be written as,

\begin{equation}
    a^q(\phi_{ij},\theta_{ij}) = \frac{1}{\sqrt{N_t}}[1, \ldots, e^{j\frac{2 \pi}{\lambda}d(p\sin{\phi_{ij}\sin{\theta_{ij}}}+q\cos{\theta_{ij}})}, \ldots, e^{j\frac{2 \pi}{\lambda}d((\sqrt{N}-1)\sin{\phi_{ij}\sin{\theta_{ij}}}+(\sqrt{N}-1)\cos{\theta_{ij}})}]^T
\end{equation}

In this equation, \(d\) is the antenna spacing, \(\lambda\) is the wavelength of the signal, \(p\) and \(q\) are the antenna indices of the horizontal and vertical plane such that \(1 \leq p < \sqrt{N_t}\) and \(1 \leq q < \sqrt{N_t}\) as the antenna plane is considered to be square.

\subsection{Path Gain}

Spreading loss and molecular absorption loss make up Terahertz channel path loss. As channel gain is heavily dependent on molecular absorption loss at high frequencies, path gain is composed of spreading loss \(L_{spr}\) and absorption loss \(L_{abs}\).

Thus, the loss can be expressed as
\begin{equation}
    L(f,d) = L_{spr}(f,d) L_{abs}(f,d)
\end{equation}

where spreading loss is
\begin{equation}
    L_{spr}(f,d) = \left(\frac{c}{4 \pi f d}\right)^{2}
\end{equation}

and absorption loss is
\begin{equation}
    L_{abs}(f,d) = e^{-k_{abs}(f)d}
\end{equation}

Hence, the path gain for \(q^{th}\) can be formulated as
\begin{equation}
    |\alpha^q(f,d)|^2 = L(f,d) = \left(\frac{c}{4 \pi f d}\right)^{2} e^{-k_{abs}(f)d}
\end{equation}

\subsection{Azimuth $\And$ Elevation Angle}

The angular power delay profile of each individual cluster can be denoted as \(P_i(\phi,\theta)_{cluster}\), and the complete power delay profile can be computed as \cite{priebe2011aoa}
\begin{equation}
    P(\phi,\theta) = \sum_{i=1}^{N_{c}} P_i(\phi,\theta)_{cluster}
\end{equation}

If no correlation between angular power profiles is considered, then the angular power profile for each cluster can be rewritten as
\begin{equation}
    P_i(\phi,\theta)_{cluster} = P_i(\phi)_{cluster} \cdot P_i(\theta)_{cluster}
\end{equation}

The azimuth and elevation angle of each ray are assumed to be independent \cite{840194}. If the azimuth angle of the \(i^{th}\) cluster is \(\varphi_i^t\), and the azimuth angle of the \(j^{th}\) ray within the \(i^{th}\) cluster is \(\Phi_{ij}^t\), then the total azimuth angle can be written as
\begin{equation}
    \phi_{ij}^t = \varphi_i^t + \Phi_{ij}^t
\end{equation}

Likewise, if the elevation angle of the \(i^{th}\) cluster is \(\vartheta_i^t\), and the elevation angle of the \(j^{th}\) ray within the \(i^{th}\) cluster is \(\Theta_{ij}^t\), then the total elevation angle is
\begin{equation}
    \theta_{ij}^t = \vartheta_i^t + \Theta_{ij}^t
\end{equation}

Here, \(\Phi_{ij}^t\) and \(\Theta_{ij}^t\) follow a Zero-Mean Second Order Gaussian Mixture Model that can be written as \cite{7036065}
\begin{equation}
    GMM(x) = \frac{a_1}{2\pi \sigma_1} e^{-\frac{1}{2}\left(\frac{x - \overline{x_1}}{\sigma_1}\right)^2} +
    \frac{a_2}{2\pi \sigma_2} e^{-\frac{1}{2}\left(\frac{x - \overline{x_2}}{\sigma_1}\right)^2},
\end{equation}

where \(\overline{x_1}, \overline{x_2} = 0\). Azimuth and elevation angles of each cluster, \(\varphi_i^t\) and \(\vartheta_i^t\), follow the uniform distribution where \(\varphi_i^t \in (-\pi,\pi]\) and \(\vartheta_i^t \in [-\frac{\pi}{2},\frac{\pi}{2}]\).

%
%

\let\textcircled=\pgftextcircled
\chapter{Problem Formulation and Solution Approach}
\label{probform}
In this section, the SINR model and optimization problem are described first. Following that, a detailed methodology for resolving the optimization problem utilizing existing approaches is provided. The proposed complexity-reduced hybrid precoding design with a redesigned objective function is presented further in this section.

\section{SINR Model}
The signal-to-interference-plus-noise ratio (SINR) is used in information theory and telecommunication engineering to set theoretical upper constraints on channel capacity in wireless communication systems such as networks. The SINR is defined as the power of a particular signal of interest divided by the total of the interference power and the power of some background noise, similar to the signal-to-noise ratio (SNR) frequently used in wired communications systems.
It is clear from the previous chapter that at the BS, $s[k], k \in \{1, 2,\cdots, K\}$, denotes the
data transmitted to the users at the $k^{th}$ subcarrier with $\mathbb{E}\left\{\mathbf{s}[k] \mathbf{s}^{H}[k]\right\}=PI$
 in which P is the transmit power for each user at each subcarrier. Here  $ \mathbb{E} [\cdot]$ denotes expectation, $ S^H $ denotes the Hermitian transpose of S, and $I$ denotes the identity matrix. n[k] denotes the additive Gaussian white noise (AWGN) at the UE with the power of $\sigma_n^2$. From equation \ref{eq3.1} it can be derived that the SINR for $ k^{th} $ subcarrier of $ q^{th} $ user is:
\\
\begin{equation}\label{4.1.2}
\gamma_{q,k} = \log_2\left(1 + \frac{\left|S_{0}\right|^2\left|h_{q}[k] W F_{q}[k]\right|^2}{\sum_{u=1}^{U} \sum_{\substack{i=0 \\ \text{when } u=q, i \neq k}}^{K}\left|S_{i-k}\right|^2\left|h_{q}[i] W F_{u}[i]\right|^2 + \psi}\right)
\end{equation}

where $\psi = \frac{\sigma_n^2}{P}$.

\section{Optimization Problem}
From equation \ref{4.1.2}, the achievable rate of the considered system is derived as:

\begin{equation}\label{4.2}
R = \frac{1}{U}\sum_{q=1}^U\sum_{k=1}^K \gamma_{q,k}
\end{equation}

Constraints:
\begin{equation}\label{4.3}
\begin{aligned}
& \{\left\{(w_n)_m\right\}_{n=1}^{N_{RF}},_{m=1}^{N_t}\;,\left\{(F_k)_u\right\}_{k=1}^K,_{u=1}^U \} \\
& = \text{argmax}\; R, \\
& \in |(w_n)_m| = \frac{1}{\sqrt{N_t}}, \quad \forall 1\leq n\leq N_{RF}, \;1\leq m\leq N_t, \\
& \left\| \|W F_u[k]\|_F^2 \right\| = L, \quad 1\leq k\leq K.
\end{aligned}
\end{equation}

To address this issue, we suggest the following design strategies. In order to investigate the hybrid beamforming architecture's performance constraints, perfect channel state information (CSI) is assumed to be available. Solving the maximizing problem across two series of variables is the most challenging part of this problem. This part presents a hybrid precoding maximization approach based on an iterative analog precoder design to optimize system spectral efficiency. Rather than addressing the original optimization problem with two series of variables, the suggested maximizing method divides the issue into two problems, each with just one series of variables. We will maximize the system spectral efficiency with respect to W and F, respectively, while keeping the other fixed.
\section{Solution with Existing Techniques}
\subsection{Digital RF Precoding Design}
In this work, we directly follow zero-forcing (ZF) beamforming. In a multi-user MIMO wireless communication system, zero-forcing precoding is a spatial signal processing approach that allows many antenna transmitters to eliminate multi-user interference. The ZF precoder is supplied by the pseudo-inverse of the channel matrix when the channel state information is adequately known at the transmitter. The effective channels, rather than the original channels, are used to build the digital beamformer. The effective channel matrix for all users for each subcarrier is calculated as follows:
\begin{equation}\label{4.4}
He[k]^H=H[k]^H W. 
\end{equation}

The base station calculates ZF precoder based on effective channels as:
\begin{equation}\label{4.5}
F_u[k]=(He[k]He[k]^H)^{-1}He[k]. 
\end{equation}

The digital precoder is finally normalized as: $F_u[k]=\frac{F_u[k]}{||WF_u[k]||_f^2}$
to fulfill the power constraints.

\subsection{Analog RF Precoding Design}
Consider designing the analog precoder W with fixed digital precoders $F_u[k]$. So, the problem from the \ref{4.2} can be reformulated as follows:

\begin{equation}\label{4.6}
\begin{aligned}
\max_W & \sum_{k=1}^K \sum_{u=1}^U R_u[k] \\
\text{s.t.} & \quad |(w_n)_m| = \frac{1}{\sqrt{N_t}}, \quad \forall 1 \leq n \leq N_{RF}, \; 1 \leq m \leq N_t, \\
& \quad \left\| \|W F_u[k]\|_F^2 \right\| = 1, \quad 1 \leq k \leq K.
\end{aligned}
\end{equation}

Because of the constant amplitude limitations, we cannot directly solve this issue in Euclidean space. Because the constant amplitude constraints establish a Riemannian manifold, we might use a solution based on manifold optimization. Because the neighborhood of each point on a manifold resembles Euclidean space, optimization methods developed for Euclidean space might be used in manifolds. To enhance spectral efficiency, we want to create the conjugate gradient method for analog precoder design in this study \cite{yu2016alternating}.
We transfer W into a vector  $  x = vec[W].$ Then, we can represent the manifold of the analog beamforming as
\begin{equation}\label{4.7}
M=\{x\in \mathbb{C}^{N_t U}:|x_i|=\frac{1}{\sqrt{N_t}},i=1,2,\cdots ,N_t U\}.     
\end{equation}

where $x_i$ is the $i^{th}$ element of x. By treating $\mathbb{C}$ as $ \mathbb{R}^2 $ with the canonical inner product, we define the Euclidean metric in the complex $ \mathbb{C} $  plane as
\begin{equation}\label{4.8}
\langle x1,x2\rangle=\Re[x1^*x2]. 
\end{equation}

which enables us to denote the tangent space of x as
\begin{equation}\label{4.9}
\tau_x M=\{Z \in \mathbb{C}^{MU}:\Re[Z\; o\; x]=0\}.
\end{equation}

According to x, we define the cost function as
\begin{equation}\label{4.10}
f(x)=R_{sum}=\sum_{k=1}^K \sum_{u=1}^U R_u[k].
\end{equation}

Moreover, since the inner product for the Riemannian manifold is defined on the tangent space of the given point x, the optimization algorithm should also be conducted on the tangent space of x. In other words, when calculating factors for the current point x, we need to cancel the parts not in the
tangent space of x by the orthogonal projection as
\begin{equation}\label{4.11}
T_x(d)=d-\Re[d \;o\; x^*]\; o\; x. 
\end{equation}

However, we use the algorithm on the tangent space of x, which makes the new point be on its tangent space but not 
guaranteed on the Riemannian manifold. Therefore, we should
also map the new point onto the Riemannian manifold by retraction, which can be stated as
\begin{equation}\label{4.12}
R(x)=\frac{1}{\sqrt{N_t}}\;vec\left[\frac{x_1}{|x_1|},\frac{x_2}{|x_2|},\cdots,\frac{x_{N_t U}}{|x_{N_t U}|}\right]. 
\end{equation}

Note that, in the proposed algorithm, the retraction restricts the constant amplitude constraints to the vector again.

\begin{algorithm}
	\caption{Conjugate Gradient Algorithm for Analog Based on Manifold Optimization}
	\label{decen_algo}
	\begin{algorithmic}[1]
	
		\State $Input: F[k]_{k=1}^K , x_0, s , \epsilon_1 $
	    
	     \State$\mathbf{d}_{0}=\mathrm{T}_{\mathbf{x}_{0}}\left(\left.\frac{\mathrm{d} f(\mathbf{x})}{\mathrm{d} \mathbf{x}}\right|_{\mathbf{x}=\mathbf{x}_{0}}\right), t=0$ and $f\left(\mathbf{x}_{-1}\right)=0;$
	   
	    \While ${f(x_t)-f(x_{t-1})>\epsilon_1}$
	    
		    \State Choose the step size:$\alpha_t=\frac{s}{||d_t||_F};$
		
		    \State Update the position as $x_{t+1} = x_t + \alpha_t d_t ;$
		
		    \State Map the vector onto the manifold: $x_{t+1} = R(x_{t+1})$;
		
		    \State Compute Riemannian gradient:$\mathbf{g}_{t+1}=\mathrm{T}_{\mathbf{x}_{t+1}}\left(\left.\frac{\mathrm{d} f(\mathbf{x})}{\mathrm{d} \mathbf{x}^{*}}\right|_{\mathbf{x}=\mathbf{x}_{t+1}}\right) $;

            \State Calculate Polak-Ribiere parameter as:
        $\beta_{t+1}=\frac{\mathbf{g}_{t+1}^{H}\left(\mathbf{g}_{t+1}-\mathrm{T}_{\mathbf{x}_{t+1}}\left(\mathbf{g}_{t}\right)\right)}{\left\|\mathrm{T}_{\mathbf{x}_{t+1}}\left(\mathbf{g}_{t}\right)\right\|_{F}^{2}};$  
        
		    \State Determine conjugate direction: $d_{t+1}=g_{t+1}+\beta_{t+1} T_{x_{t+1}}\left(d_{t}\right);$
		
		    \State $	t \leftarrow t+1$;
		
	    \EndWhile
		
	\end{algorithmic} 
\end{algorithm}
It is notable that, in Step 6, $\frac{df(x)}{dx*}$ is the Euclidean gradient.Specifically,$ \frac{df(x)}{dx*} $ is the vectorized Euclidean gradient of $W*$,i.e.,$\frac{df(x)}{dx*}=vec\left[ \frac{df(x)}{dW*} \right]$

\subsection{Hybrid Precoding Design}
With Algorithm \ref{decen_algo}, the hybrid precoding design via alternating maximization still requires digital precoding while
fixing analog precoding. This is achieved by Digital precoding, ZF. Combining Algorithm \ref{decen_algo} and ZF digital precoding design, we finally characterize the alternating maximization
algorithm for hybrid precoding in Algorithm \ref{algo_2}.
\begin{algorithm}
	\caption{Manifold Optimization-based Hybrid Precoding}
	\label{algo_2}
	\begin{algorithmic}[1]
	
	\State Input:$W^{(0)},F[k]^{(0)},\epsilon $
    \State Set t=0;
    \State Generate W from the continuous uniform distribution.
	
	\While {$R_{sum}^t-R_{sum}^{t-1} >\epsilon$}
			
		\State Calculate effective channels and  ZF digital precoding $F[k]^{t+1},k=1,2,\ldots,k$ according to\; \;\;\;\;( \ref{4.4}) $\And$  (\ref{4.5});

	    \State Optimize $W^{t+1}$ using Algorithm \ref{decen_algo} when $F[k]^{t+1}$ is fixed;

	\EndWhile
		
	\end{algorithmic} 
\end{algorithm}
\\
\\
\\

\newcommand*\algoframe[5]{
\begin{algorithm}[H]
  \caption{#1}
  \SetAlgoRefName{alg:#1}
  \SetKwProg{Fn}{Function}{}{end}
  \SetKwData{Left}{left}\SetKwData{This}{this}\SetKwData{Up}{up}
  \SetKwFunction{Union}{Union}\SetKwFunction{FindCompress}{FindCompress}
  \LinesNumbered
  \SetKwInOut{Input}{input }
  \SetKwInOut{Output}{output }
  \DontPrintSemicolon
  \Input{#2}
  \Output{#3}
  \SetKwFunction{FMain}{#1}
  \SetKwProg{Fn}{Function}{:}{}
  \Fn{\FMain{#4}}{
      \BlankLine
      #5
  }
\end{algorithm}}
\vspace{-15mm}
\subsection{Complexity Reduction  for Hybrid Precoding Design}

In the previous section, we suggested an iterative technique for designing OFDM-based multiuser MIMO hybrid beamformers. Although it is intended to improve system performance, it necessitates a high level of computational complexity that is not appropriate for actual systems. As a result, this section has included several basic, low-complexity ideas for multiuser MIMO hybrid beamforming across wideband channels. The OFDM-based hybrid precoding issue, like the technique, is divided into two distinct domains, each with its own set of restrictions. The suggested algorithm's fundamental concept is to break the precoder computation into two parts. The analog precoder is built in the first stage to reduce interference and noise interference by selecting the analog precoder, W, and the digital precoder, F, is discovered in the second stage using a similar procedure as previously since it is not as complicated. So, the new objective function is:
\begin{equation}\label{4.13}
\text{Interference} = \sum_{u=1}^U \sum_{\substack{i=0 \\ \text{when } u=q, i\neq k}}^K |S_{i-k}|^2 \; |h_q[i] W F_u[i]|^2.
\end{equation}

\begin{equation}\label{4.14}
\begin{aligned}
\min_W & \sum_{k=1}^K \sum_{u=1}^U \text{Interference} \\
\text{s.t.} & \quad |(w_n)_m| = \frac{1}{\sqrt{N_t}}, \quad \forall 1 \leq n \leq N_{RF}, \; 1 \leq m \leq N_t, \\
& \quad \left\| \|W F_u[k]\|_F^2 \right\| = 1, \quad 1 \leq k \leq K.
\end{aligned}
\end{equation}

Consequently, we are not changing our procedures. Using the Riemannian manifold technique and ZF methodology, we are finding the W and F parameters. However, instead of maximizing data rate R, we minimize interference, which maximizes rate R in the end.
So, the basic concept is to first find the digital precoder with the ZF optimization approach and then the analog precoder with algorithm 1. In analog beamforming design, however, the change is in the objection function, f(x). Instead of considering spectral efficiency as our target function, we focus on interference. Moreover, we have identified the analog precoder, W by minimizing the interference objection function.

%
%
\chapter{Numerical Results} \label{num}
We specified our system parameters and simulation settings in this section. Then, we compare the computational time between the conventional and our suggested method and demonstrate the performance disparity resulting from time minimization. The response of spectral efficiency to variable SNR is then illustrated by varying the ICI coefficient, number of antennas, number of RF chains, and number of users. The significance of ICI, spectral efficiency deterioration induced by ICI, was then demonstrated.
\section{System Parameters $\And$ Simulation Set-up}
The operating frequency is assumed to be 0.35 THz, as higher frequencies than this will result in a larger atmospheric attenuation. A total of 64 subcarriers is assumed. the molecular absorption coefficient is taken $0.0033 m^{-1}$ as of\; \cite{7036065}.  The antenna gain of the transmitter and receiver is taken as identical, which is 20dBi. We have used the Saleh Valenzuela channel model, where cluster-based data streams are transmitted across the subcarriers. We assumed 10 rays embedded in each cluster for generating results. As we are considering an indoor scenario, the distance is set at 5m. In UE, we assumed that just one antenna is available, as UE has power and size constraints.
\\

Matlab is used for all simulations since it is a powerful numerical analysis and optimization-based simulation tool. In a real-world setting, channels change over time; hence, channels are regarded as random and a random channel matrix is constructed in the Matlab environment. Multiple realizations are taken and averaged to get more meticulous outcomes. 
\begin{table*}[ht]
\caption{\large Simulation Parameters}
\centering
\begin{tabular}{lc}
\hline
\textbf{Parameters} & \textbf{Values} \\
\hline
Operating Frequency $f$ & $0.35\; \text{THz}$ \\
Subcarrier $K$ & $64$ \\
Absorption Coefficient $K_{abs}$ & $0.0033 \; \text{m}^{-1}$ \\
No of Data Streams $N_s$ & $9$ \\
Transmit Antenna Gain $G_t$ & $20\; \text{dBi}$ \\
Receive Antenna Gain $G_r$ & $20\; \text{dBi}$ \\
No of Rays in Each Cluster $N_{ray}$ & $10$ \\
Transmit Antennas $N_t$ & $64 \le N_t \le 256$ \\
Receive Antenna $R_x$ & $1$ \\
No of Users $U$ & $1 \le N_t \le 16$ \\
No of RF Chain $N_{RF}$ & $9 \le N_{RF} \le 15$ \\
ICI Coefficient $S$ & $0.1 \le S \le 0.3$ \\
Distance $d$ & $5$ \\
\hline
\end{tabular}
\label{overall_res}
\end{table*}

\newpage

\section{Computational Time Comparison between Proposed and Typical Approaches:}
\begin{figure}[H]
\centerline{\includegraphics[scale=0.9]{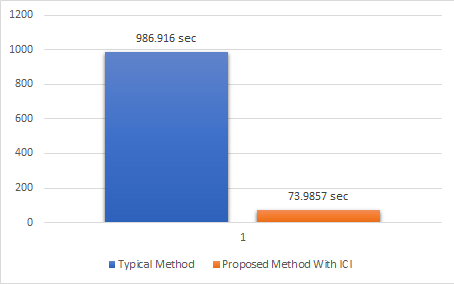}}
\caption{Time variation for the conventional and proposed methods when ICI is taken into account}
\label{time}
\end{figure}

Figure \ref{time} demonstrates that the traditional optimization approach requires almost 13 times longer than the suggested method to solve beamforming for each new channel gain. In this context, conventional optimization refers to the technique optimized across the bit rate equation. In contrast, the suggested approach optimizes by reducing the interference portion of the bit rate equation. This implicitly increases the bit rate and decreases computational complexity, significantly reducing optimization time. For the conventional method, beamforming for each new channel gain required 986.916 seconds, but our updated system requires 73.9857 seconds. Therefore, our system minimizes latency, and we must check the system's performance before adopting it.\\
\newpage

\section{Performance Comparison between typical and proposed method with ICI:}
The proposed solution reduces the time by 13 times, despite minimal performance degradation. In the following figure \ref{performance}, it can be understood that the difference between our suggested's spectral efficiency system and the standard technique with varying SNR is not very significant; both are near. At low SNR, the difference is too small to see, while at high SNR, it is discernible yet near enough to allow the system to settle down. Therefore, We will use this system as it effectively decreases latency without significantly hurting performance.
\\
For generating figure \ref{performance}, transmit antennas and number of RF chains are taken as 64 and 10 respectively as shown in table \ref{performance_table}. Moreover for generating multi-user environment number of users is taken as 9. And ICI coefficient is taken as 0.3.

\begin{table*}[ht]
\caption{Simulation Parameters for figure \ref{performance}}
\centering
\begin{tabular}{lcc}
\hline
Parameters          &             Values
\\ 
\hline
Transmit Antennas $N_t$ & ${64}$
\\
No of Users $U$ & ${9}$
\\
No of RF Chain $N_{RF}$ & ${10}$
\\
ICI Coefficent $S$ & ${0.3}$
\\
\hline
\end{tabular}

\label{performance_table}
\end{table*}

\begin{figure}[H]
\centerline{\includegraphics[width=0.8\textwidth]{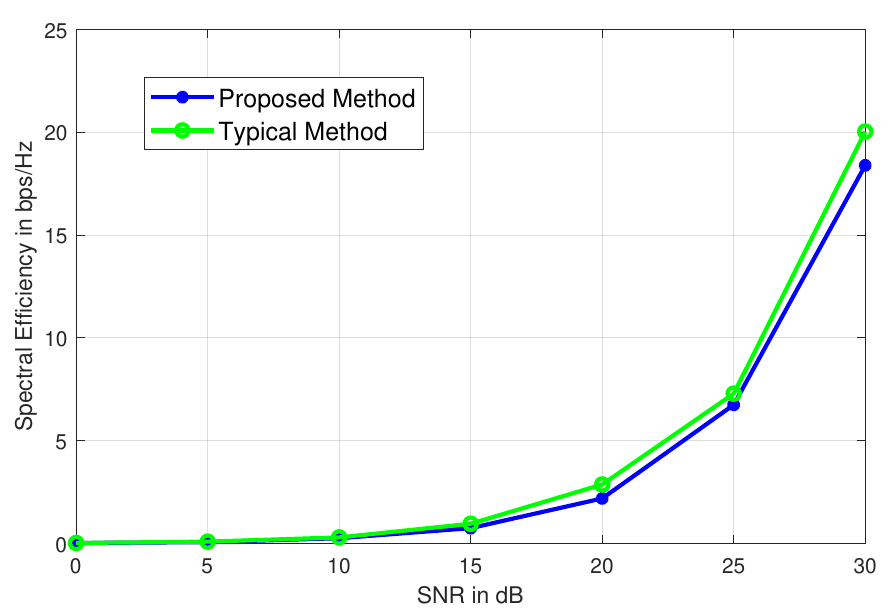}}
\caption{Performance comparison between typical and proposed method}
\label{performance}
\end{figure}

\clearpage 

\section{Significance of ICI Consideration}

\begin{figure}[H]
\centerline{\includegraphics[width=0.8\textwidth]{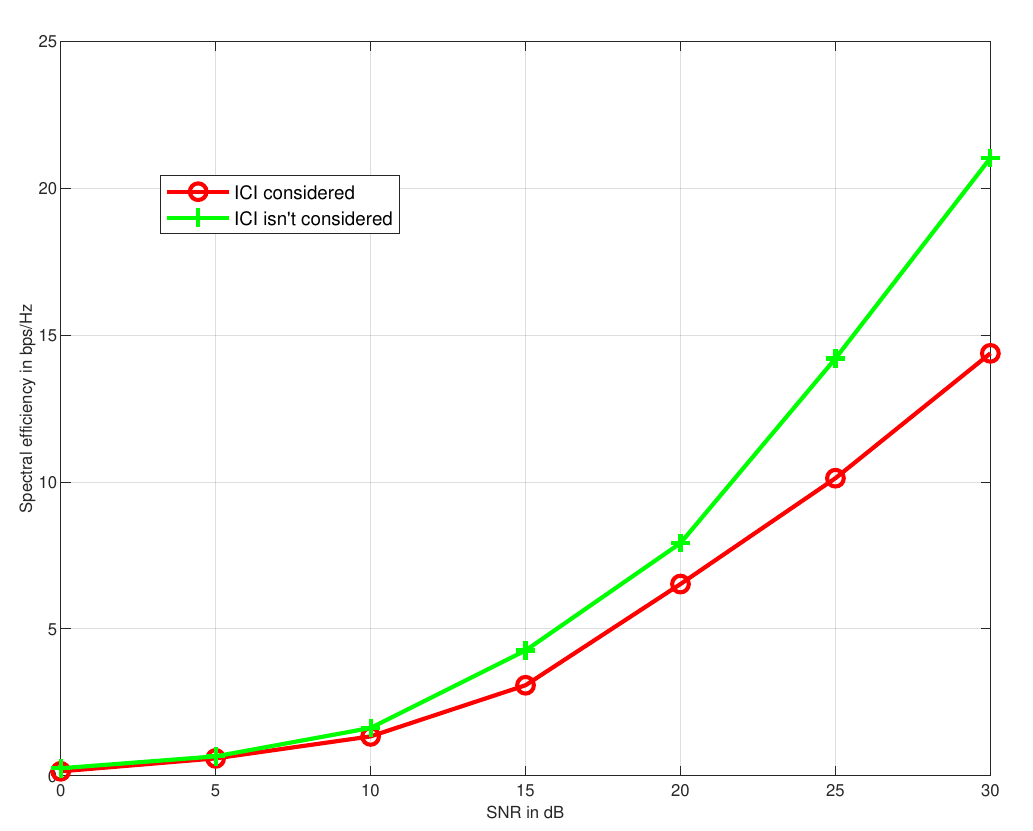}}
\caption{Effect of ICI}
\label{icivary}
\end{figure}

The graphic \ref{icivary} illustrates how much of a difference it makes to consider ICI versus not taking it into account. The ICI coefficient is considered 0.3 in this figure. We get 14 bps/Hz spectrum efficiency without ICI at an SNR of 25 dB and 10 bps/Hz spectral efficiency when ICI is taken into account. This means that at the 25 dB SNR level, the difference is 4 bps/Hz. The bigger the SNR, the greater the disparity. In the absence of ICI, spectral efficiency seems to be better as SNR changes; however, the true spectral efficiency is lower, and we shall witness the difference between simulation and practical implementation when we implement it. This gap will only impact ICI if the other parameters are assumed to be optimal. Consequently, this distinction must be considered and ICI must be considered.
\\
The specific simulation parameters for generating figure \ref{icivary} is given in table \ref{icivary_table}. Here number of transmit antennas taken as 169 and number of RF chain considered as 9.
\\
\begin{table*}[ht]
\caption{Simulation Parameters for figure \ref{icivary}}
\centering
\begin{tabular}{lcc}
\hline
Parameters          &             Values
\\ 
\hline
Transmit Antennas $N_t$ & ${169}$
\\
No of Users $U$ & ${9}$
\\
No of RF Chain $N_{RF}$ & ${8}$
\\
ICI Coefficent $S$ & ${0.3}$
\\
\hline
\end{tabular}

\label{icivary_table}

\end{table*}

\section{Importance of ICI coefficient}

\begin{figure}[H]
\centerline{\includegraphics[width=0.8\textwidth]{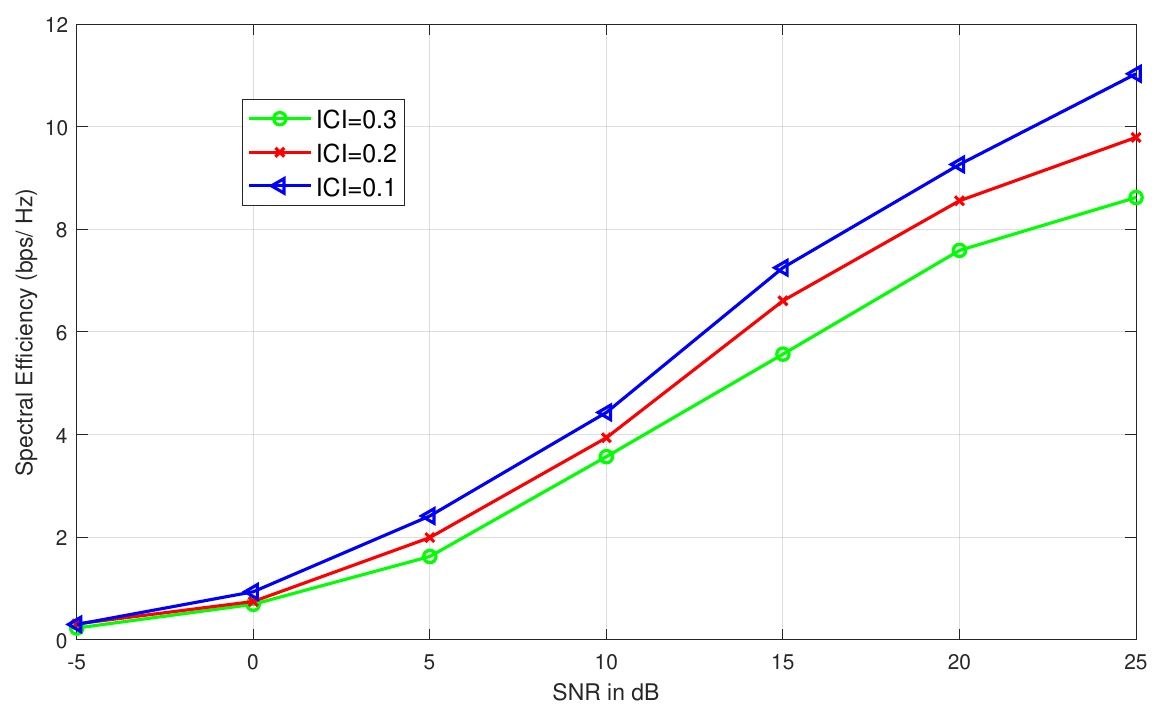}}
\caption{Varying ICI}
\label{ici}
\end{figure}

The impact of the ICI coefficient is seen in figure \ref{ici}. An increase in the ICI coefficient indicates an increase in the ICI effect, and as a result, the spectrum efficiency decreases. Conversely, a decrease in the ICI coefficient indicates a modest ICI impact, so the spectra efficiency increases. The best spectral efficiency is achieved when the ICI co-efficient is as low as possible (0.1 for this figure). The spectrum efficiency is tense to 11 bps/Hz at a 25 dB SNR level, and the ICI co-efficient of 0.3 results in the lowest spectral efficiency, which is  8.2 bps/Hz. If ICI=0, it means that ICI consideration is not necessary. If ICI is near zero, the result is closer to spectral efficiency without taking ICI into account. The difference between ICI consideration and non-consideration grows as ICI grows.
\\
The specified simulation parameter for generating figure \ref{ici} is given in table \ref{ici_table}. The ICI coefficient is varied from 0.1 to 0.3 to generate the figure as shown in table \ref{ici_table}.

\begin{table*}[ht]
\caption{Simulation Parameters for Figure \ref{ici_table}}
\label{ici_table}
\centering
\begin{tabular}{lcc}
\hline
Parameters & Values \\
\hline
Transmit Antennas $N_t$ & $169$ \\
Number of Users $U$ & $9$ \\
Number of RF Chains $N_{RF}$ & $8$ \\
ICI Coefficient $S$ & $0.1 \leq S \leq 0.3$ \\
\hline
\end{tabular}
\end{table*}

\clearpage 


\section{Effect of varying Antennas}

\begin{figure}[H]
\centerline{\includegraphics[width=0.8\textwidth]{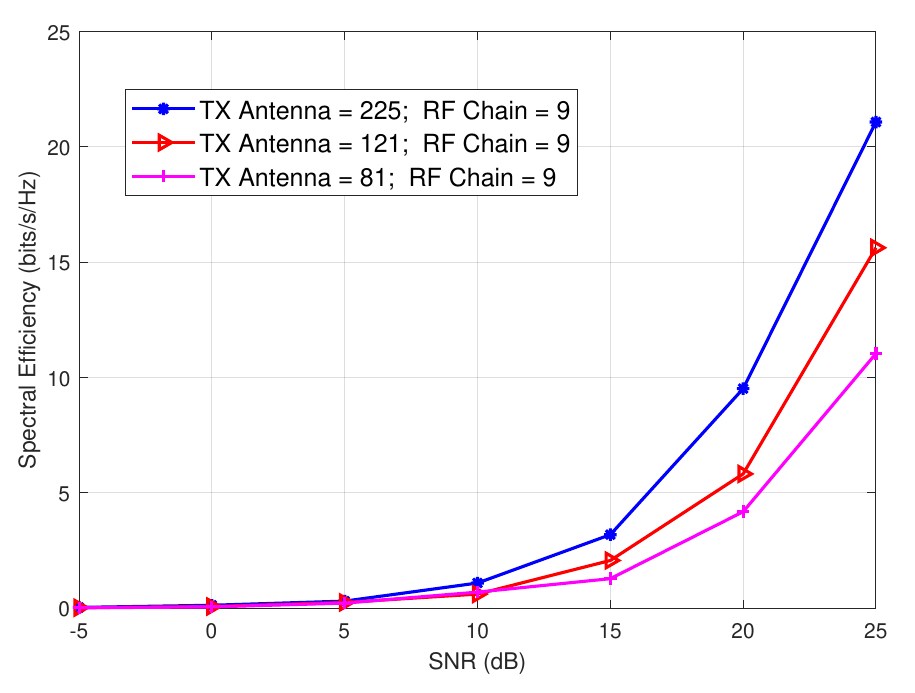}}
\caption{Varying Transmit Antenna}
\label{antennavary}
\end{figure}

It is feasible to measure the impact of antennas by establishing the other settings and then modifying the number of antennas. All other parameters in figure (\ref{antennavary}) have been maintained constant except for the number of antennas. Spectral efficiency improves as the number of antennas increases, but spectral efficiency decreases as the number decreases. The figure shows the best spectral efficiency for 225 antennas and the lowest spectral efficiency for 81 antennas. At a 25 dB SNR level, 225 antennas show about 21 bps/Hz, 121 antennas show around 15.5 bps/Hz, and 81 antennas show around 11 bps/Hz. The spectral efficiency will rise in direct proportion to the increase in the number of antennas.
\\
the specified simulation parameter for generating figure \ref{antennavary} is given in table \ref{antennavary_table}. The number of antennas ranged from 81 to 255, and the figure \ref{antennavary} represents the response of spectral efficiency at various antennas.
\\
\begin{table*}[ht]
\caption{Simulation Parameters for Figure \ref{antennavary}}
\label{antennavary_table}
\centering
\begin{tabular}{lcc}
\hline
Parameters & Values \\
\hline
Transmit Antennas $N_t$ & $81 \leq N_t \leq 255$ \\
Number of Users $U$ & $9$ \\
Number of RF Chains $N_{RF}$ & $9$ \\
ICI Coefficient $S$ & $0.3$ \\
\hline
\end{tabular}
\end{table*}

\section{Effect of varying RF chain}

\begin{figure}[H]
\centerline{\includegraphics[width=0.8\textwidth]{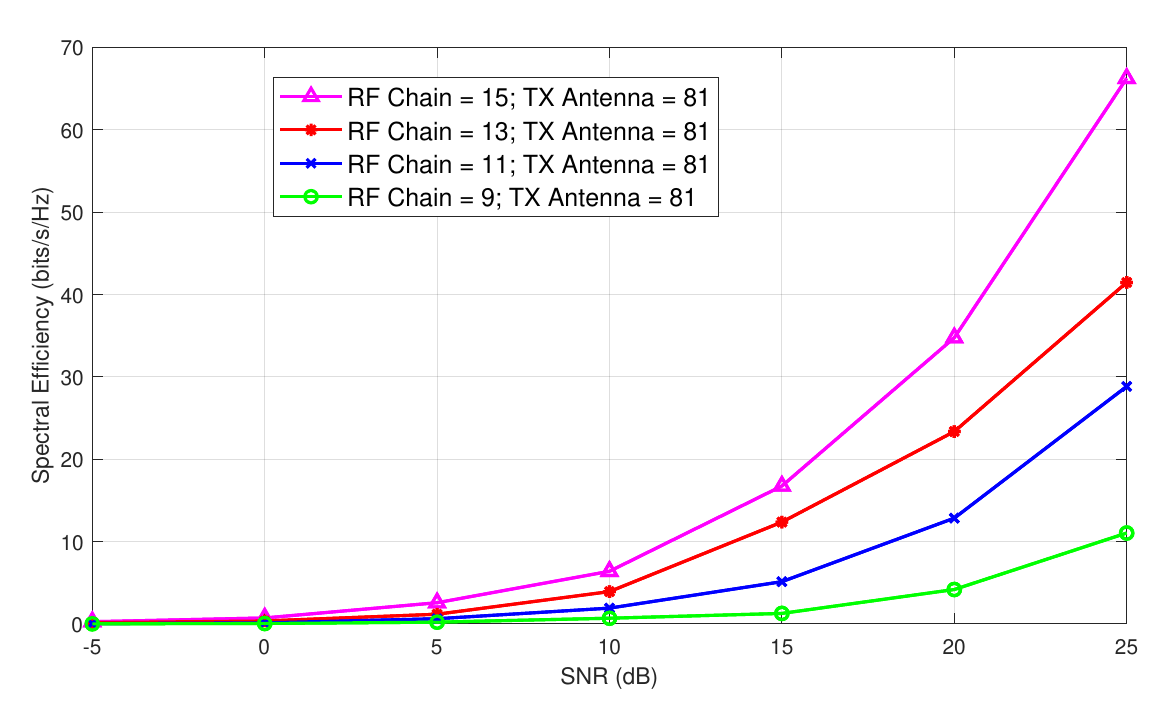}}
\caption{Varying RF Chain}
\label{rfvary}
\end{figure}

It is feasible to determine the impact of the RF chain by fixing all other parameters and then modifying the RF chain. Except for the number of RF chains, all other factors in the accompanying figure remain constant. In this illustration, 15 RF chains have the maximum spectral efficiency, whereas 9 RF chains have the lowest. 15 RF chains display approximately 35 bps/Hz, 13 RF chains display approximately 24 bps/Hz, 11 RF chains display approximately 13 bps/Hz, and 9 RF chains display approximately 5 bps/Hz. Increasing the number of radio frequency (RF) chains improves spectral efficiency. The image illustrates that the RF chain has a greater influence than the number of antennas. The issue, though, is the price. RF is far more expensive than antennas. So, RF chain selection is required to trade off between performance and cost.
\\
the specified simulation parameter for generating figure \ref{rfvary} is given in table \ref{rfvary_table}. The number of RF chains ranged from 81 to 255, and the figure \ref{rfvary} represents the response of spectral efficiency at various number of RF chains.
\\
\begin{table*}[ht]
\caption{Simulation Parameters for Figure \ref{rfvary}}
\label{rfvary_table}
\centering
\begin{tabular}{lcc}
\hline
Parameters & Values \\
\hline
Transmit Antennas $N_t$ & $81$ \\
Number of Users $U$ & $9$ \\
Number of RF Chains $N_{RF}$ & $9 \leq N_{RF} \leq 15$ \\
ICI Coefficient $S$ & $0.3$ \\
\hline
\end{tabular}
\end{table*}

\newpage

\section{Combined effect of RF chain and the number of antennas}

\begin{figure}[H]
\centerline{\includegraphics[width=0.8\textwidth]{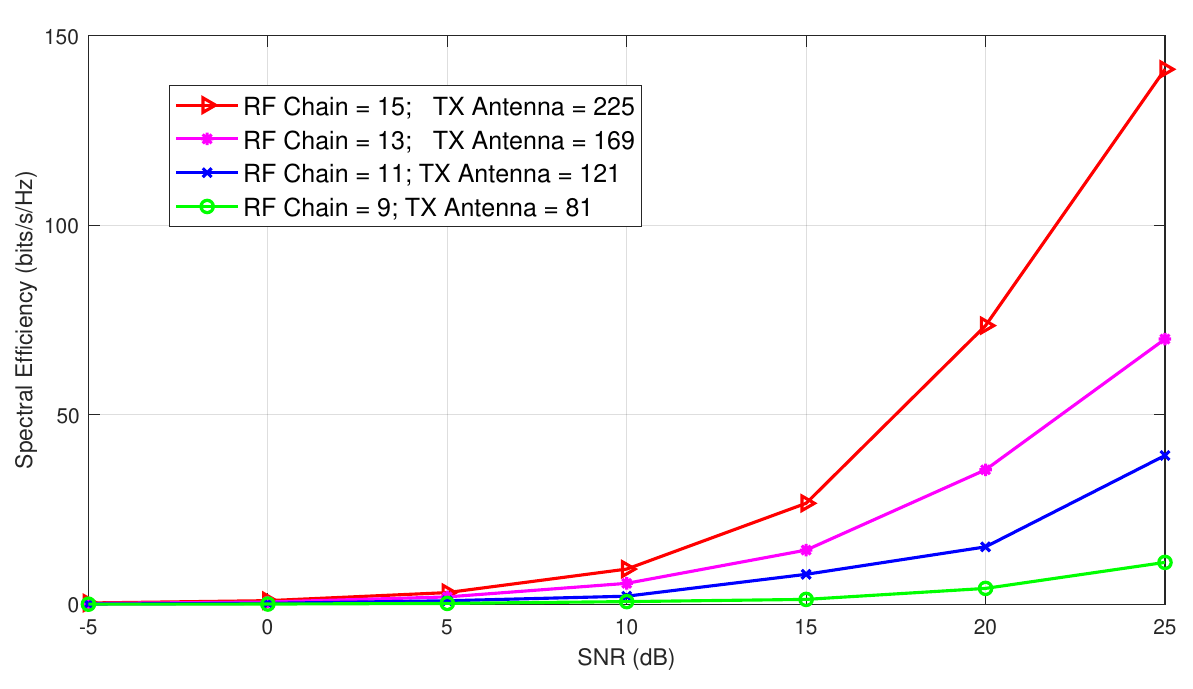}}
\caption{Varying RF Chain and antenna}
\label{varyrfant}
\end{figure}

As base stations can be employed with lots of antennas in massive MIMO downlink transmission and the RF chain is costly, desirable spectral efficiency can be achieved by modifying the RF chain and antenna count. Variable RF chains and the number of TX antennas allow more spectral efficiency flexibility. Figure \ref{varyrfant} demonstrates that by increasing the number of RF chains and antennas, spectral efficiency can be significantly enhanced, which is not possible by only altering the number of RF chains or antennas.
\\
the specified simulation parameter for generating figure \ref{varyrfant} is given in table \ref{varyrfant_table}.
The number of antennas and RF chains varied from 81 to 225 and 9 to 15 respectively to generate figure \ref{varyrfant}. Figure \ref{varyrfant} demonstrates that instead of altering either antenna or only RF chain, simultaneously varying RF chain and antennas resulted in greater spectral efficiency and a smoother SNR vs Spectral Efficiency curve. 
\\

\begin{table*}[ht]
\caption{Simulation Parameters for Figure \ref{varyrfant}}
\label{varyrfant_table}
\centering
\begin{tabular}{lcc}
\hline
Parameters & Values \\
\hline
Transmit Antennas $N_t$ & $81 \leq N_t \leq 255$ \\
Number of Users $U$ & $9$ \\
Number of RF Chains $N_{RF}$ & $9 \leq N_{RF} \leq 15$ \\
ICI Coefficient $S$ & $0.3$ \\
\hline
\end{tabular}
\end{table*}

\newpage

\section{Effect of Varying Number of Users}

\begin{figure}[H]
\centerline{\includegraphics[width=0.8\textwidth]{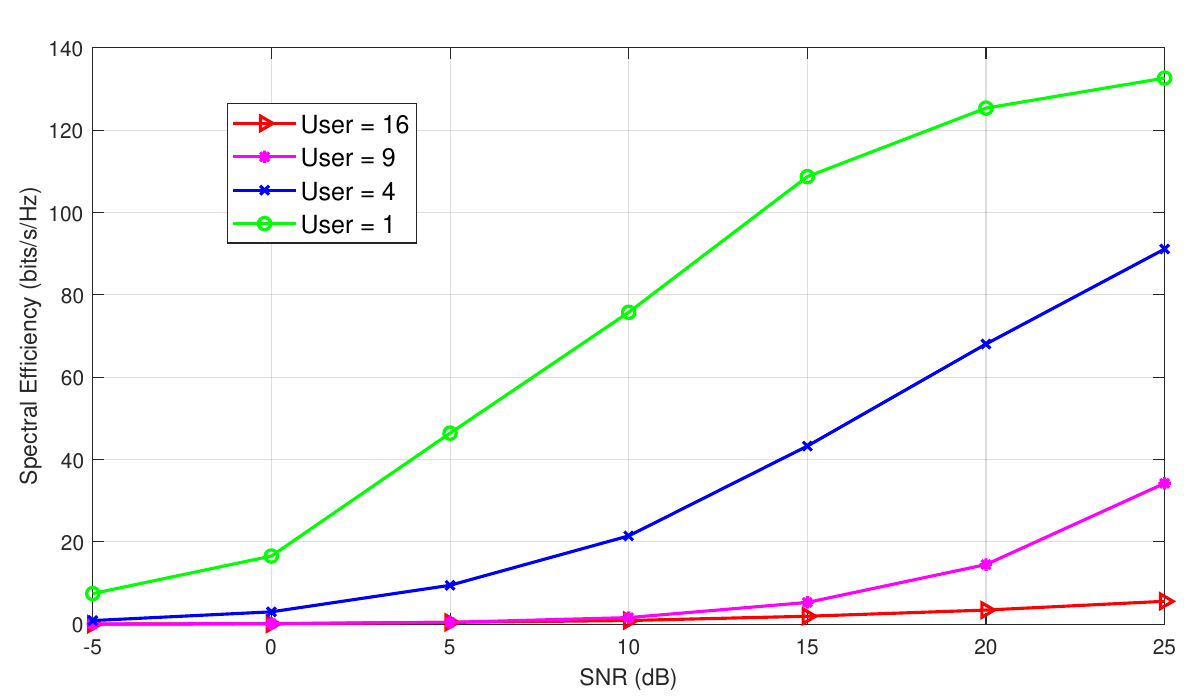}}
\caption{Varying RF Chain and antenna}
\label{uservary}
\end{figure}

By holding the other variables constant and modifying the users, it is feasible to quantify the impact of the users. Except for users, all parameters in the accompanying figure \ref{uservary} have been maintained constant. There will be less interference if there are fewer users; hence spectral efficiency will be high. As the number of users increases, interference also increases, resulting in a decrease in spectral efficiency. In the case of a single user, interference is limited to other subcarriers of the same user. Therefore, interference is minimal and spectral efficiency is optimal. The spectral efficiency for a single user is the highest, while the spectral efficiency for 16 is the lowest. For a 20 dB SNR level, a single user will display roughly 125 bps/Hz, four users will display around 70 bps/Hz, nine users will display around 18 bps/Hz, and sixteen users will display around 4 bps/Hz.
\\
The specified simulation parameter for generating figure \ref{uservary} is given in table \ref{uservary_table}. The number of users varied from 1 to 16. For single-user cases, the spectral efficiency value is maximum than the other cases. The transmit antennas and number of RF chains are taken as 81 and 9, respectively.

\begin{table*}[ht]
\caption{Simulation Parameters for Figure \ref{uservary}}
\label{uservary_table}
\centering
\begin{tabular}{lcc}
\hline
Parameters & Values \\
\hline
Transmit Antennas $N_t$ & $81$ \\
Number of Users $U$ & $1 \leq U \leq 16$ \\
Number of RF Chains $N_{RF}$ & $9$ \\
ICI Coefficient $S$ & $0.3$ \\
\hline
\end{tabular}
\end{table*}

\newpage

\section{Performance gap if the problem is solved without considering ICI at different ICI coefficient: }

\begin{figure}[h]
\centerline{\includegraphics[width=0.7\textwidth]{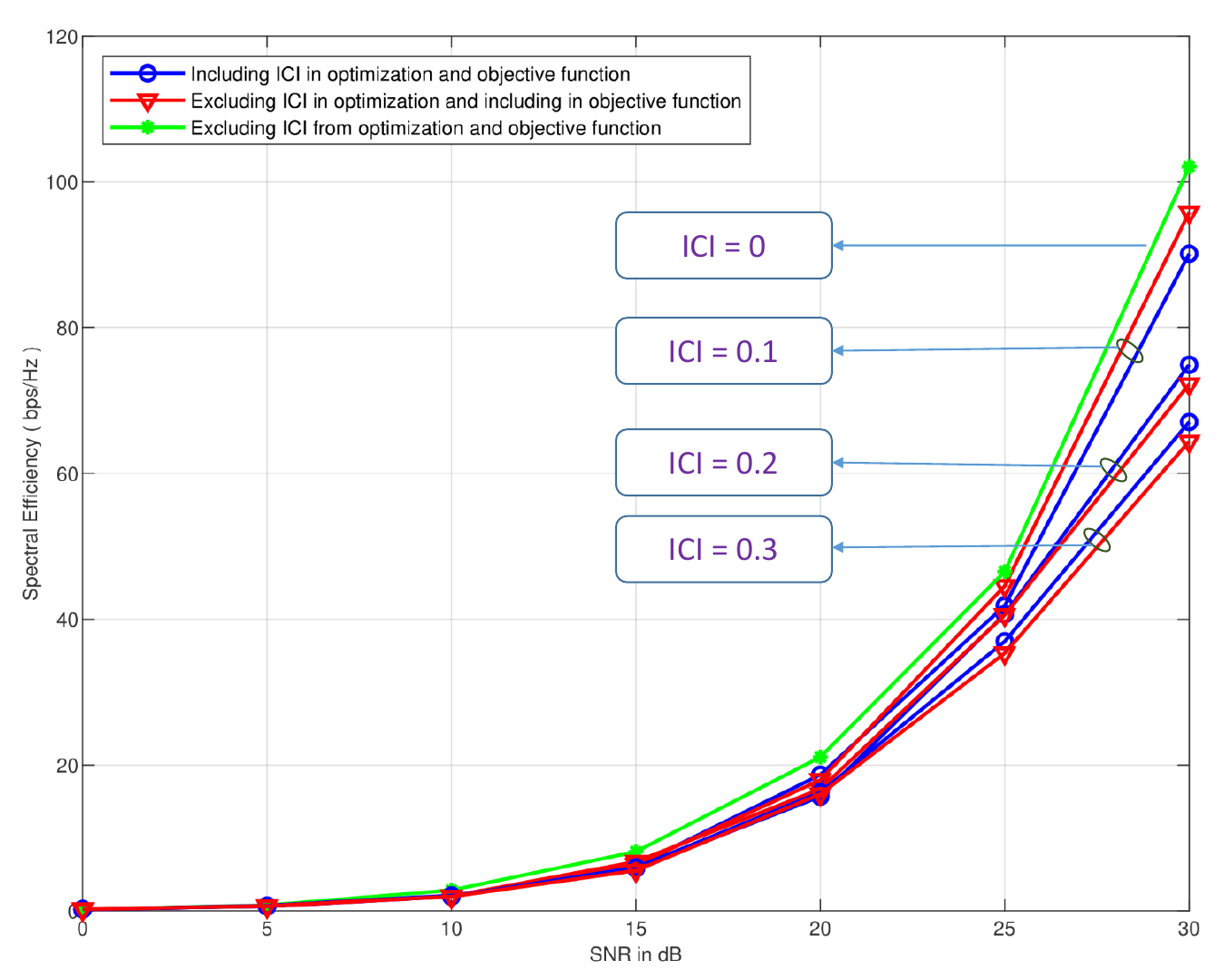}}
\caption{Varying ICI for three cases}
\label{icimul}
\end{figure}

With the ICI coefficient equal to zero, the spectral efficiency is also equal to what it would be if the ICI situation was not considered. As the coefficient of ICI grows, spectral efficiency drops and the difference from the situation without ICI consideration increases. Excluding ICI in optimization and including it in the objection function scenario is significantly closer to the case without ICI for lower ICI levels. With rising ICI, however, the Including ICI in optimization and objection function and Excluding ICI in optimization but Including ICI in objection function case situations become tensely similar.
\\
The specified simulation parameter for generating figure \ref{icimul} is given in table \ref{icimul_table}. Here, ICI coefficient is varied form 0 to 0.3. Zero ICI coefficient indicates a system with no Inter Carrier Interference.

\begin{table*}[ht]
\caption{Simulation Parameters for Figure \ref{icimul}:}
\label{icimul_table}
\centering
\begin{tabular}{lcc}
\hline
Parameters & Values \\
\hline
Transmit Antennas $N_t$ & $64$ \\
Number of Users $U$ & $9$ \\
Number of RF Chains $N_{RF}$ & $15$ \\
ICI Coefficient $S$ & $0 \leq \text{ICI} \leq 0.3$ \\
\hline
\end{tabular}
\end{table*}

\newpage

\section{Performance gap if the problem is solved without
considering ICI with varying TX antennas:}

\begin{figure}[h]
\centerline{\includegraphics[trim= 0 0 0 0cm, width=0.7\textwidth,clip]{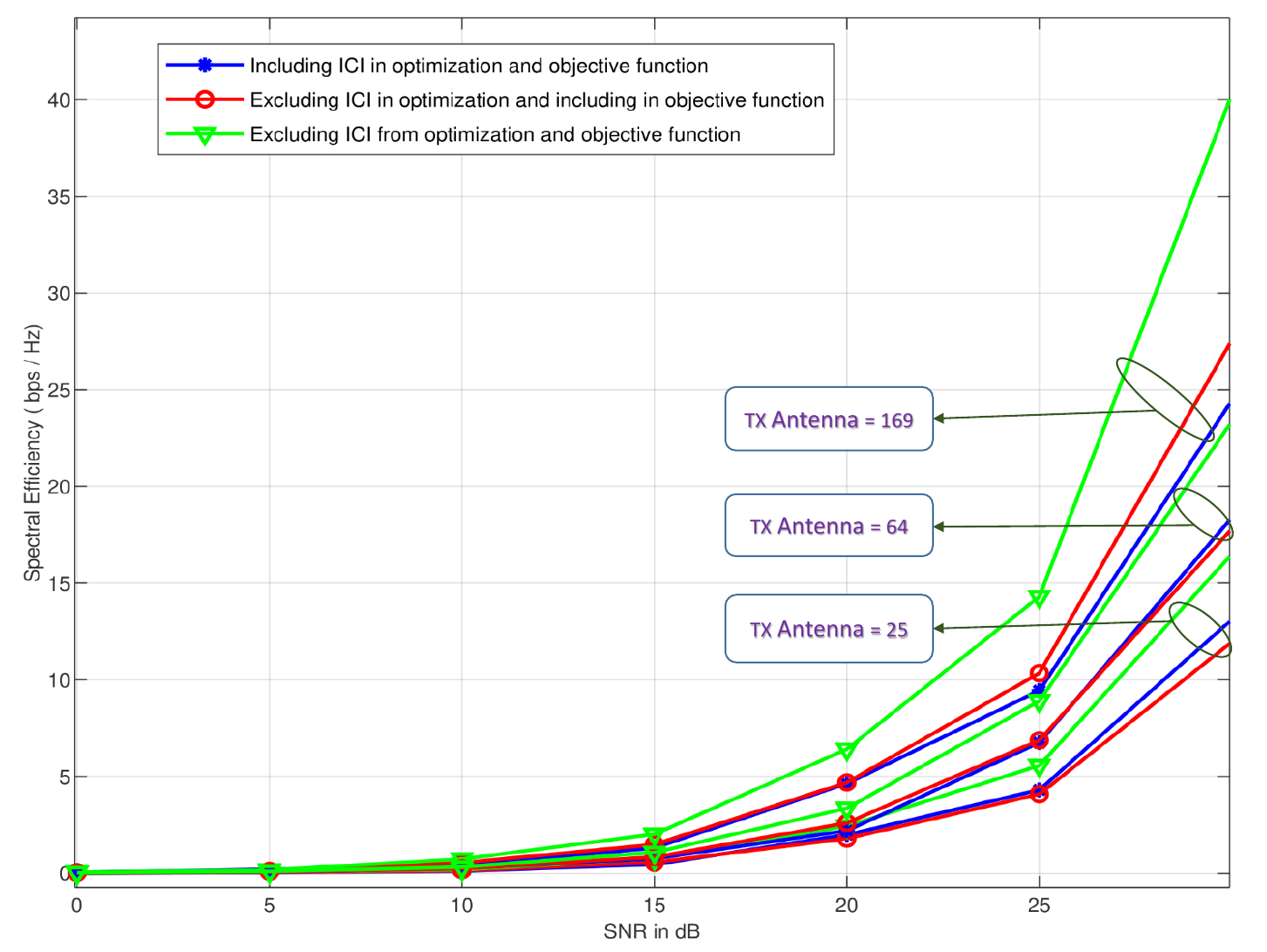}}
\caption{Varying transmit antenna for three cases}
\label{varyingTX_2ndpg}
\end{figure}

In \ref{varyingTX_2ndpg} for each antenna, we have created three graphs (Including ICI in optimization and objection function, Excluding ICI in optimization but including in objection function and Excluding ICI in optimization and objective function). Increasing the number of antennas enhances spectral efficiency in every circumstance. For lower SNRs, there is minimal difference in spectral efficiency between including ICI in the optimization case and excluding ICI from optimization but including it in the objective function case. Also, the difference between the instances is insignificant for small numbers of antennas. Still, as the number of antennas grows, the optimization case without ICI in optimization but including in objective function has greater spectral efficiency than the one with ICI in the optimization case for larger SNRs scenarios. As SNR grows, the differences between spectral efficiency with and without ICI increase. Moreover, the disparity between the result with ICI and the result without ICI increases with the number of antennas.
The specified simulation parameter for generating figure \ref{varyingTX_2ndpg} is given in table \ref{varyingTX_2ndpg_t}. Here, the number of 
transmit antennas is varied from 25 to 169.

\begin{table*}[ht]
\caption{Simulation Parameters for Figure \ref{varyingTX_2ndpg}}
\label{varyingTX_2ndpg_t}
\centering
\begin{tabular}{lc}
\hline
Parameters & Values \\
\hline
ICI coefficient $S$ & $0.3$ \\
Number of Users $U$ & $9$ \\
Number of RF Chains $N_{RF}$ & $10$ \\
Transmit Antennas $N_t$ & $25 \leq N_t \leq 169$ \\
\hline
\end{tabular}
\end{table*}

\newpage

\section{Performance gap if the problem is solved without
considering ICI with varying RF chain:}

\begin{figure}[h]
\centering
\includegraphics[width=0.8\textwidth]{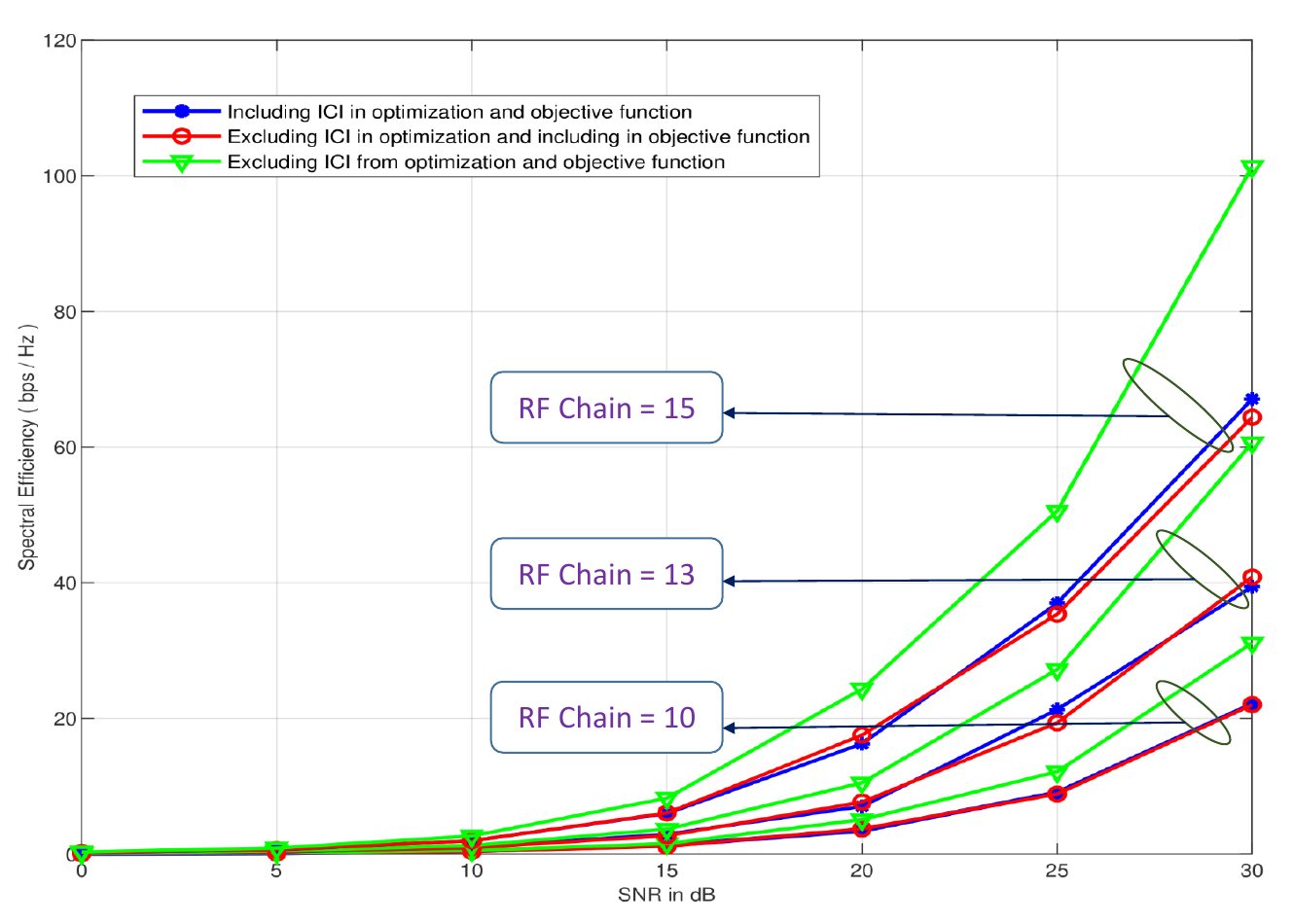}
\caption{Varying RF chain for three cases}
\label{varying_RF_2ndfg}
\end{figure}

In figure \ref{varying_RF_2ndfg} for each RF chain, we have created three graphs (Including ICI in optimization and objection function, Excluding ICI in optimization but including in objection function and Excluding ICI in optimization and objective function). According to this graph, the most significant difference between ICI and non-ICI is seen when the number of RF chains is the greatest and the lowest when the number of RF chains is at its minimum. Increasing the number of RF chains increases spectral efficiency in each case. For lower SNRs, there is minimal difference in spectral efficiency between Including ICI in optimization and Excluding ICI from optimization but including it in objective function scenarios. However, higher RF chains and higher SNR situations suggest that integrating ICI in optimization produces marginally better results than the other. Moreover, the difference between them for lower RF chains is minimal. But Increasing the RF chain significantly increases the disparity between the ICI and non-ICI results.
The specified simulation parameter for generating figure \ref{varying_RF_2ndfg} is given in table \ref{overall_res01}. Here, the number of RF chains is varied from 10 to 15.

\begin{table*}[ht]
\caption{Simulation Parameters for Figure \ref{varying_RF_2ndfg}}
\centering
\begin{tabular}{lc}
\hline
\textbf{Parameters} & \textbf{Values} \\
\hline
ICI Coefficient $S$ & $0.3$ \\
No of Users $U$ & $9$ \\
Transmit Antennas $N_t$ & $64$ \\
No of RF Chains $N_{RF}$ & $10 \le N_{RF} \le 15$ \\
\hline
\end{tabular}
\label{overall_res01}
\end{table*}

\section{Performance gap if the problem is solved without
considering ICI with varying number of users:}

\begin{figure}[H]
\centerline{\includegraphics[trim= 0 0 0 0.4cm, width=0.8\textwidth,height=0.8\textheight,keepaspectratio,clip]{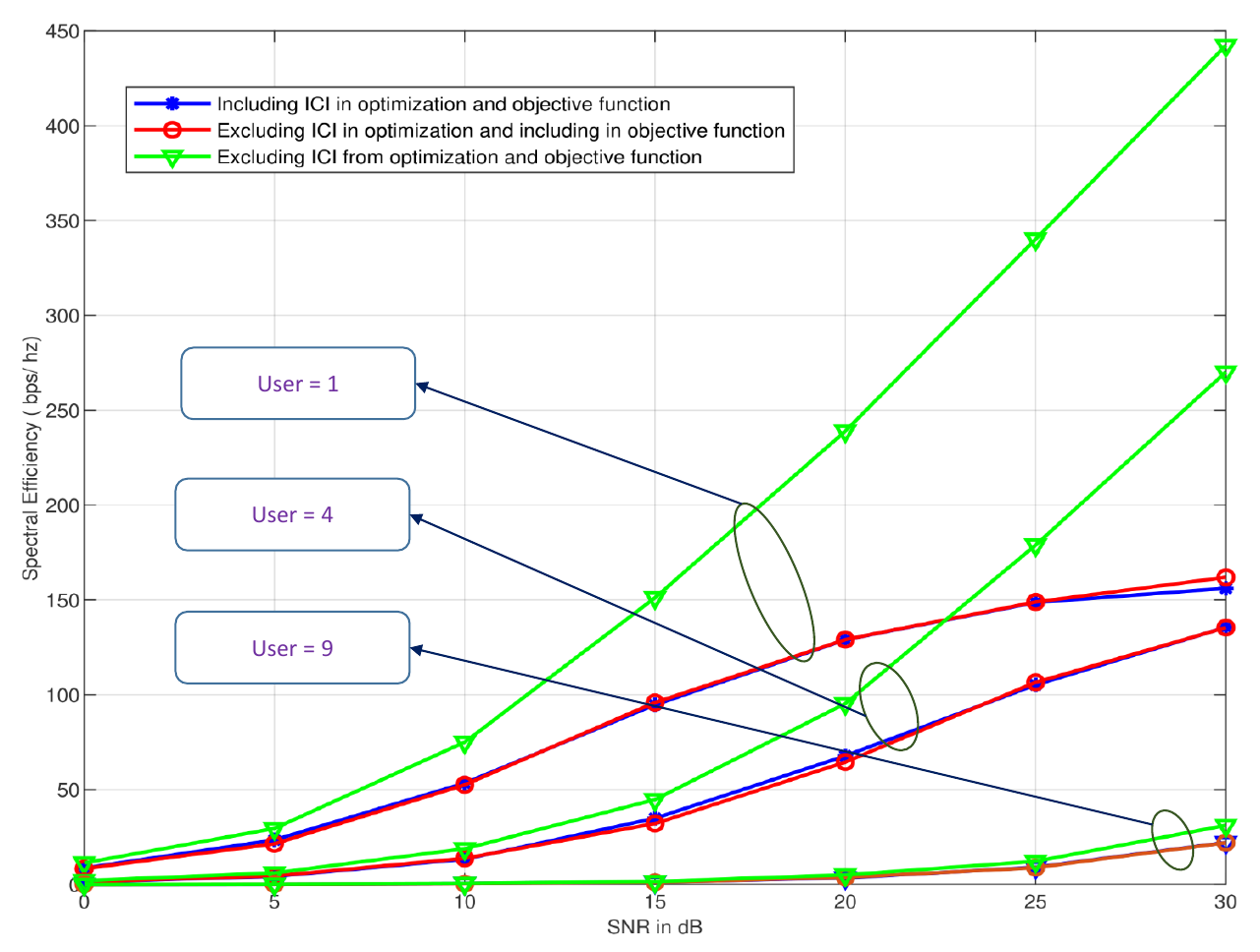}}
\caption{Varying user for three cases}
\label{varying_user_2ndfg}
\end{figure}

If there are fewer users, there will be less interference; hence, spectral efficiency will be very high. As the number of users increases, interference also increases, leading to a decrease in spectral efficiency. The diagram \ref{varying_user_2ndfg} illustrates the number of users affected by three distinct scenarios. There is no substantial difference in spectral efficiency between incorporating ICI in the optimization case and removing ICI from optimization but including it in the objection function case for lower SNRs, as shown by the graph. For higher users, the difference between them is insignificant. However, for lower users, it is evident. For single-user cases excluding ICI in optimization but including in objective function case is slightly higher for higher SNRs than including ICI in optimization case. As the number of users increases, interference and spectral efficiency decrease in all cases, but the discrepancy between the scenarios narrows. When the number of users drops, the performance gap between those who consider ICI and those who do not increase significantly. The specified simulation parameter for generating figure \ref{varying_user_2ndfg} is given in table \ref{overall_res02}. Here, the number of users is varied from 1 to 9.

\begin{table*}[ht]
\caption{Simulation Parameters for Figure \ref{varying_user_2ndfg}}
\label{overall_res02}
\centering
\begin{tabular}{lcc}
\hline
Parameters & Values \\
\hline
ICI coefficient $S$ & $0.3$ \\
Number of RF Chains $N_{RF}$ & $10$ \\
Transmit Antennas $N_{t}$ & $64$ \\
Number of Users $U$ & $1 \leq U \leq 9$ \\
\hline
\end{tabular}
\end{table*}

\section{Deterioration of performance with increasing distance:}

\begin{figure}[h]
\centerline{\includegraphics[width=0.8\textwidth]{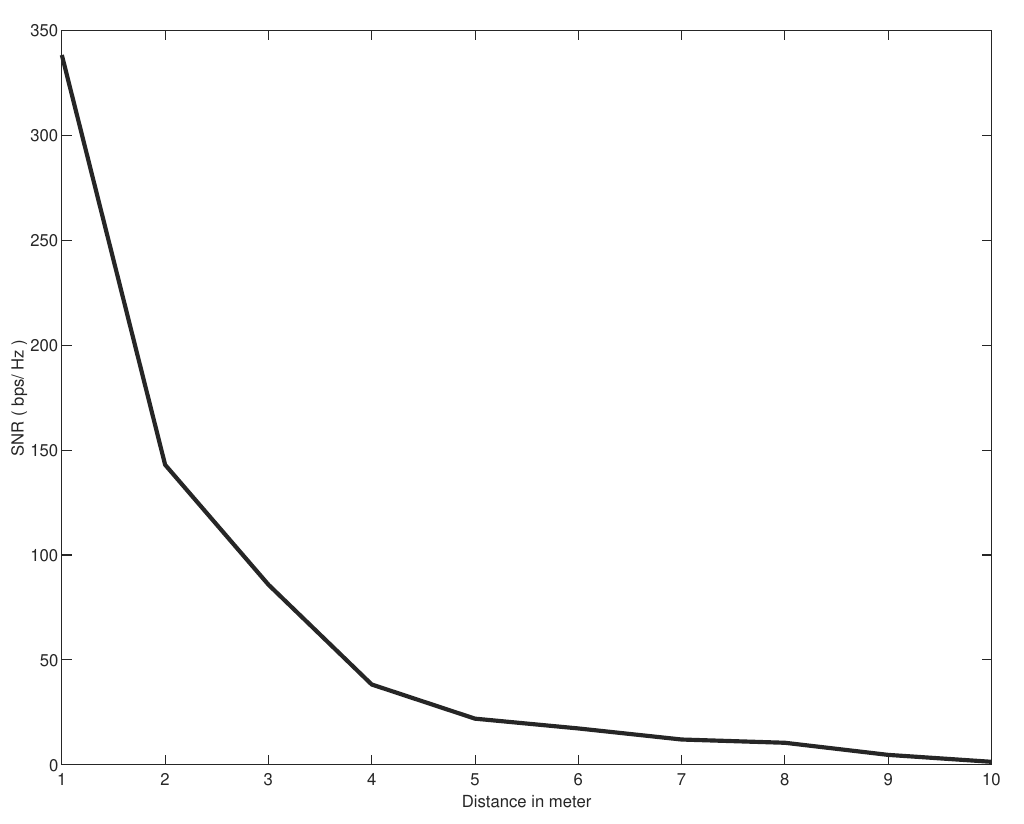}}
\caption{Varying distance with SNR}
\label{varying_distance_2ndfg}
\end{figure}

Due to its extremely short wavelength, the THz channel band has more molecular loss over long distances. As a result of introducing the THz channel band into our system, it exhibits more attenuation at greater distances. Figure \ref{varying_distance_2ndfg} demonstrates that the spectral efficiency falls dramatically with distance and is excessively descending at distances higher than 5m. Therefore, it performs well in interior scenarios when a high data rate is required at a short distance and the distance is not too much. The specified simulation parameter for generating figure \ref{varying_distance_2ndfg} is
given in table \ref{overall_res}. Here, the distance is varied from 1 to 10 meters.

\begin{table*}[ht]
\caption{Simulation Parameters for Figure \ref{varying_distance_2ndfg}}
\label{overall_res}
\centering
\begin{tabular}{lcc}
\hline
Parameters & Values \\
\hline
ICI coefficient $S$ & $0.3$ \\
Number of RF Chains $N_{RF}$ & $10$ \\
Transmit Antennas $N_{t}$ & $64$ \\
SNR(dB) & $15$ \\
Distance $d$ & $1 \leq d \leq 10$ \\
\hline
\end{tabular}
\end{table*}

\newpage
\chapter{Conclusion and Future Work} \label{conclu}
\section{Conclusion}
The globe is now working on deploying 5G. However, more than 5G will be required to meet our needs in the next ten years. As a result, we will require 6G by 2030. Deep learning and big data analytics will be integrated into the sixth generation of cellular networks, bringing together hitherto unrelated technologies. A significant driver of 6G is the requirement to deploy edge computing to provide overall throughput and low latency for ultrareliable, low-latency communications solutions. Another driving force will be the Internet of Things (IoT), which will require support for machine-to-machine connectivity. While edge computing resources may handle some IoT and mobile technology data, much of it will require more centralized High-Performance Computing (HPC) resources to process. We are therefore developing 6G technology.

In chapter \ref{model}, we have developed a model that incorporates massive MIMO, THz channel, and hybrid beamforming techniques for simultaneously transmitting bits to multiple users. For hybrid beamforming design, we have employed the fully-connected technique, while the indoor THz model is utilized for the channel model. These methods are required for 6G technology.

In chapter \ref{probform}, we have created the SINR model, from which we have derived the problem statement we must solve. This issue can be resolved using Riemannian manifold optimization and the Zero Forcing technique. However, to reduce complexity and latency, we have changed the conventional method for determining the parameters of hybrid beamforming with a negligible drop in performance. Instead of raising the achievable rate, interference is reduced, which leads to a slight drop in performance. Our suggested method produces exceptional spectrum efficiency, particularly at high SNR. In chapter \ref{num}, we have demonstrated through numerical findings why ICI is vital for MIMO-OFDM wireless communication and why our method considers it. We also have demonstrated how spectral efficiency increases as the number of RF chains and antennas grows but falls when the number of users increases. We have also shown how considering ICI versus not considering ICI affects the RF chain and antenna count. Finally, the influence of distance on spectral efficiency is also shown.

Consequently, we conclude that considering ICI is essential for massive MIMO-OFDM THz wireless systems with hybrid beamforming for multiple users.

\section{ Future Work}
Inappropriate latency is one of the major concerns with our method. We have drastically reduced latency by altering the system, yet more is needed for practical deployment. Reduce latency to a few microseconds for practical implementation. One of the 6G internet objectives is facilitating communications with one-microsecond latency. We will endeavor to reduce latency and increase the system's realism in the future. We have demonstrated how ICI has affected our results in the results section. In order to increase spectral efficiency, we will therefore create a method to reduce the ICI effect for THz transmission. We have poor spectral efficiency at low SNR. Assuming a constant power allocation for each subcarrier, we have yet to achieve the expected spectral efficiency at lower SNRs. Using adaptive power allocation approaches, we will strive to increase spectral efficiency at lower SNRs in the future. As artificial intelligence (AI) is essential for the next generation, we will be aiming to incorporate machine learning and deep learning (ML/DL) soon. Our task has been completed in the physical layer. We will integrate the Data Link Layer and attempt to design the system with scheduling and modulation approaches. We have utilized the Shannon capacity formula to calculate spectral efficiency, but it is impossible in practice; we must incorporate real-world circumstances. We will work on this process in the future.

\appendix

\bibliographystyle{ieeetr}
\bibliography{bibliography}
\end{document}